\documentclass[twoside]{elsart_mm}

\usepackage{fancyhdr}
\headsep=\baselineskip
\usepackage{ARAstroBib}
\usepackage{amssymb}
\usepackage{amsmath}
\usepackage[section]{placeins}
\usepackage[expert,z-times,vw-times]{tmsfnts}

\usepackage{graphicx}
\usepackage[breaklinks,hypertex]{hyperref}  

\newcommand{\Msun}{M_{\odot}}
\renewcommand{\vec}[1]{\bmath{#1}}
\newcommand{\be}{\begin{equation}}
\newcommand{\ee}{\end{equation}}
\newcommand{\ba}{\begin{eqnarray}}
\newcommand{\ea}{\end{eqnarray}}
\newcommand{\brr}{\begin{array}}
\newcommand{\err}{\end{array}}
\newcommand{\bc}{\begin{center}}
\newcommand{\ec}{\end{center}}
\newcommand{\hm}{\,h^{-1}{\rm Mpc}}

\newcommand{\bx}{{\bf x}}

\newcommand{\Rdc}{\mbox{$R_{\Delta_c}$}}
\newcommand{\Dc}{\mbox{$\Delta_c$}}
\newcommand{\dc}{\mbox{$\delta_c$}}
\newcommand{\Dm}{\mbox{$\Delta_{\rm m}$}}
\newcommand{\Mdc}{\mbox{$M_{\Delta_c}$}}

\newcommand{\mincir}{\raise
  -2.truept\hbox{\rlap{\hbox{$\sim$}}\raise5.truept \hbox{$<$}\ }}
\newcommand{\magcir}{\raise
  -2.truept\hbox{\rlap{\hbox{$\sim$}}\raise5.truept \hbox{$>$}\ }}
\newcommand{\siml}{\raise
  -2.truept\hbox{\rlap{\hbox{$\sim$}}\raise5.truept \hbox{$<$}\ }}
\newcommand{\simg}{\raise
  -2.truept\hbox{\rlap{\hbox{$\sim$}}\raise5.truept \hbox{$>$}\ }}
\newcommand{\Mg}{M_{\rm g}}
\newcommand{\Mnl}{M_{\rm NL}}
\newcommand{\Mgas}{\mbox{$M_{\rm g}$}}
\newcommand{\fgas}{\mbox{$f_{\rm g}$}}

\newcommand{\Cg}{\mbox{$C_{\rm g}$}}

\newcommand{\CT}{\mbox{$C_{\rm T}$}}
\newcommand{\ag}{\alpha_{\rm g}}

\newcommand{\aT}{\alpha_{\rm T}}
\newcommand{\Cgo}{\mbox{$C_{\rm g0}$}}

\newcommand{\rhog}{\rho_{\rm g}}
\newcommand{\rhogs}{\rho_{\rm g\ast}}
\newcommand{\trhog}{\tilde{\rho}_{\rm g}}
\newcommand{\Ts}{T_{\ast}}
\newcommand{\Tx}{T_{\rm X}}
\newcommand{\tT}{\tilde{T}}

\newcommand{\Lx}{L_{\rm X}}
\newcommand{\Lsx}{L_{\rm Xs}}
\newcommand{\Lbol}{L_{\rm bol}}

\newcommand{\Omm}{\Omega_{\rm m}}
\newcommand{\Oml}{\Omega_{\rm \Lambda}}

\newcommand{\rhom}{\rho_{\rm m}}
\newcommand{\rhoc}{\rho_{\rm cr}}

\submitted{Annual Reviews of Astrononomy and Astrophysics 2012, in press\hfill}

\begin{document}
\input epsf.tex    

\input psfig.sty

\begin{frontmatter}
\title{Formation of Galaxy Clusters}

\author{Andrey V. Kravtsov$^1$ and Stefano Borgani$^2$}

\markboth{Kravtsov \& Borgani}{Formation of Galaxy Clusters}

\address{$^1$Department of Astronomy \& Astrophysics, Kavli
  Institute for Cosmological Physics, The University of Chicago,
  Chicago, IL 60637\\email: {\tt andrey@oddjob.uchicago.edu}}
  
\address{$^2$Dipartimento di Fisica dell’Universit\`a di Trieste,
  Sezione di Astronomia, I-34131 Trieste, Italy;\\INAF – Osservatorio Astronomico di Trieste, Italy;\\ 
and INFN – Istituto Nazionale di Fisica Nucleare, Trieste, Italy; \\email: {\tt borgani@oats.inaf.it}}

\begin{abstract}
  Formation of galaxy clusters corresponds to the collapse of the
  largest gravitationally bound overdensities in the initial density
  field and is accompanied by the most energetic phenomena since the
  Big Bang and by the complex interplay between gravity--induced
  dynamics of collapse and baryonic processes associated with galaxy
  formation. Galaxy clusters are, thus, at the cross-roads of cosmology
  and astrophysics and are unique laboratories for testing models of
  gravitational structure formation, galaxy evolution, thermodynamics
  of the intergalactic medium, and plasma physics. At the same time,
  their large masses make them a useful probe of growth of structure
  over cosmological time, thus providing cosmological constraints that
  are complementary to other probes. In this review, we describe our
  current understanding of cluster formation: from the general picture
  of collapse from initial density fluctuations in an expanding
  Universe to detailed simulations of cluster formation including the
  effects of galaxy formation. We outline both the areas in which
  highly accurate predictions of theoretical models can be obtained
  and areas where predictions are uncertain due to uncertain physics
  of galaxy formation and feedback. The former includes the
  description of the structural properties of the dark matter halos
  hosting cluster, their mass function and clustering
  properties. Their study provides a foundation for cosmological
  applications of clusters and for testing the fundamental assumptions
  of the standard model of structure formation. The latter includes
  the description of the total gas and stellar fractions, the
  thermodynamical and non-thermal processes in the intracluster
  plasma. Their study serves as a testing ground for galaxy formation
  models and plasma physics.  In this context, we identify a
  suitable radial range where the observed thermal properties of the
  intra-cluster plasma exhibit the most regular behavior and thus can
  be used to define robust observational proxies for the total cluster
  mass. Finally, we discuss the formation of clusters in
  non-standard cosmological models, such as non-Gaussian models for
  the initial density field and models with modified gravity, along
  with prospects for testing these alternative scenarios with large
  cluster surveys in the near future.
\end{abstract}


\end{frontmatter}


\clearpage
\vspace*{-0.8cm}
\tableofcontents
\clearpage

\section{Introduction}

Tendency of nebulae to cluster has been ddiscovered by
Charles Messier and William Herschel, who have constructed the first
systematic catalogs of these objects. This tendency has become more
apparent as larger and larger samples of galaxies were compiled in the
19th and early 20th centuries. Studies of the most prominent
concentrations of nebulae, the clusters of galaxies, were
revolutionized in the 1920s by Edwin Hubble's proof that spiral
and elliptical nebulae were bona fide galaxies like the Milky Way
located at large distances from us \citep{hubble25,hubble26}, which
implied that clusters of galaxies are systems of enormous size. Just a
few years later, measurements of galaxy velocities in regions of
clusters made by \citet{hubble_humason31} and assumption of the virial
equilibrium of galaxy motions were used to show that the total
gravitating cluster masses for the Coma \citep[][and see also
\citealt{zwicky37}]{zwicky33} and Virgo clusters \citep{smith36} were
enormous as well.

The masses implied by the measured velocity dispersions were found to
exceed combined mass of all the stars in clusters galaxies by factors
of $\sim 200-400$, which prompted Zwicky to postulate the existence of
large amounts of ``dark matter'' (DM), inventing this widely used term
in the process. Although the evidence for dark matter in clusters was
disputed in the subsequent decades, as it was realized that stellar
masses of galaxies were underestimated in the early studies, dark
matter was ultimately confirmed by the discovery of extended hot
intracluster medium (ICM) emitting at X--ray energies by thermal
bremsstrahlung that was found to be smoothly filling intergalactic
space within the Coma cluster
\citep{gursky_etal71,meekins_etal71,kellogg_etal72,forman_etal72,cavaliere_etal71}. The
X--ray emission of the ICM has not only provided a part of the missing
mass \citep[as was conjectured on theoretical grounds
by][]{limber59,van_albada60}, but also allows the detection of
clusters out to $z>1$ \cite{rosati_etal02}. Furthermore, measurement
of the ICM temperature has provided an independent confirmation that
the depth of gravitational potential of clusters requires additional
dark component.  It was also quickly realized that inverse Compton
scattering of the cosmic microwave background (CMB) photons off
thermal electrons of the hot intergalactic plasma should lead to
distortions in the CMB spectrum, equivalent to black body temperature
variations of about $10^{-4}$--$10^{-5}$ [the Sunyaev--Zel'dovich (SZ)
effect;\citealt{sz70,sz72,sz80}]. This effect has now been measured in
hundreds of clusters \citep[e.g.,][]{carlstrom_etal02}.

Given such remarkable properties, it is no surprise that the quest to
understand the formation and evolution of galaxy clusters has become
one of the central efforts in modern astrophysics over the past
several decades. Early pioneering models of collapse of initial
density fluctuations in the expanding Universe have shown that systems
resembling the Coma cluster can indeed form
\citep{van_albada60,van_albada61,peebles70,white76}.  \citet[][see
also \citealt{sunyaev_zeldovich72}]{gott_gunn71} showed that hot gas
observed in the Coma via X-ray observations can be explained within
such a collapse scenario by heating of the infalling gas by the strong
accretion shocks. Subsequently, emergence of the hierarchical model of
structure formation
\citep{press_schechter74,gott_rees75,white_rees78}, combined with the
cold dark matter (CDM) cosmological scenario
\citep{bond_etal82,blumenthal_etal84}, provided a powerful framework
for interpretation of the multi-wavelength cluster observations.  At
the same time, rapid advances in computing power and new, efficient
numerical algorithms have allowed fully three-dimensional {\it ab
  initio} numerical calculations of cluster formation within
self-consistent cosmological context in both dissipationless regime
\citep{klypin_shandarin83,efstathiou_etal85} and including
dissipational baryonic component \citep{evrard88,evrard90}.

In the past two decades, theoretical studies of cluster formation have
blossomed into a vibrant and mature scientific field. As we detail in
the subsequent sections, the standard scenario of cluster formation has
emerged and theoretical studies have identified the most important
processes that shape the observed properties of clusters and their evolution, which has enabled usage of clusters as powerful cosmological probes \citep[see, e.g.,][for a recent review]{allen_etal11}. At the same time, observations of clusters at different redshifts have highlighted several key discrepancies between models and observations, which are particularly salient in the central regions (cores) of clusters. 

In the current paradigm of structure formation clusters are thought to
form via an hierarchical sequence of mergers and accretion of smaller
systems driven by gravity and DM that dominates the
gravitational field.  Theoretical models of clusters employ a variety
of techniques determined by a particular aspect of cluster formation
they aim to understand. Many of the bulk properties of clusters are
thought to be determined solely by the initial conditions,
dissipationless DM that dominates cluster mass budget, and
gravity. Thus, cluster formation is often approximated in models as
DM--driven dissipationless collapse from cosmological initial
conditions in an expanding Universe. Such models are quite successful
in predicting the existence and functional form of correlations
between cluster properties, as well as their abundance and clustering,
as we discuss in detail in Section~\ref{sec:clusform}. One of the most
remarkable models of this kind is a simple self-similar model of
clusters \citep[][see \S~\ref{sec:selfsimilar}
below]{kaiser86}. Despite its simplicity, the predictions of this
model are quite close to results of observations and have, in fact,
been quite useful in providing baseline expectations for evolution of
cluster scaling relations. Studies of abundance and spatial
distribution of clusters using dissipationless cosmological
simulations show that these statistics retain remarkable memory of the
initial conditions.

The full description of cluster formation requires detailed modeling
of the non--linear processes of collapse and the dissipative physics
of baryons.  The gas is heated to high, X-ray emitting temperatures by
adiabatic compression and shocks during collapse and settles in
hydrostatic equilibrium within the cluster potential well. Once the
gas is sufficiently dense, it cools, the process that can feed both
star formation and accretion onto supermassive black holes (SMBHs)
harbored by the massive cluster galaxies. The process of cooling and
formation of stars and SMBHs can then result in energetic feedback due
to supernovae (SNe) or active galactic nuclei (AGN), which can inject
substantial amounts of heat into the ICM and spread heavy elements
throughout the cluster volume.

Galaxy clusters are therefore veritable crossroads of astrophysics and
cosmology: While abundance and spatial distribution of clusters bear
indelible imprints of the background cosmology, gravity law, and
initial conditions, the nearly closed--box nature of deep cluster
potentials makes them ideal laboratories to study processes operating
during galaxy formation and their effects on the surrounding
intergalactic medium.

In this review we discuss the main developments and results in the
quest to understand the formation and evolution of galaxy
clusters. Given the limited space available for this review and the
vast amount of literature and research directions related to galaxy
clusters, we have no choice but to limit the focus of our review, as
well as the number of cited studies. Specifically, we focus on the
most basic and well-established elements of the standard paradigm of
DM-driven hierarchical structure formation within the framework of
$\Lambda$CDM cosmology as it pertains to galaxy clusters.  We focus
mainly on the theoretical predictions of the properties of the total
cluster mass distribution and properties of the hot intracluster gas,
and only briefly discuss results pertaining to the evolution of
stellar component of clusters, understanding of which is still very
much a work in progress. Comparing model predictions to real
clusters, we mostly focus on comparisons with X-ray observations, which
have provided the bulk of our knowledge of ICM properties so
far. In \S~\ref{sec:cosmo}, we briefly discuss the differences in
formation of clusters in models with the non-Gaussian initial
conditions and modified gravity. Specifically, we focus on the
information that statistics sensitive to the cluster formation
process, such as cluster abundance and clustering, can provide about
the primordial non-Gaussianity and possible deviations of gravity from
General Relativity. We refer readers to recent extensive
reviews on cosmological uses of galaxy clusters by
\citet{allen_etal11} and \citet{weinberg_etal12} for a more extensive
discussion of this topic.

\section{The observed properties of galaxy clusters}
\label{sec:obsprop}

Observational studies of galaxy clusters have now developed into a
broad, multi-faceted and multi-wavelength field. Before we embark on
our overview of different theoretical aspects of cluster formation, we
briefly review the main observational properties of clusters and, in
particular, the basic properties of their main matter constituents.

\begin{figure}
\centerline{ 
\vspace{-3cm}
\psfig{file=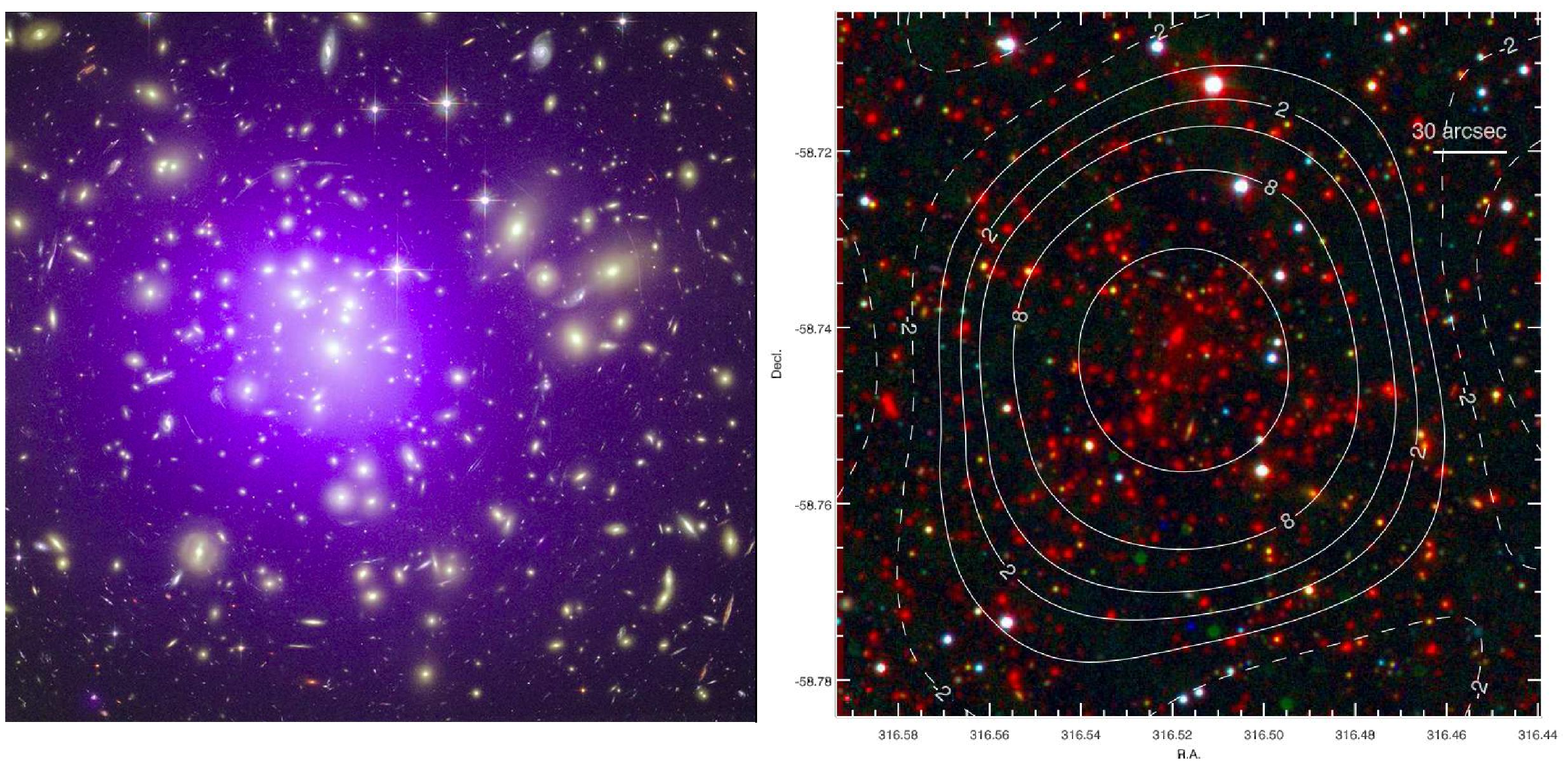,width=17truecm,angle=-90}
}
\caption{Left panel: the composite X-ray/optical image (556 kpc on a
  side) of the galaxy cluster Abell 1689 at redshift $z=0.18$. The
  purple haze shows X-ray emission of the $T\sim 10^8$ K gas, obtained
  by the {\em Chandra X-ray Observatory}. Images of galaxies in the optical band,
  colored in yellow, are from observations performed with the Hubble
  Space Telescope. The long arcs in the optical image are caused by
  the gravitational lensing of background galaxies by matter in the
  galaxy cluster, the largest system of such arcs ever found
  (Credit:X-ray: NASA/CXC/MIT; Optical: NASA/STScI). Right panel: the
  galaxy cluster SPT-CL J2106-5844 at $z = 1.133$, the most massive
  cluster known at $z>1$ discovered via its Sunyaev--Sel'dovich (SZ) signal ($M_{200}\approx
  1.3\times 10^{15}\rm\,\Msun$). The color image shows the
  Magellan/LDSS3 optical and Spitzer/IRAC mid-infrared measurements
  (corresponding to the blue-green-red color channels). The frame subtends
  $4.8\times 4.8$~ arcmin, which corresponds to $2.4\times
  2.4$~Mpc at the redshift of the cluster. The white contours
  correspond to the South Pole Telescope SZ significance values, as labeled, where
  dashed contours are used for the negative significance
  values. (Adapted from \protect\citealt{foley_etal11}).  }
\label{fig:a1689}
\end{figure}

Figure \ref{fig:a1689} shows examples of the multiwavelength
observations of two massive clusters at two different cosmic epochs:
the Abell 1689 at $z=0.18$ and the SPT-CL J2106-5844 at $z = 1.133$.
It illustrates all of the main components of the clusters: the
luminous stars in and around galaxies (the intracluster light or ICL),
the hot ICM observed via its X-ray emission and the Sunyaev-Zel'dovich
effect and, in the case of Abell 1689, even the presence of invisible
DM manifesting itself through gravitational lensing of background
galaxies distorting their images into extended, cluster-centric arcs
\citep[][and references therein]{bartelmann10}. At larger radii, the
lensing effect is weaker. Although not easily visible by eye, it can
still be reliably measured by averaging the shapes of many background
galaxies and comparing the average with the expected value for an
isotropic distribution of shapes. The gravitational lensing is a
direct probe of the total mass distribution in clusters, which makes
it both extremely powerful in its own right and a very useful check of
other methods of measuring cluster masses. The figure shows several
bright elliptical galaxies that are typically located near the cluster
center. A salient feature of such central galaxies is that they show
little evidence of ongoing star formation, despite their extremely
large masses.

\begin{figure}
\centerline{ 
\psfig{file=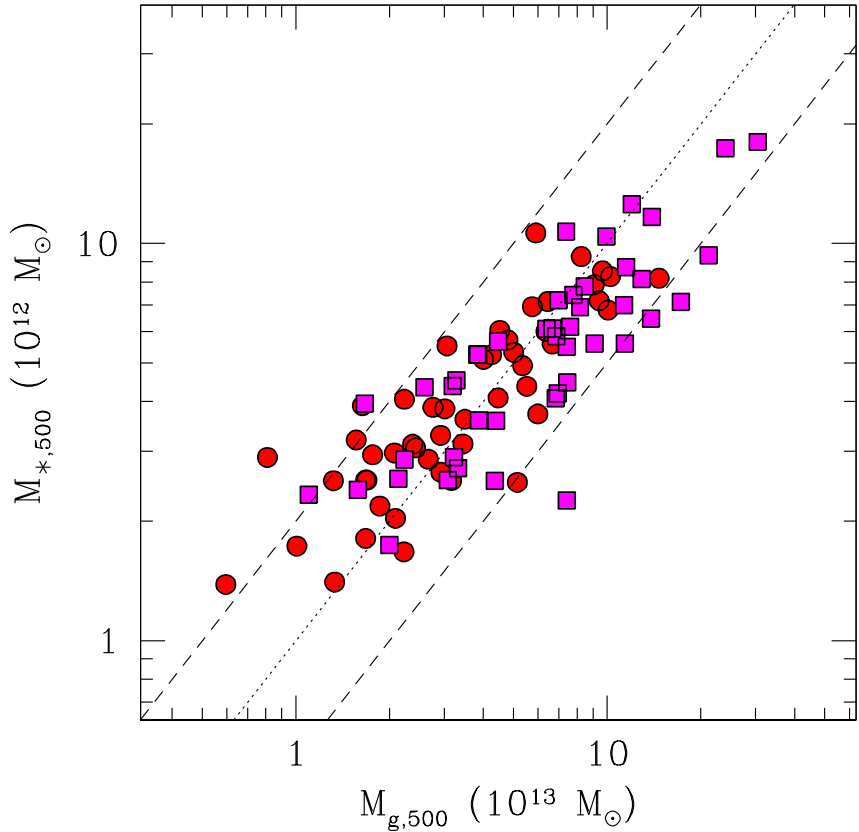,width=13.truecm}
}
\caption{The mass in stars vs. the mass of hot, X-ray emitting
  gas. Both masses are measured within the radius $R_{500}$ estimated
  from the observationally calibrated $Y_{\rm X}-M_{500}$ relation,
  assuming flat $\Lambda$CDM cosmology with $\Omega_{\rm
    m}=1-\Omega_{\Lambda}=0.26$ and $h=0.71$. Red circles show local
  clusters located at $z<0.1$, whereas magenta squares show
  higher-redshift clusters: $0.1<z<0.6$ \protect\citep[see][for
  details]{lin_etal12}. The dotted line corresponds to the constant
  stellar-to-gas mass ratio $M_{\ast,500}/M_{\rm g,500}=0.1$, whereas
  the dashed lines correspond to the values of $0.05$ and $0.2$ for
  this ratio.}
\label{fig:mgms}
\end{figure}

The diffuse plasma is not associated with individual galaxies and
constitutes the intra-cluster medium, which contains the bulk of the
normal baryonic matter in massive clusters. Although the hot ICM is
not directly associated with galaxies, their properties are
correlated. For example, Fig.~\ref{fig:mgms} shows the mass of the ICM
gas within the radius $R_{500}$, defined as the radius enclosing mean
overdensity of $\Delta_c=500\rhoc$, versus stellar mass in galaxies
within the same radius for a number of local ($z\lesssim 0.1$) and
distant ($0.1<z<0.6$) clusters \citep{lin_etal12}. Here
$\rhoc(z)=3H(z)^2/(8\pi G$ is the critical mean density of the
Universe, defined in terms of the Hubble function $H(z)$. The figure
shows a remarkably tight, albeit non-linear, correlation between these
two baryonic components. It also shows that the gas mass in clusters
is on average about ten times larger than the mass in stars, although
this ratio is systematically larger for smaller mass clusters, ranging
from $M_{\ast}/M_{\rm g}\approx 0.2$ to $\approx 0.05$, as mass
increases from group scale ($M_{500}\sim {\rm few\ }\times 10^{13}\rm\
M_{\odot}$) to massive clusters ($M_{500}\sim 10^{15}\rm\ M_{\odot}$).

The temperature of the ICM is consistent with velocities of galaxies and indicates that both
galaxies and gas are nearly in equilibrium within a common gravitational
potential well.  The mass of galaxies and hot gas is not sufficient to
explain the depth of the potential well, which implies that
most of the mass in clusters is in a form of DM.  Given that
hydrogen is by far the most abundant element in the Universe, most of
the plasma particles are electrons and protons, with a smaller number
of helium nuclei.  There are also trace amounts of heavier nuclei some
of which are only partially ionized. The typical average abundance of the
heavier elements is about one-third of that found in the Sun or a
fraction of one per cent by mass; it decreases with increasing radius
and can be quite inhomogeneous, especially in merging systems
\citep[][for a review]{werner_etal08}.

\begin{figure}
\centerline{ 
\psfig{file=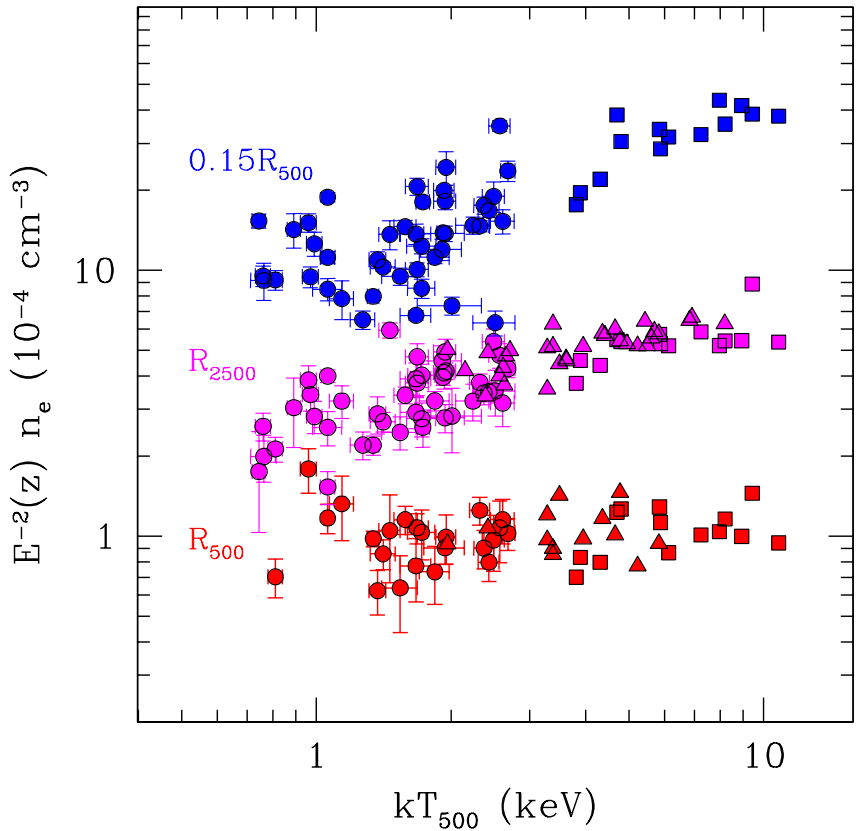,width=13.truecm}
}
\caption{The observed electron number density, $n_e$, in galaxy
  clusters and groups, measured at different radii (from top to
  bottom: $0.15R_{500}$, $R_{2500}$, $R_{500}$; see labels) as a
  function of the intracluster medium temperature at $R_{500}$. The values of $n_e$
  are rescaled by $E^{-2}(z)$, the scaling expected from the
  definition of the radii at which densities are measured. Squares and circles show
  systems observed with the {\it Chandra X-ray Observatory} from the
  studies by \protect\cite{vikhlinin_etal09} and
  \protect\cite{sun_etal09}, triangles show systems observed with the
  {\em XMM--Newton} telescope by \protect\cite{pratt_etal10}. Note
  that electron densities at large radii are independent of
  temperature, as expected from the self-similar model,
  whereas at small radii the rescaled densities increase with temperature. Note
  also that the scatter from cluster to cluster increases with decreasing
  radius, especially for low-temperature groups \protect\citep[after][]{sun12}. }
\label{fig:obsICM}
\end{figure}

Thermodynamic properties of the ICM are of utmost importance, because
comparing such properties to predictions of baseline models without
cooling and heating can help to isolate the impact of these physical
processes in cluster formation.  The most popular baseline model is
the self-similar model of clusters developed by \citet[][]{kaiser86},
which we consider in detail in Section \ref{sec:selfsimilar} below. In its
simplest version, this model assumes that clusters are scaled versions
of each other, so that gas density at a given fraction of the
characteristic radius of clusters, defined by their mass, is
independent of cluster mass. Figure~\ref{fig:obsICM} shows the
electron density in clusters as a function of ICM temperature (and
hence mass) at different radii. It is clear that density is independent of temperature only outside
cluster core at $r\sim R_{500}$, although there is an indication that
density is independent of temperature at $r=R_{2500}$ for 
$k_BT\gtrsim 3$~keV. This indicates that processes associated with
galaxy formation and feedback affect the properties of clusters at
$r\lesssim R_{2500}$, but their effects are mild at larger radii.

\begin{figure}
\centerline{ 
\psfig{file=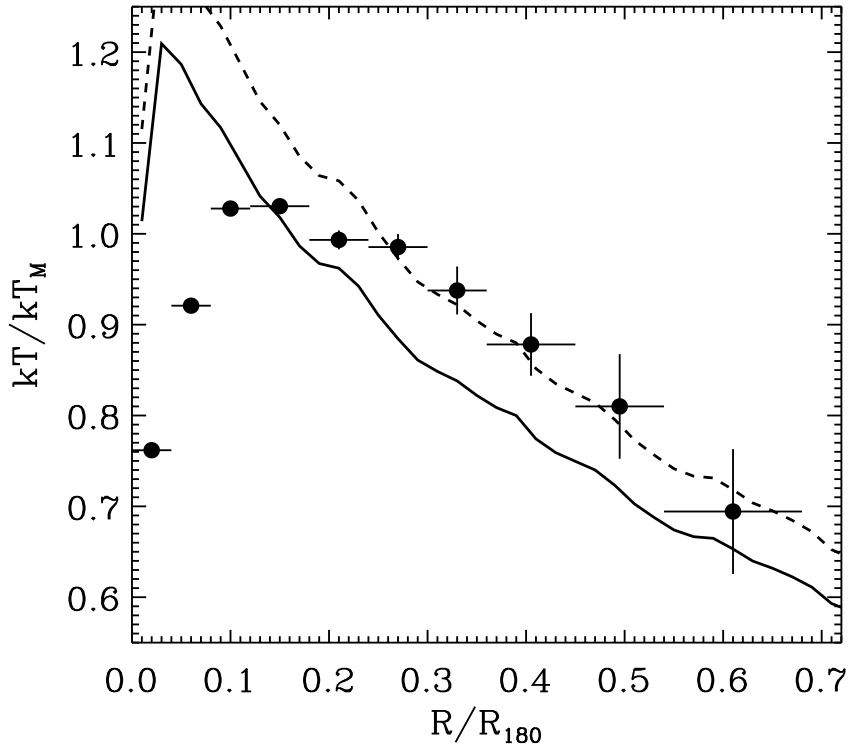,width=13.truecm}}
\caption{Comparison between temperature profiles, normalized to the
  global temperature measured within $R_{180}$ by
  \protect\cite{leccardi_molendi08}, for a set of about 50 nearby
  clusters with $z\mincir 0.3$ and with temperature $k_BT_X> 3$ keV,
  observed with the {\em XMM--Newton (dots with errorbars)} and
  results from cosmological hydrodynamical simulations including the
  effect of radiative cooling, star formation and supernova feedback
  in the form of galactic winds ({\em solid curve};
  \protect\citealt{borgani_etal04}).  From
  \protect\cite{leccardi_molendi08}.}
\label{fig:Tprofs}
\end{figure}

During the past two decades, it has been established that the core
regions of the relaxed clusters are generally characterized by a
strongly peaked X--ray emissivity, indicating efficient cooling of the
gas \citep[e.g.,][]{fabian94}. Quite interestingly, spectroscopic
observations with the Chandra and XMM--Newton satellites have
demonstrated that, despite strong X-ray emission of the hot gas, only
a relatively modest amount of this gas cools down to low temperatures
\citep[e.g.,][]{peterson_etal01,boehringer_etal01}. This result is
generally consistent with the low levels of star formation observed in
the brightest cluster galaxies \citep[BCGs;
e.g.,][]{mcdonald_etal11}. It implies that a heating mechanism should
compensate for radiative losses, thereby preventing the gas in cluster
cores to cool down to low temperature.  The presence of cool cores is
also reflected in the observed temperature profiles \citep[e.g.][see
also Figure
\protect\ref{fig:Tprofs}]{leccardi_molendi08,pratt_etal07,vikhlinin_etal06},
which exhibit decline of temperature with decreasing radius in the
innermost regions of relaxed cool--core clusters.

\begin{figure}
\vspace{-6cm}
\hspace{1cm}
\centerline{ 
\psfig{file=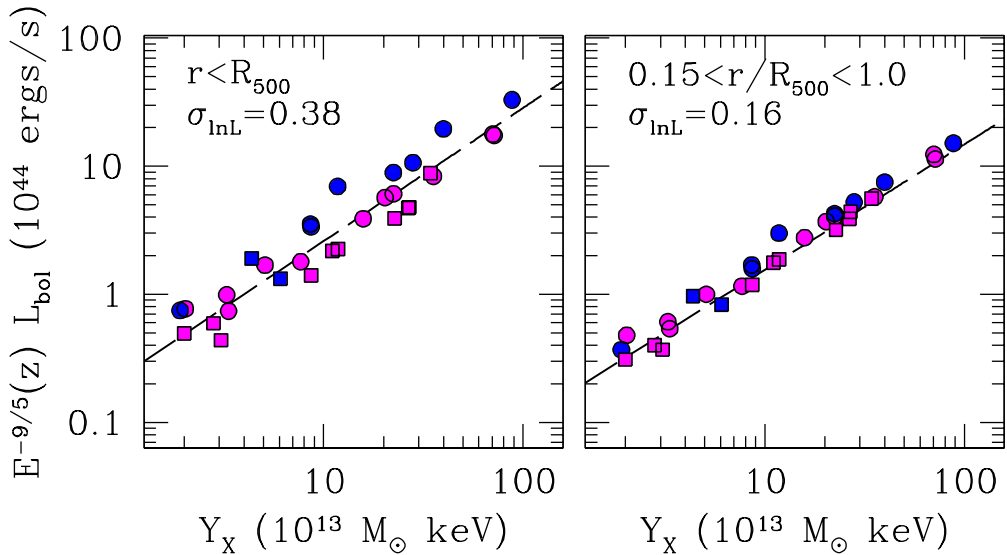,width=15.truecm} }
\caption{Correlation of {\it bolometric} luminosity of intracluster gas and
  $Y_{\rm X}\equiv M_{\rm gas}T_X$, where $M_{\rm gas}$ is the mass of
  the gas within $R_{500}$ and $T_X$ is temperature derived from the
  fit to gas spectrum accounting only for emission from radial range
  $(0.15-1)R_{500}$. Results are shown for the local clusters from the
  Representative XMM--Newton Cluster Structure Survey sample of
  \protect\citet{pratt_etal08}. The left panel shows total luminosity
  integrated within radius $R_{500}$, whereas the right panel shows
  bolometric luminosity calculated with the central $0.15R_{500}$ of the cluster
  excised. 
  Labels in the top left corner indicate the radial range used in
  computing the luminosity and logarithmic scatter of luminosity at
  fixed $Y_{\rm X}$. The blue points show cool core clusters, whereas
  magenta points are non--cool core clusters. Clusters classified as
  relaxed and disturbed are shown by circles and squares,
  respectively. Note that exclusion of the cluster cores reduces the
  scatter between luminosity and $Y_{\rm X}$ by more than a factor of
  two.}
\label{fig:ly_rexcess}
\end{figure}

One of the most important and most widely studied aspects of ICM
properties are correlations between its different observable
integrated quantities and between observable quantities and total
mass. Such {\it scaling relations\/} are the key ingredient in
cosmological uses of clusters, where it is particularly desirable that
the relations are characterized by small scatter and are independent
of the relaxation state and other properties of clusters. Although
clusters are fascinatingly complex systems overall, they do exhibit
some remarkable regularities. As an example,
Figure~\ref{fig:ly_rexcess} shows the correlation between 
the bolometric luminosity
emitted from within $R_{500}$ and the $Y_{\rm X}$ parameter defined as
a product of gas mass within $R_{500}$ and ICM temperature derived
from the X--ray spectrum within the radial range $(0.15-1)R_{500}$
\citep{kravtsov_etal06} for theRepresentative XMM--Newton Cluster
Structure Survey (REXCESS) sample of clusters studied by
\citet{pratt_etal08}. Different symbols indicate clusters in different
states of relaxation, whereas clusters with strongly peaked central gas
distribution (the cool core clusters) and clusters with less centrally
concentrated gas distribution are shown with different colors. The left
panel shows total luminosity integrated within radius $R_{500}$, wheareas
the right panel shows luminosity calculated with the central region
within $0.15R_{500}$ excised. Quite clearly, the core-excised X-ray
luminosity exhibits remarkably tight correlation with $Y_{\rm X}$,
which, in turn, is expected to correlate tightly with total cluster
mass \citep{kravtsov_etal06,stanek_etal10,fabjan_etal11}. This figure
illustrates the general findings in the past decade that clusters
exhibit strong regularity and tight correlations among X-ray
observable quantities and total mass, provided that relevant
quantities are measured after excluding the emission from cluster
cores.

\section{Understanding the formation of galaxy clusters}
\label{sec:clusform}

\subsection{Initial density perturbation field and its linear evolution}
\label{sec:collbasics}

In the currently standard hierarchical structure formation scenario,
objects are thought to form via gravitational collapse of peaks in the
initial primordial density field characterized by the density contrast
(or overdensity) field: $\delta({\bf x})=(\rho({\bf
  x})-\bar{\rho}_{\rm m})/{\bar{\rho}_{\rm m}}$, where
$\bar{\rho}_{\rm m}$ is the mean mass density of the Universe.
Properties of the field $\delta({\bf x})$ depend on specific details
of the processes occurring during the earliest inflationary stage of
evolution of the Universe 
\citep{guth_pi82,starobinsky82,bardeen_etal83} and the subsequent
stages prior to recombination
\citep{peebles82,bond_efstathiou84,bardeen_etal86,eisenstein_hu99}. A
fiducial assumption of most models that we discuss is that
$\delta({\bf x})$ is a homogeneous and isotropic Gaussian random
field.  We briefly discuss non-Gaussian models in Section
\ref{sec:nong}.

Statistical properties of a uniform and isotropic Gaussian field can
be fully characterized by its power spectrum, $P(k)$, which depends
only on the modulus $k$ of the wavevector, but not on its direction. A
related quantity is the variance of the density contrast field {\it
  smoothed\/} on some scale $R$: $\delta_{R}({\bf x})\equiv \int
\delta({\bf x}-{\bf r})W({\bf r}, R)d^3r$, where
\be
\langle\delta_{R}^2\rangle\equiv\sigma^2(R)=\frac{1}{(2\pi)^3}\int
P(k)\vert \tilde{W}({\bf k},R)\vert^2d^3k,
\label{eq:variance}
\ee 
where $\tilde{W}({\bf k},R)$ is the Fourier transform of the window
(filter) function $W({\bf r},R)$, such that $\delta_{R}({\bf
  k})=\delta({\bf k})\tilde{W}({\bf k},R)$ [see, e.g.,
\cite{zentner07} or \cite{mo_etal10} for details on the definition of
$P(k)$ and choices of window function]. For the cases, when one is
interested in only a narrow range of $k$ the power spectrum can be
approximated by the power--law form, $P(k)\propto k^n$, and the
variance is $\sigma^2(R)\propto R^{-(n+3)}$.

At a sufficiently high redshift $z$, for the spherical top--hat window
function mass and radius are interchangeable according to the relation
$M=4\pi/3\rho_{\rm m}(z)R^3$.  We can think about the density field
smoothed on the scale $R$ or the corresponding mass scale $M$. The
characteristic amplitude of peaks in the $\delta_R$ (or $\delta_{M}$)
field smoothed on scale $R$ (or mass scale $M$) is given by
$\sigma(R)\equiv\sigma(M)$.  The smoothed Gaussian density field is,
of course, also Gaussian with the probability distribution function
(PDF) given by \be
p(\delta_{M})=\frac{1}{\sqrt{2\pi}\sigma(M)}\exp\left[-\frac{\delta^2_{M}}{2\sigma^2(M)}\right].
\label{eq:pGaussian}
\ee 

During the earliest linear stages of evolution in the standard
structure formation scenario the initial Gaussianity of the
$\delta({\bf x})$ field is preserved, as 
different Fourier modes $\delta({\bf k})$ evolve independently and grow
at the same rate, described by the {\it linear growth factor},
$D_+(a)$, as a function of expansion factor $a=(1+z)^{-1}$, which for a
$\Lambda$CDM cosmology is given by \cite{heath77}: 
\be
\delta(a)\propto
D_+(a)=\frac{5\Omm}{2}E(a)\int_0^a\frac{da^{\prime}}{\left[a^{\prime}E(a^{\prime})\right]^{3}},
\label{eq:Dgrowth}
\ee
where $E(a)$ is the normalized expansion rate, which is given by 
\be
E(a)\equiv\frac{H(a)}{H_0}=\left[\Omm a^{-3}+(1-\Omm-\Oml)a^{-2}+\Oml\right]^{1/2},
\label{eq:Ea}
\ee if the contribution from relativistic species, such as radiation
or neutrinos, to the energy-density is neglected. Growth rate and the
expression for $E(a)$ in more general, homogeneous dark energy (DE)
cosmologies are described by \citet{percival05}.  Note that in models
in which DE is clustered \citep{alimi_etal10} or gravity deviates from
General Relativity
(GR, see Section~\ref{sec:nongr}), the growth factor can be scale
dependent.

Correspondingly, the linear evolution of the root mean square (rms) amplitude of fluctuations is given by $\sigma(M,a)=\sigma(M,a_i)D_+(a)/D_+(a_i)$, which is often useful to recast in terms of linearly extrapolated rms amplitude $\sigma(M,a=1)$ at $a=1$ (i.e., $z=0$):
\be
\sigma(M,a)=\sigma(M,a=1)D_{+0}(a),\ \ \ {\rm where\ } D_{+0}(a)\equiv D_+(a)/D_+(a=1).
\ee

Once the amplitude of typical fluctuations approaches unity,
$\sigma(M,a)\sim 1$, the linear approximation breaks down. Further
evolution must be studied by means of nonlinear models or direct
numerical simulations. We discuss results of numerical
simulations extensively below. However, we consider first the
simplified, but instructive, spherical collapse model and associated
concepts and terminology. Such model can be used to gain physical
insight into the key features of the evolution and is used as a basis
for both definitions of collapsed objects (see
Section~\ref{sec:massdef}) and quantitative models for halo abundance
and clustering (Section~\ref{sec:mf} and \ref{sec:bias}).

\subsection{Non--linear evolution of spherical perturbations and non--linear mass scale}
\label{sec:sphcoll}

The simplest model of non--linear collapse assumes that density peak can be characterized as constant overdensity spherical perturbation of radius $R$. Despite its simplicity and limitations discussed below, the model provides a useful insight into general features and timing of non--linear collapse. Its results are commonly used in analytic models for halo abundance and clustering and motivate mass definitions for collapsed objects. Below we briefly describe the model and non--linear mass scale that is based on its predictions. 
 
\subsubsection{Spherical collapse model.}
The {\it spherical collapse model\/} considers a
spherically-symmetric density fluctuation of initial radius $R_{\rm
  i}$, amplitude $\delta_{\rm i}>0$, and mass
$M=(4\pi/3)(1+\delta_{\rm i})\bar{\rho}R_{\rm i}^3$, where $R_{\rm i}$ is physical radius of the perturbation and $\bar{\rho}$ is the mean density of the Universe at the initial time. Given the
symmetry, the collapse of such perturbation is a one-dimensional
problem and is fully specified by evolution of the top-hat radius
$R(t)$ \citep[][]{gunn_gott72,lahav_etal91}. It consists of an
initially decelerating increase of the perturbation radius, until it
reaches the maximum value, $R_{\rm ta}$, at the turnaround epoch,
$t_{\rm ta}$, and subsequent decrease of $R(t)$ at $t>t_{\rm ta}$
until the perturbation collapses, virializes, and settles at the final
radius $R_{\rm f}$ at $t=t_{\rm coll}$.  Physically, $R_{\rm f}$ is
set by the virial relation between potential and kinetic energy and is
$R_{\rm f}=R_{\rm ta}/2$ in cosmologies with $\Oml=0$.  The turnaround
epoch and the epoch of collapse and virialization are defined by
initial conditions.

The final mean internal density of a collapsed object can be estimated
by noting that in a $\Oml=0$ Universe the time interval $t_{\rm
  coll}-t_{\rm ta}=t_{\rm ta}$ should be equal to the free-fall time
of a uniform sphere $t_{\rm ff}=\sqrt{3\pi/(32G\rho_{\rm ta})}$, which
means that the mean density of perturbation at turnaround is
$\rho_{\rm ta}=3\pi/(32Gt^2_{\rm ta})$ and $\rho_{\rm coll}=8\rho_{\rm
  ta}=3\pi/(Gt_{\rm coll}^2)$. These densities can be compared with
background mean matter densities at the corresponding times to get
mean internal density contrasts: $\Delta=\rho/\bar{\rho}_{\rm m}$. In
the Einstein-de Sitter model ($\Omm=1$, $\Oml=0$), background density
evolves as
$\bar{\rho}_{\rm m}=1/(6\pi G t^2)$, which means that density contrast
after virialization is
\be
\Delta_{\rm vir}\equiv \frac{\rho_{\rm coll}}{\bar{\rho}_{\rm m}}=18\pi^2=177.653.
\ee

For general cosmologies, density contrast can be computed by
estimating $\rho_{\rm coll}$ and $\bar{\rho}_m(t_{\rm coll})$ in a
similar fashion. For lower $\Omm$ models, fluctuation of the same mass
$M$ and $\delta$ has a larger initial radius and smaller physical
density and, thus, takes longer to collapse. The density contrasts of
collapsed objects therefore are larger in lower density models because
the mean density of matter at the time of collapse is
smaller. Accurate (to $\lesssim 1\%$ for $\Omm=0.1-1$) approximations
for $\Delta_{\rm vir}$ in open ($\Oml=0$) and flat $\Lambda$CDM
($1-\Oml-\Omm=0$) cosmologies are given by \citet[][their equation
6]{bryan_norman98}. For example, for the concordance $\Lambda$CDM
cosmology with $\Omm=0.27$ and $\Oml=0.73$ \citep{komatsu_etal11},
density contrast at $z=0$ is $\Delta_{\rm vir}\approx 358$.

Note that if the initial density contrast $\delta_{\rm i}$ would grow
only at the linear rate, $D_+(z)$, then the density contrast at the time of
collapse would be more than a hundred times smaller.  Its value can be
derived starting from the density contrast linearly extrapolated to
the turn around epoch, $\delta_{\rm ta}$. This epoch corresponds to
the time at which perturbation enters in the non--linear regime and
detaches from the Hubble expansion, so that $\delta_{\rm ta}\sim 1$
is expected. In fact, the exact calculation in the case of 
$\Omm(z)=1$ at the redshift of turn--around gives $\delta_{ta}=1.062$
\citep{gunn_gott72}. Because $t_{\rm coll}=2t_{\rm ta}$, further linear
evolution for $\Omm(z)=1$ until the collapse time gives
$\dc=\delta_{\rm ta} D_+(t_c)/D_+(t_{\rm ta})\approx 1.686$. In the
case of $\Omm\ne 1$ we expect that $\delta_{\rm ta}$ should have
different values. For instance, for $\Omm< 1$ density contrast at
turn--around should be higher to account for the higher rate of the
Hubble expansion. However, linear growth from $t_{\rm ta}$ to $t_{\rm
  coll}$ is smaller due to the slower redshift dependence of
$D_+(z)$. As a matter of fact, these two factors nearly cancel, so that 
$\delta_c$ has a weak dependence on
$\Omm$ and $\Oml$ \citep[e.g.,][]{percival05}. For the concordance
$\Lambda$CDM cosmology at $z=0$, for example, $\delta_c\approx 1.675$.

Additional interesting effects may arise in models with DE
characterized by small or zero speed of sound, in which structure
growth is affected not only because DE influences linear
growth, but also because it participates non-trivially in the collapse
of matter and may slow down or accelerate the formation of clusters of
a given mass depending on DE equation of state
\citep{abramo_etal07,creminelli_etal10}. DE in such models
can also contribute non-trivially to the gravitating mass of clusters.

\subsubsection{The nonlinear mass scale $\Mnl$.}
\label{sec:Mnl}
The linear value of the collapse overdensity $\delta_c$ is useful in
predicting whether a given initial perturbation $\delta_i\ll 1$ at
initial $z_{\rm i}$ collapses by some later redshift $z$. The collapse
condition is simply $\delta_iD_{+0}(z)\geq \delta_c(z)$ and is used
extensively to model the abundance and clustering of collapsed
objects, as we discuss below in \S~\ref{sec:mf}.  The distribution of
peak amplitudes in the initial Gaussian overdensity field smoothed
over mass scale $M$ is given by a Gaussian PDF with a rms value of
$\sigma(M)$ (Equation~\ref{eq:pGaussian}). The peaks in the initial
Gaussian overdensity field smoothed at redshift $z_i$ over mass scale
$M$ can be characterized by the ratio $\nu=\delta_i/\sigma(M,z_i)$
called the {\it peak height}. For a given mass scale $M$, the peaks
collapsing at a given redshift $z$ according to the spherical collapse
model have the peak height given by: 
\be
\nu\equiv\frac{\delta_c(z)}{\sigma(M,z)}.
\label{eq:nu}
\ee
Given that $\delta_c(z)$ is a very weak function of $z$ (changing by
$\lesssim 1-2\%$ typically), whereas
$\sigma(M,z)=\sigma(M,z=0)D_{+0}(z)$ decreases strongly with
increasing $z$, the peak height of collapsing objects of a given mass
$M$ increases rapidly with increasing redshift.

Using Equation~\ref{eq:nu} we can define the characteristic mass scale for
which a typical peak ($\nu=1$) collapses 
at redshift $z$:
\be
\sigma(\Mnl,z)=\sigma(\Mnl,z=0)D_{+0}(z)=\delta_c(z).
\label{eq:Mnl}
\ee This {\it nonlinear mass}, $\Mnl(z)$, is a key quantity in the
self-similar models of structure formation, which we consider in
Section~\ref{sec:selfsimilar}.

\subsection{Nonlinear collapse of real density peaks}
\label{sec:realcoll}

The spherical collapse model provides a useful approximate guideline
for the time scale of halo collapse and has proven to be a very useful
tool in developing approximate statistical models for the formation
and evolution of halo populations. Such a simple model and its
extensions (e.g., ellipsoidal collapse model) do, however, miss many
important details and complexities of collapse of the real density
peaks. Such complexities are usually explored using three-dimensional
numerical cosmological simulations. Techniques and numerical details
of such simulations are outside the scope of this review and we refer
readers to recent reviews on this subject
\citep[][]{bertschinger98,dolag_etal08b,norman10,borgani_kravtsov11}. Here,
we simply discuss the main features of gravitational collapse
learned from analyses of such simulations.

\begin{figure}
\centerline{ 
\psfig{file=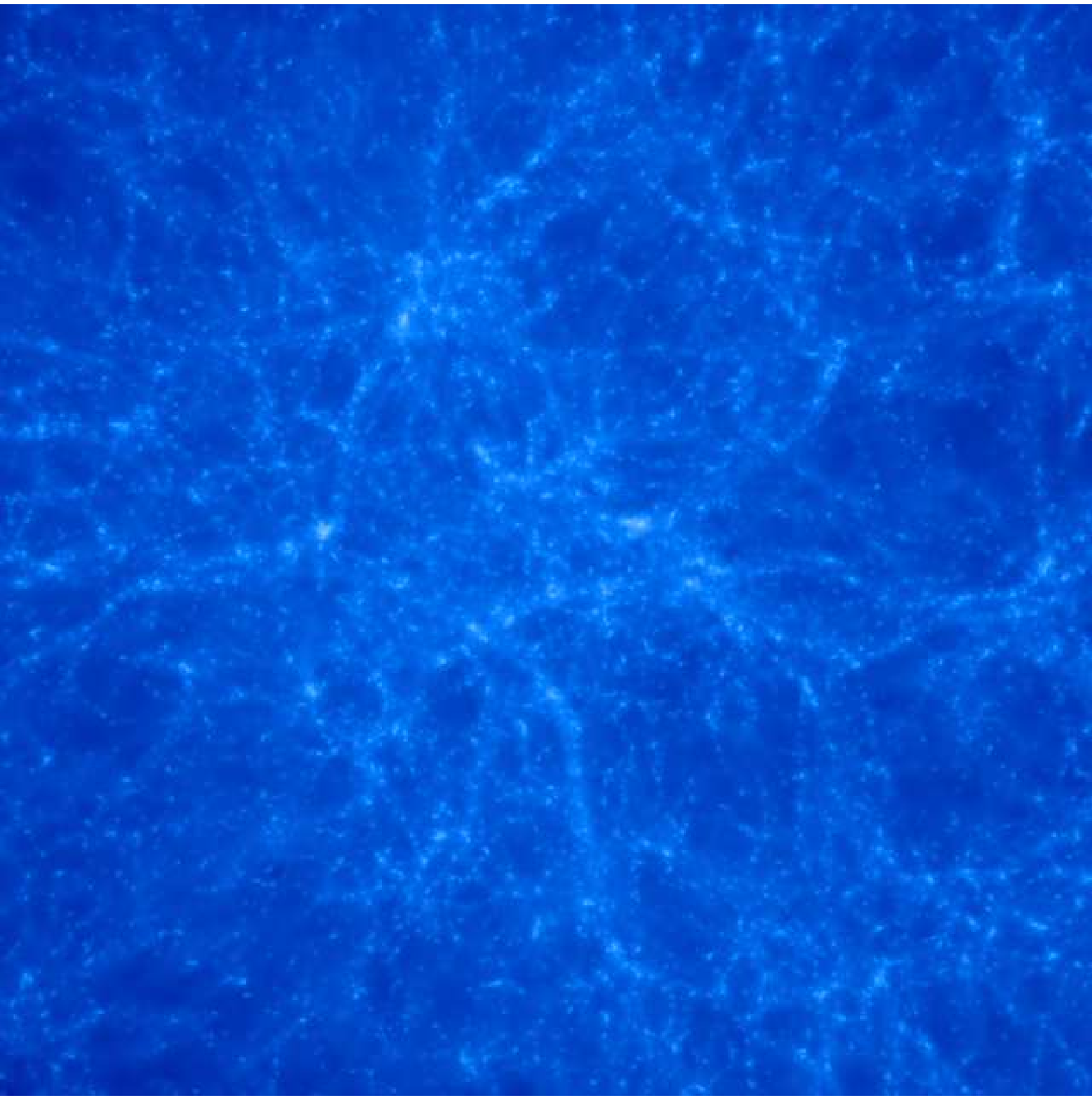,width=0.5\linewidth}
\psfig{file=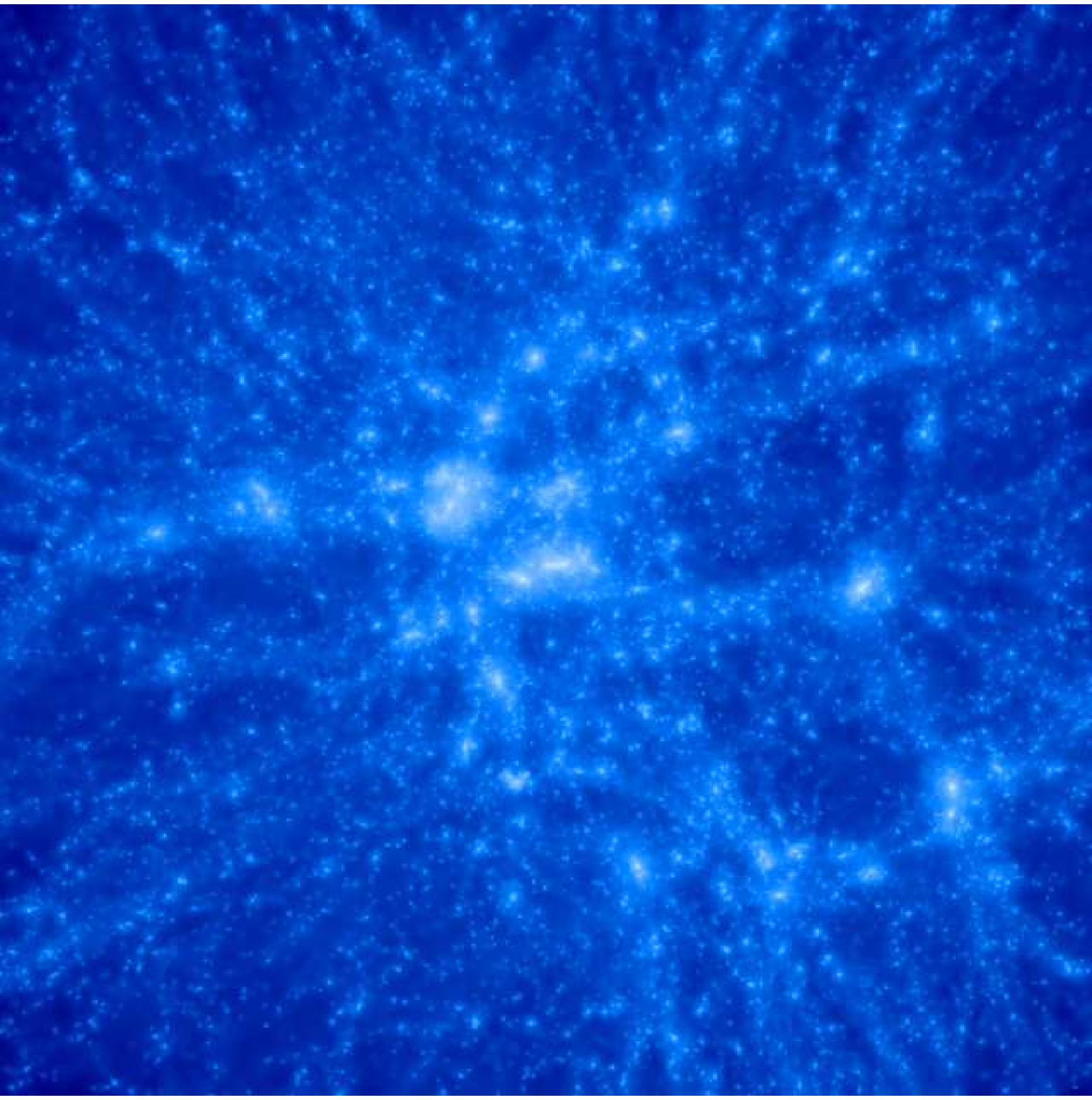,width=0.5\linewidth}
}
\vspace{-2.75cm}
\centerline{ 
\psfig{file=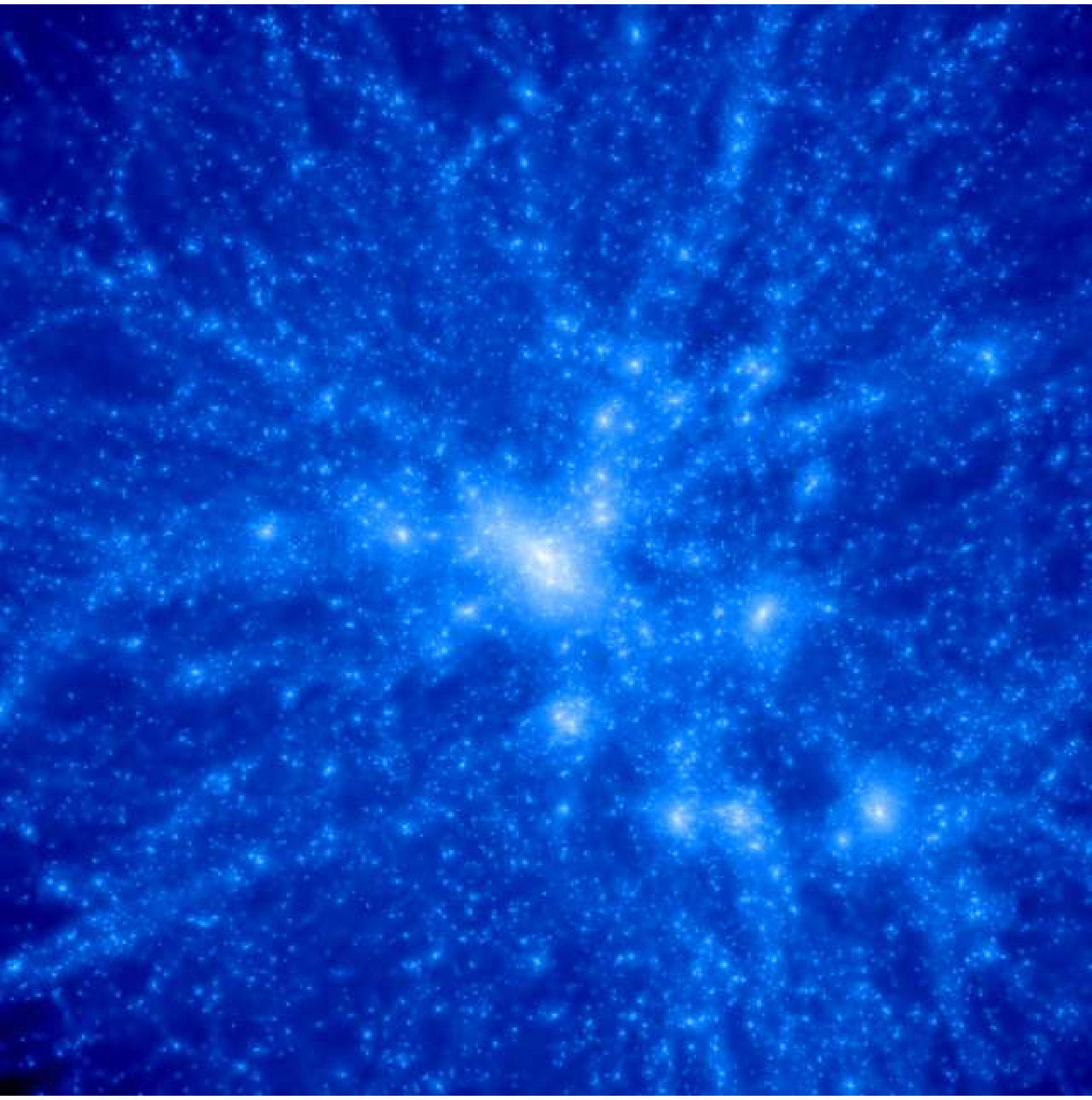,width=0.5\linewidth}
\psfig{file=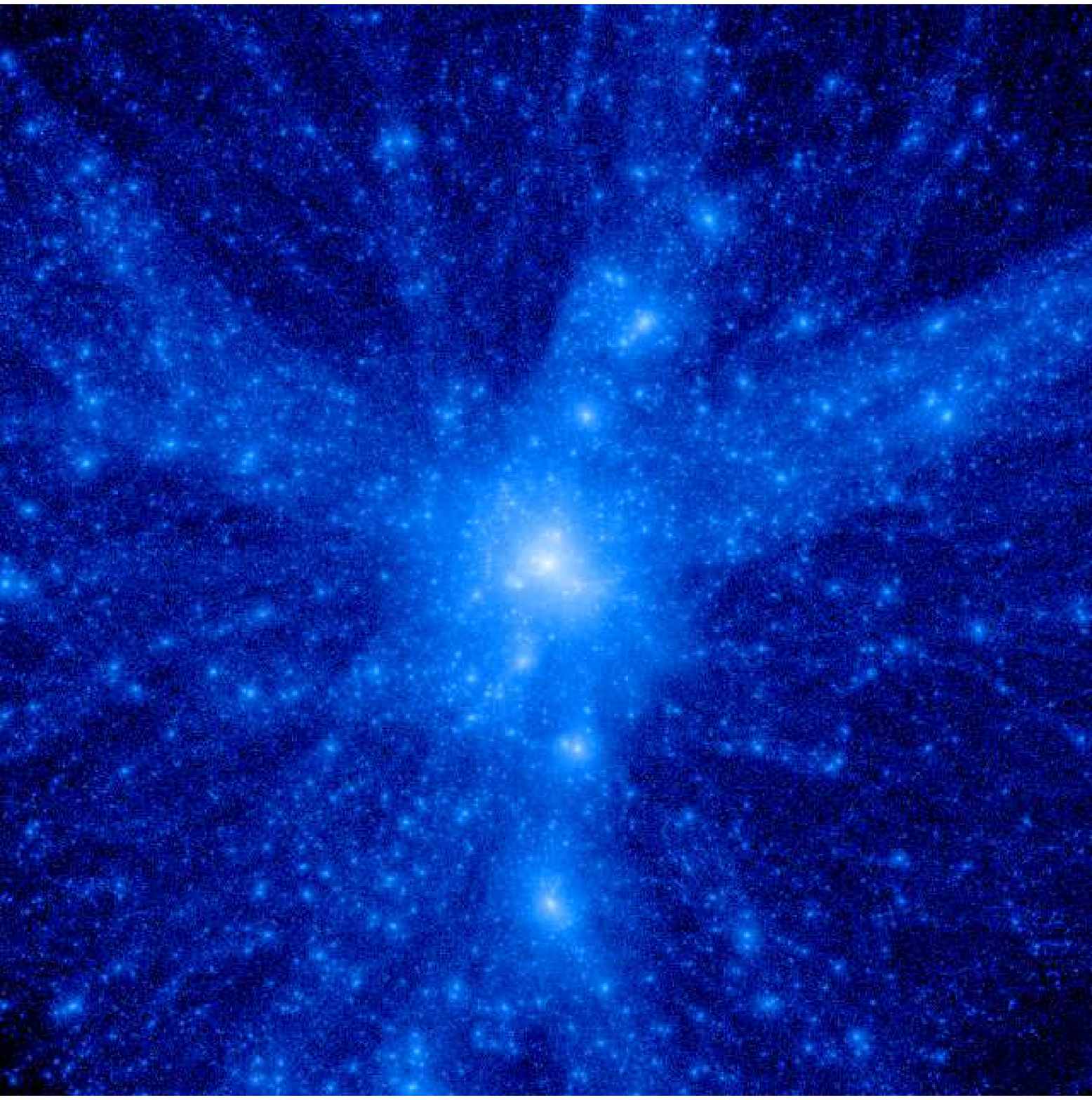,width=0.5\linewidth}
}
\vspace{-1cm}
\caption{Evolution of a dark matter density field in a comoving region
  of $15h^{-1}$~Mpc on a side around cluster mass density peak in the
  initial perturbation field. The four panels, from top left to
  bottom right, show redshifts $z=3$, $z=1$, $z=0.5$ and $z=0$. The
  forming cluster has a mass $M_{200}\simeq 1.2\times
  10^{15}h^{-1}M_\odot$ at $z=0$. The figure illustrates the complexities
  of the actual collapse of real density peaks: strong deviations from
  spherical symmetry, accretion of matter along filaments, and the 
  presence of smaller-scale structure within the collapsing
  cluster-scale mass peak.}
\label{fig:clevol}
\end{figure}

Figure~\ref{fig:clevol} shows evolution of the DM density field in a
cosmological simulation of a comoving region of $15h^{-1}$~Mpc on a
side around cluster mass-scale density peak in the initial
perturbation field from $z=3$ to the present epoch. The overall
picture is quite different from the top-hat collapse. First of all,
real peaks in the primordial field do not have the constant density or
sharp boundary of the top-hat, but have a certain radial profile and
curvature \citep{bardeen_etal86,dalal_etal08}. As a result, different
regions of a peak collapse at different times so that the overall
collapse is extended in time and the peak does not have a single
collapse epoch \citep[e.g.,][]{diemand_etal07}. Consequently, the
distribution of matter around the collapsed peak can smoothly extend
to several virial radii for late epochs and small masses
\citep{prada_etal06,cuesta_etal08}. This creates ambiguity about the
definition of halo mass and results in a variety of mass definitions
adopted in practice, as we discuss in Section~\ref{sec:massdef}.

Second, the peaks in the smoothed density field, $\delta_{\rm R}({\bf
  x})$, are not isolated but are surrounded by other peaks and density
inhomogeneities. The tidal forces from the most massive and rarest
peaks in the initial density field shepherd the surrounding matter
into  massive filamentary structures that connect them 
\citep{bond_etal96}. Accretion of matter onto clusters at late epochs occurs
preferentially along such filaments, as can be clearly seen in
Figure~\ref{fig:clevol}.

Finally, the density distribution within the peaks in the actual
density field is not smooth, as in the smoothed field $\delta_{\rm
  R}({\bf x})$, but contains fluctuations on all scales. Collapse of
density peaks on different scales can proceed almost simultaneously,
especially during early stages of evolution in the CDM models when
peaks undergoing collapse involve small scales, over which the power
spectrum has an effective slope $n\approx -3$. Figure~\ref{fig:clevol}
shows that at high redshifts the proto-cluster region contains mostly
small-mass collapsed objects, which merge to form a larger and larger
virialized system near the center of the shown region at later
epochs. Nonlinear interactions between smaller-scale peaks within a
cluster-scale peak during mergers result in relaxation processes and
energy exchange on different scales, and mass redistribution. Although
the processes accompanying major mergers are not as violent as
envisioned in the violent relaxation scenario
\citep[][]{valluri_etal07}, such interactions lead to significant
redistribution of mass \citep{kazantzidis_etal06} and angular momentum
\citep{vitvitska_etal02}, both within and outside of the virial
radius.

\subsection{Equilibrium}
\label{sec:hse}
Following the collapse, matter settles into an equilibrium
configuration. For collisional baryonic component this configuration
is approximately described by the hydrostatic equilibrium (HE
hereafter) equation, in which the pressure gradient $\nabla p(\bx)$ at
point $\bx$ is balanced by the gradient of local gravitational
potential $\nabla\phi(\bx)$: $\nabla\phi(\bx) = -\nabla
p(\bx)/\rho_{\rm g}(\bx)$, where $\rho_{\rm g}(\bx)$ is the gas
density. Under the further assumption of spherical symmetry, the HE
equation can be written as $\rho_{\rm g}^{-1}dp/dr=-GM(<r)/r^2$, where
$M(<r)$ is the mass contained within the radius $r$. Assuming the
equation of state of ideal gas, $p=\rho_{\rm g}k_BT/\mu m_p$ where
$\mu$ is the mean molecular weight and $m_p$ is the proton mass;
cluster mass within $r$ can be expressed in terms of the density and
temperature profiles, $\rho_{\rm g}(r)$ and $T(r)$, as
\be M_{HE}(
<r)\,=\,-{rk_BT(r)\over G\mu m_p}
\left[\frac{d\ln\rho_{\rm g}(r)}{d\ln r}+ \frac{d\ln T(r)}{d\ln
    r}\right]\,.
\label{eq:m_he}
\ee Interestingly, the slopes of the gas density and temperature
profiles that enter the above equation exhibit correlation that
appears to be a dynamical attractor during cluster formation
\citep{juncher_etal12}.

For a collisionless system of particles, such as CDM, the
condition of equilibrium is given by the Jeans equation
\citep[e.g.,][]{binney_tremaine08}. For a non--rotating spherically
symmetric system, this equation can be written as
\be
M_J(<r)\,=\,-{r\sigma_r^2\over G}\,\left[{d\ln \nu(r) \over d\ln r}+{d\ln
  \sigma_r(r)^2\over d\ln r} +2\beta(r)\right]\,,
\label{eq:j_sphe2}
\ee
where $\beta=1-{\sigma_t^2\over 2\sigma_r^2}$ is the orbit anisotropy
parameter defined in terms of the radial ($\sigma_r$) and tangential
($\sigma_t$) velocity dispersion components ($\beta=0$ for isotropic
velocity field). We consider equilibrium density and velocity
dispersion profiles, as well as anisotropy profile $\beta(r)$ in
\S~\ref{sec:veldisp}.  Equation \ref{eq:j_sphe2} is also commonly used
to describe the equilibrium of cluster galaxies. Although, in
principle, galaxies in groups and clusters are not strictly
collisionless, interactions between galaxies are relatively rare and
the Jeans equation should be quite accurate.

Note that the difference between equilibrium configuration of
collisional ICM and collisionless DM and galaxy systems is
significant. In HE, the iso-density surfaces of the ICM should trace
the iso-potential surfaces. The shape of the iso-potential surfaces in
equilibrium is always more spherical than the shape of the underlying
mass distribution that gives rise to the potential. Given that the
potential is dominated by DM at most of the cluster-centric
radii, the ICM distribution (and consequently the X-ray isophotes and
SZ maps) will be more spherical than the underlying DM distribution.

As we noted in the previous section, the gravitational collapse of a
halo is a process extended in time. Consequently, a cluster may not
reach complete equilibrium over the Hubble time due to ongoing
accretion of matter and the occurrence of minor and major mergers. The
ICM reaches equilibrium state following a major merger only after $\approx
3-4$ Gyr \citep[e.g.,][]{nelson_etal12}. Deviations from equilibrium
affect observable properties of clusters and cause systematic errors
when equations~\ref{eq:m_he} and \ref{eq:j_sphe2} are used to estimate
cluster masses
\citep[e.g.,][]{rasia_etal04,nagai_etal07b,ameglio_etal09,piffaretti_valda08,lau_etal09}.

\subsection{Internal structure of cluster halos}
\label{sec:structure}
Relaxations processes establish the equilibrium internal structure of
clusters. Below we review our current undertstanding of the
equilibrium radial density distribution, velocity dispersion, and
triaxiality (shape) of the cluster DM halos.

\subsubsection{Density Profile.}
\label{sec:denpro}
Internal structure of collapsed halos may be expected to depend both
on the properties of the initial density distribution around
collapsing peaks \citep{hoffman_shaham85} and on the processes
accompanying hierarchical collapse
\citep[e.g.,][]{syer_white98,valluri_etal07}.  The fact that
simulations have demonstrated that the characteristic form of the
spherically averaged density profile arising in CDM models,
characterized by the logarithmic slope steepening with increasing
radius
\citep{dubinski_carlberg91,katz91,navarro_etal95,navarro_etal96}, is
virtually independent of the shape of power spectrum and background
cosmology \citep{katz91,cole_lacey96,navarro_etal97,huss_etal99b} is
non trivial. Such a generic form of the profile also arises when
small-scale structure is suppressed and the collapse is smooth, as is
the case for halos forming at the cut-off scale of the power spectrum
\citep{moore_etal99,diemand_etal05,wang_white09} or even from
non-cosmological initial conditions \citep{huss_etal99}.

The density profiles measured in dissipationless simulations are most
commonly approximated by the ``NFW'' form proposed by
\citet{navarro_etal95} based on their simulation of cluster formation:
\be
\rho_{\rm NFW}(r)=\frac{4\rho_s}{x(1+x)^2},\ \ \ x\equiv r/r_s, 
\ee
where $r_s$ is the scale radius, at which the logarithmic slope of the
profile is equal to $-2$ and $\rho_s$ is the characteristic density at
$r=r_s$. Overall, the slope of this profile varies with radius as
$d\ln\rho/d\ln r=-[1+2x/(1+x)]$, i.e., from the asymptotic slope of
$-1$ at $x\ll 1$ to $-3$ at $x\gg 1$, where the enclosed mass diverges
logarithmically: $M(<r)=M_{\Delta}f(x)/f(c_{\Delta})$, where
$M_{\Delta}$ is the mass enclosing a given overdensity $\Delta$,
$f(x)\equiv \ln(1+x)-x/(1+x)$ and $c_{\Delta}\equiv R_{\Delta}/r_s$ is
the concentration parameter.  Accurate formulae for the conversion of mass of the NFW halos defined for different values of $\Delta$ are given in the appendix of \citet{hu_kravtsov03}.

Subsequent simulations
\citep{navarro_etal04,merritt_etal06,graham_etal06} showed that the
\citet{einasto65} profile and other similar models designed to
describe de-projection of the S\'ersic profile \citep{merritt_etal06}
provide a more accurate description of the DM density profiles arising
during cosmological halo collapse, as well as profiles of bulges and
elliptical galaxies \citep{cardone_etal05}. The Einasto profile is
characterized by the logarithmic slope that varies as a power law with
radius: 
\be \rho_{\rm
  E}(r)=\rho_s\exp\left[\frac{2}{\alpha}(1-x^{\alpha})\right],\ \ \
x\equiv r/r_s\,, 
\ee 
where $r_s$ is again the scale radius at which the
logarithmic slope is $-2$, but now for the Einasto profile,
$\rho_s\equiv\rho_{\rm E}(r_s)$, and $\alpha$ is an additional
parameter that describes the power-law dependence of the logarithmic
slope on radius: $d\ln\rho_{\rm E}/d\ln r=-2x^{\alpha}$.

Note that unlike the NFW profile and several other profiles discussed
in the literature, the Einasto profile does not have an asymptotic
slope at small radii. The slope of the density profile becomes
increasingly shallower at small radii at the rate controlled by
$\alpha$. The parameter $\alpha$ varies with halo mass and redshift:
at $z=0$ galaxy-sized halos are described by $\alpha\approx 0.16$,
whereas massive cluster halos are described by $\alpha\approx
0.2-0.3$; these values increase by $\sim 0.1$ by $z\approx 3$
\citep{gao_etal08}. Although $\alpha$ depends on mass and redshift
(and thus also on the cosmology) in a non-trivial way, \citet[][and see
also \citeauthor{duffy_etal08} \citeyear{duffy_etal08}]{gao_etal08}
showed that these dependencies can be captured as a universal
dependence on the peak height $\nu=\delta_c/\sigma(M,z)$ (see
Section~\ref{sec:Mnl} above): $\alpha=0.0095\nu^2+0.155$.  Finally,
unlike the NFW profile, the total mass for the Einasto profile is
finite due to the exponentially decreasing density at large radii. A
number of useful expressions for the Einasto profile, such as mass
within a radius, are provided by \citet{cardone_etal05},
\citet{mamon_lokas05}, and \citet{graham_etal06}.

The origin of the generic form of the density profile has recently
been explored in detail by \citet{lithwick_dalal11}, who show that it
arises due to two main factors: {\em (a)} the density and triaxiality
profile of the original peak and {\em (b)} approximately adiabatic
contraction of the previously collapsed matter due to deepening of the
potential well during continuing collapse. Without adiabatic
contraction the profile resulting from the collapse would reflect the
shape of the initial profile of the peak. For example, if the initial
profile of mean linear overdensity within radius $r$ around the peak
can be described as $\bar{\delta}_{\rm L}\propto r^{-\gamma}_{\rm L}$,
it can be shown that the resulting differential density profile after
collapse without adiabatic contraction behaves as $\rho(r) \propto
r^{-g}$, where $g=3\gamma/(1+\gamma)$ \citep[][]{fillmore_goldreich84}. Typical profiles of
initial density peaks are characterized by shallow slopes, $\gamma\sim
0-0.3$ at small radii, and very steep slopes at large radii
\citep[e.g.,][]{dalal_etal08}, which means that resulting
profiles after collapse should have slopes varying from $g\approx
0-0.7$ at small radii to $g\approx 3$ at large radii.

However, \citet{lithwick_dalal11} showed that contraction of particle
orbits during subsequent accretion of mass interior to a given radius
$r$ leads to a much more gradual change of logarithmic slope with
radius, such that the regime within which $g\approx 0-0.7$ is shifted
to very small radii ($r/r_{\rm vir}\lesssim 10^{-5}$), whereas at the
radii typically resolved in cosmological simulations the logarithmic
slope is in the range of $g\approx 1-3$, so that the radial dependence of
the logarithmic slope $g(r)=d\ln\rho/d\ln r$ is in good qualitative
agreement with simulation results.  This contraction occurs because
matter that is accreted by a halo at a given stage of its evolution
can deposit matter over a wide range of radii, including small
radii. The orbits of particles that accreted previously have to
respond to the additional mass, and they do so by contracting. For
example, for a purely spherical system in which mass is added slowly
so that the adiabatic invariant is conserved, radii $r$ of spherical
shells must decrease to compensate an increase of $M(<r)$. This model
thus elegantly explains both the qualitative shape of density profiles
observed in cosmological simulations and their universality. The
latter can be expected because the contraction process crucial to
shaping the form of the profile should operate under general collapse
conditions, in which different shells of matter collapse at different
times.

Although the model of \citet{lithwick_dalal11} provides a solid physical picture of halo profile formation, it also neglects some of the processes that may affect details of the resulting density profile, most notably the effects of mergers. Indeed, major mergers lead to resonant dynamical heating of a certain fraction of collapsed matter due to the potential fluctuations and tidal forces that they induce. The amount of mass that is affected by such heating is significant \citep[e.g.,][]{valluri_etal07}. In fact, up to $\sim 40\%$ of mass within the virial radii of merging halos may end up outside of the virial radius of the merger. This implies, for example, that virial mass is not additive in major mergers. Nevertheless, in practice the merger remnant retains the functional form of the density profiles of the merger progenitors \citep{kazantzidis_etal06}, which means that major mergers do not lead to efficient violent relaxation. 

Although the functional form of the density profile arising during
halo collapse is generic for a wide variety of collapse conditions and
models, initial conditions and cosmology do significantly affect
the physical properties of halo profiles
such as its characteristic density and scale radius
\citep[][]{navarro_etal97}. These dependencies are often discussed in
terms of halo concentrations, $c_{\Delta}\equiv R_{\Delta}/r_s$.
Simulations show that the scale radius is approximately constant
during late stages of halo evolution
\citep[][]{bullock_etal01,wechsler_etal02}, but evolves as $r_s=c_{\rm
  min}\,R_{\Delta}$ during early stages, when a halo quickly increases
its mass through accretion and mergers
\citep{zhao_etal03b,zhao_etal09}. The minimum value of concentration
is $c_{\rm min}=\rm const\approx 3-4$ for $\Delta=200$. For massive
cluster halos, which are in the fast growth regime at any redshift,
the concentrations are thus expected to stay approximately constant
with redshift or may even increase after reaching a minimum
\citep{klypin_etal11,prada_etal11}.

The characteristic time separating the two regimes can be identified
as the formation epoch of halos. This time approximately determines
the value of the scale radius and the subsequent evolution of halo
concentration.  The initial conditions and cosmology determine the
formation epoch and the typical mass accretion histories for halos of
a given mass \citep{navarro_etal97,bullock_etal01,zhao_etal09}, and
therefore determine the halo concentrations. Although these
dependences are non-trivial functions of halo mass and redshift, they
can also be encapsulated by a universal function of the peak height
$\nu$ \citep{zhao_etal09,prada_etal11}.

Baryon dissipation and feedback are expected to affect the density
profiles of halos appreciably, although predictions for these effects
are far less certain than predictions of the DM distribution in the
purely dissipationless regime. The main effect is contraction of DM in
response to the increasing depth of the central potential during
baryon cooling and condensation, which is often modelled under the
assumption of slow contraction conserving adiabatic invariants of
particle orbits
\citep[e.g.,][]{zeldovich_etal80,barnes_white84,blumenthal_etal86,ryden_gunn87}. The
standard model of such {\it adiabatic contraction} assumes that DM
particles are predominantly on circular orbits, and for each shell of
DM at radius $r$ the product of the radius and the enclosed mass
$rM(r)$ is conserved \citep{blumenthal_etal86}. The model makes a
number of simplifying assumptions and does not take into account
effects of mergers. Nevertheless, it was shown to provide a reasonably
accurate description of the results of cosmological simulations
\citep{gnedin_etal04}. Its accuracy can be further improved by
relaxing the assumption of circular orbits and adopting an empirical
ans{\"a}tz, in which the conserved quantity is $rM(\bar{r})$, where
$\bar{r}$ is the average radius along the particle orbit, instead of
$rM(r)$ \citep{gnedin_etal04}. At the same time, several recent
studies showed that no single set of parameters of such simple models
describes all objects that form in cosmological simulations equally
well
\citep{gustafsson_etal06,abadi_etal10,tissera_etal10,gnedin_etal11}.

A more subtle but related effect is the increase of the overall
concentration of DM within the virial radius of halos due to
re-distribution of binding energy between DM and baryons during the
process of cluster assembly \citep{rudd_etal08}. The larger range of
radii over which this effect operates makes it a potential worry for
the precision constraints from the cosmic shear power spectrum
\citep{jing_etal06,rudd_etal08}. This effect depends primarily on the
fraction of baryons that condense into the central halo galaxies and
may be mitigated by the blow-out of gas by efficient AGN or SN
feedback \citep{van_daalen_etal11}. The effects of baryons on the
overall concentration of mass distribution in clusters are thus
uncertain, but can potentially increase halo concentration and thereby
significantly enhance the cross section for strong lensing
\citep{puchwein_etal05,rozo_etal08,mead_etal10} and affect statistics
of strong lens distribution in groups and clusters
\citep[e.g.,][]{more_etal11}.

A number of studies have derived observational constraints on density
profiles of clusters and their concentrations
\citep{pointecouteau_etal05,vikhlinin_etal06,schmidt_allen07,buote_etal07,mandelbaum_etal08,wojtak_lokas10,okabe_etal10,ettori_etal10,umetsu_etal11a,umetsu_etal11b,sereno_zitrin11}.
Although most of these studies find that the concentrations of galaxy
clusters predicted by $\Lambda$CDM simulations are in the ballpark of
values derived from observations, the agreement is not perfect and
there is tension between model predictions and observations, which may
be due to effects of baryon dissipation
\citep[e.g.,][]{rudd_etal08,fedeli12}.
 
Some studies do find that the concordance cosmology predictions of the
average cluster concentrations are somewhat lower than the average
values derived from X-ray observations
\citep{schmidt_allen07,buote_etal07,duffy_etal08}. Moreover, lensing
analyses indicate that the slope of the density profile in central
regions of some clusters may be shallower than predicted
\citep{tyson_etal98,sand_etal04,sand_etal08,newman_etal09,newman_etal11},
whereas concentrations are considerably higher than both theoretical
predictions and most other observational determinations from X-ray and
WL analyses
\citep{comerford_natarajan07,oguri_etal09,oguri_etal11,zitrin_etal11}.

At this point, it is not clear whether these discrepancies imply
serious challenges to the $\Lambda$CDM structure formation paradigm,
unknown baryonic effects flattening the profiles in the centers, or
unaccounted systematics in the observational analyses
\citep[e.g.,][]{dalal_keeton03,hennawi_etal07}. When considering such
comparisons, it is important to remember that density profiles in
cosmological simulations are always defined with respect to the center
defined as the global density peak or potential minimum, whereas in
observations the corresponding location is not as unambiguous as
in simulations and the choice of center may affect the derived slope.

It should be noted that improved theoretical predictions for
cluster-sized systems generally predict larger concentrations for the
most massive objects than do extrapolations of the concentration-mass
relations from smaller mass objects
\citep{zhao_etal09,prada_etal11,bhattacharya_etal11a}. In addition, as
we noted above, the evolution predicted for the concentrations of
these rarest objects is much weaker than $c\propto (1+z)$ found for
smaller mass halos, so rescaling the concentrations of high-redshift
clusters by $(1+z)$ factor, as is often done, could lead to an
overestimate of their concentrations.

\subsubsection{Velocity dispersion profile and velocity anisotropy.} 
\label{sec:veldisp}

Velocity dispersion profile is a halo property related to its density
profile. Simulations show that this profile generally increases from
the central value to a maximum at $r\approx r_s$ and slowly decreases
outward \citep[e.g.,][]{cole_lacey96,rasia_etal04}.  One remarkable
result illustrating the close connection between density and velocity
dispersion is that for collapsed halos in dissipationless simulations
the ratio of density to the cube of the rms velocity dispersion can be
accurately described by a power law over at least three decades in
radius \citep{taylor_navarro01}: $Q(r)\equiv \rho/\sigma^3\propto
r^{-\alpha}$ with $\alpha\approx 1.9$.

An important quantity underlying the measured velocity dispersion
profile is the profile of the mean velocity, and the mean radial
velocity, $\bar{v}_{\rm r}$, in particular. For a spherically
symmetric matter distribution in HE, we expect $\bar{v}_{\rm
  r}=0$. Therefore, the profile of $\bar{v}_{\rm r}$ is a useful
diagnostic of deviations from equilibrium at different
radii. Simulations show that clusters at $z=0$ generally have zero
mean radial velocities within $r\approx R_{\rm vir}$ and turn sharply
negative between $1$ and $\approx 3R_{\rm vir}$, where density is
dominated by matter infalling onto cluster
\citep{cole_lacey96,eke_etal98,cuesta_etal08}.

The distinguishing characteristic between gas and DM is the fact that
gas has an isotropic velocity dispersion tensor on small scales,
whereas DM in general does not. On large scales, however, both gas and
DM may have velocity fields that are anisotropic. The degree of
velocity anisotropy is commonly quantified by the anisotropy profile,
$\beta(r)$ (see \S~\ref{sec:hse}).  DM anisotropy is mild:
$\beta\approx 0-0.1$ near the center and increases to $\beta\approx
0.2-0.4$ near the virial radius
\citep{cole_lacey96,eke_etal98,colin_etal00,rasia_etal04,lemze_etal11}. Interestingly,
velocities exhibit substantial tangential anisotropy outside the
virial radius in the infall region of clusters
\citep{cuesta_etal08,lemze_etal11}. Another interesting finding is
that the velocity anisotropy correlates with the slope of the density
profile \citep{hansen_moore06}, albeit with significant scatter
\citep{lemze_etal11}.

The gas component also has some residual motions driven by
mergers and gas accretion along filaments. Gas velocities tend to have
tangential anisotropy \citep[][]{rasia_etal04}, because radial motions
are inhibited by the entropy profile, which is convectively stable in general.

\subsubsection{Shape.}
\label{sec:shape}

Although the density structure of mass distribution in clusters is
most often described by spherically averaged profiles, clusters are
thought to collapse from generally triaxial density peaks
\citep{doroshkevich70,bardeen_etal86}. The distribution of matter within halos resulting from hierarchical collapse is triaxial as well \citep{frenk_etal88,dubinski_carlberg91,warren_etal92,cole_lacey96,jing_suto02,kasun_evrard05,allgood_etal06}, with triaxiality predicted by {\it dissipationless} simulations increasing with decreasing distance from halo center \citep{allgood_etal06}. Triaxiality of halos decreases with decreasing mass and redshift \citep{kasun_evrard05,allgood_etal06} in a way that again can be parameterized in a universal form as a function of peak height \citep{allgood_etal06}.  
The major axis of the triaxial distribution of clusters is generally aligned with the filament connecting a cluster with its nearest neighbor of comparable mass \citep[e.g.,][]{west_etal00,lee_etal08}, which reflects the fact that a significant fraction of mass and mergers is occurring along such filaments \citep[e.g.,][]{onuora_thomas00,lee_evrard07}. 

\citet{jing_suto02} showed how the formalism of density distribution
as a function of distance from cluster center can be extended to the
density distribution in triaxial shells. Accounting for such
triaxiality is particularly important in theoretical predictions and
observational analyses of weak and strong lensing
\citep{dalal_keeton03,clowe_etal04,oguri_etal05,corless_king07,hennawi_etal07,becker_kravtsov11}.
At the same time, it is important to keep in mind that, as with many
other results derived mainly from dissipationless simulations, the
physics of baryons may modify predictions substantially.

The shape of the DM distribution in particular is quite
sensitive to the degree of central concentration of mass. As baryons
condense towards the center to form a central galaxy within a halo,
the DM distribution becomes more spherical
\citep{dubinski94,evrard_etal94,tissera_etal98,kazantzidis_etal04}. The
effect increases with decreasing radius, but is substantial even at
half of the virial radius \citep{kazantzidis_etal04}. The main
mechanism behind this effect lies in adiabatic changes of the shapes
of particle orbits in response to more centrally concentrated mass
distribution after baryon dissipation
\citep{dubinski94,debattista_etal08}.

In considering effects of triaxiality, it is important to remember
that triaxiality of the hot intracluster gas and DM distribution
are different \citep[gas is rounder, see, e.g., discussion in][and references
therein]{lau_etal11}. This is one of the reasons why mass proxies
defined within spherical aperture using observable properties of gas (see \S~\ref{sec:regul} below)
exhibit small scatter and are much less sensitive to cluster orientation. 

The observed triaxiality of the ICM can be used as a probe of the
shape of the underlying potential \citep{lau_etal11} and as a powerful
diagnostic of the amount of dissipation that is occurring in cluster
cores \citep{fang_etal09} and of the mass of the central cluster galaxy \citep{lau_etal12}. 

\subsection{Mass definitions}
\label{sec:massdef}
 
As we discussed above, the existence of a particular density contrast
delineating a halo boundary is predicted only in the limited context of
the spherical collapse of a density fluctuation with the top-hat profile
(i.e., uniform density, sharp boundary). Collapse in such a case
proceeds on the same time scale at all radii and the collapse time and
``virial radius'' are well defined. However, the
peaks in the initial density field are not uniform in density, are not
spherical, and do not have a sharp boundary. Existence of a density
profile results in different times of collapse for different radial
shells. Note also that even in the spherical collapse model the virial density contrast
formally applies only at the time of collapse; after a given density
peak collapses its internal density stays constant while the
reference 
(i.e., either the mean or critical)
density changes merely due to cosmological expansion. The actual
overdensity of the collapsed top-hat initial fluctuations will therefore
grow larger than the initial virial overdensity at $t>t_{\rm collapse}$.

The triaxiality of the density peak makes the tidal effects of the 
surrounding mass distribution important. Absence of a sharp boundary,
along with the effects of non-uniform density, triaxiality and nonlinear
effects during the collapse of smaller scale fluctuations within each
peak, result in a continuous, smooth outer density profile without a
well-defined radial boundary. Although one can identify a radial
range, outside of which a significant fraction of mass is still infalling, this
range is fairly wide and does not correspond to a single well-defined
radius \citep{eke_etal98,cuesta_etal08}. The boundary based on the virial density contrast is, thus, only loosely motivated by theoretical considerations. 

The absence of a well-defined boundary of collapsed objects makes
the definition of the halo boundary and the associated enclosed mass ambiguous. 
This explains, at least partly,
the existence of various halo boundary and mass definitions in the
literature. 
Below we describe the main two such definitions: the Friends-of-Friends (FoF) and spherical overdensity \citep[SO, see also][]{white01}. The FoF mass definition is used almost predominantly in analyses of cosmological simulations of cluster formation, whereas the SO halo definition
is used both in observational and simulation analyses, as well as in analytic models, such as
the Halo Occupation Distribution (HOD) model. Although other definitions of the halo mass are discussed, theoretical mass determinations often have to conform
to the observational definitions of mass. Thus, for example, although
it is possible to define the entire mass that will ever collapse onto
a halo in simulations \citep[][]{cuesta_etal08,anderhalden_diemand11},
it is impossible to measure this mass in observations, which makes
it of interest only from the standpoint of the theoretical models
of halo collapse.

\subsubsection{The Friends-of-Friends mass.}
\label{sec:fof}

Historically, the FoF algorithm was used to define groups and clusters
of galaxies in observations
\citep{huchra_geller82,press_davis82,einasto_etal84} and was adopted
to define collapsed objects in simulations of structure formation
\citep{einasto_etal84,davis_etal85}. The FoF algorithm considers two
particles to be members of the same group (i.e., ``friends''), if they
are separated by a distance that is less than a given linking length. Friends of
friends are considered to be members of a single group -- the
condition that gives the algorithm its name. The linking length, the
only free parameter of the method, is usually defined in units of the
mean interparticle separation: $b=l/\bar{l}$, where $l$ is the
linking length in physical units and $\bar{l}=\bar{n}^{-1/3}$ is the
mean interparticle separation of particles with mean number density of
$\bar{n}$.

Attractive features of the FoF algorithm are its simplicity (it has
only one free parameter), a lack of any assumptions about the halo center, and the fact that it does not assume any
particular halo shape. Therefore, it can better match the generally
triaxial, complex mass distribution of  halos forming in the hierarchical
structure formation models. 

The main disadvantages of the FoF algorithm are the difficulty in
theoretical interpretation of the FoF mass, and sensitivity of the FoF
mass to numerical resolution and the presence of substructure.  For
the smooth halos resolved with many particles the FoF algorithm with
$b=0.2$ defines the boundary corresponding to the fixed {\it local\/}
density contrast of $\delta_{\rm FoF}\approx 81.62$
\citep{more_etal11}. Given that halos forming in hierarchical
cosmologies have concentrations that depend on mass, redshift, and
cosmology, the {\it enclosed\/} overdensity of the FoF halos also
varies with mass, redshift and cosmology. Thus, for example, for the
current concordance cosmology the FoF halos (defined with $b=0.2$) of
mass $10^{11}-10^{15}\ \Msun$ have enclosed overdensities of $\sim
450-350$ at $z=0$ and converge to overdensity of $\sim 200$ at high
redshifts where concentration reaches its minimum value of $c\approx
3-4$ \citep{more_etal11}. For small particle numbers the boundary of
the FoF halos becomes ``fuzzier'' and depends on the resolution (and
so does the FoF mass).  Simulations most often have fixed particle
mass and the number of particles therefore changes with halo mass,
which means that properties of the boundary and mass identified by the
FoF are mass dependent. The presence of substructure in well-resolved
halos further complicates the resolution and mass dependence of the
FoF-identified halos \citep{more_etal11}. Furthermore, it is well
known that the FoF may spuriously join two neighboring distinct halos
with overlapping volumes into a single group. The fraction of such
neighbor halos that are ``bridged'' increases significantly with
increasing redshift.

\subsubsection{The Spherical Overdensity mass.}
\label{sec:so}

The spherical overdensity algorithm defines the boundary of a halo as
a sphere of radius enclosing a given density contrast $\Delta$ with
respect to the reference density $\rho$. Unlike the FoF algorithm the
definition of an SO halo also requires a definition of the halo
center. The common choices for the center in theoretical analyses are
the peak of density, the minimum of the potential, the position of the
most bound particle, or, more rarely, the center of mass. Given that
the center and the boundary need to be found simultaneously, an
iterative scheme is used to identify the SO boundary around a
given peak. The radius of the halo boundary, $\Rdc$, is defined by
solving the implicit equation \be M(<r)=\frac{4\pi}{3}\Delta\rho(z) r^3,
\label{eq:mdef}
\ee where $M(<r)$ is the total mass profile and $\rho(z)$ is the
reference physical density at redshift $z$ and $r$ is in physical (not
comoving) radius.

The choice of $\Delta$ and the reference $\rho$ may be motivated by
theoretical considerations or by observational limitations. For
example, one can choose to define the enclosed overdensity to be equal to
the ``virial'' overdensity at collapse predicted by the spherical
collapse model, $\Delta\rho=\Delta_{\rm vir,c}\rho_{\rm crit}$ (see
Section~\ref{sec:sphcoll}).  Note that in $\Omm\ne 1$ cosmologies,
there is a choice for reference density to be either the critical
density $\rhoc(z)$ or the mean matter density $\rhom(z)$ and both are
in common use. The overdensities defined with respect to these
reference densities, which we denote here as $\Dc$ and $\Dm$, are related
as $\Dm=\Dc/\Omm(z)$.  Note that $\Omm(z)=\Omega_{\rm
  m0}(1+z)^3/E^2(z)$, where $E(z)$ is given by
Equation~\ref{eq:Ea}. For concordance cosmology, $1-\Omm(z)<0.1$ at
$z\geq 2$ and the difference between the two definitions decreases at
these high redshifts. In observations, the choice may simply be determined by the extent of
the measured mass profile. Thus, masses derived from X-ray data under the 
assumption of HE are limited by the extent of the 
measured gas density and temperature profiles and are therefore often
defined for the high values of overdensity: $\Delta_{\rm c}=2,500$ or
$\Delta_{\rm c}=500$. 
  
The crucial difference from the FoF algorithm is the fact that the SO
definition forces a spherical boundary on the generally non-spherical
mass definition. In addition, spheres corresponding to different halos
may overlap, which means that a certain fraction of mass may be double
counted \citep[although in practice this fraction is very small, see,
e.g., discussion in \S~2.2 of][]{tinker_etal08}.
  
The advantage of the SO algorithm is the fact that the SO-defined mass
can be measured both in simulations and observational analyses of
clusters. In the latter the SO mass can be estimated from the total
mass profile derived from the hydrostatic and Jeans equilibrium
analysis for the ICM gas and galaxies, respectively (see
Section~\ref{sec:hse} above), or gravitational lensing analyses
\citep[e.g.,][]{vikhlinin_etal06,hoekstra07}. 
Furthermore, suitable observables that correlate with
the SO mass with scatter of $\mincir 10\%$ can be defined (see \S~\ref{sec:regul} below),
thus making this mass definition preferable in the cosmological
interpretation of observed cluster populations. The small scatter
shows that the effects of triaxiality is quite small in
practice. Note, however, that the definition of the halo center in
simulations and observations may not necessarily be identical, because
in observations the cluster center is usually defined at the position
of the peak or the centroid of X-ray emission or SZ signal, or at the
position of the BCG.

\subsection{Abundance of halos}
\label{sec:mf}

Contrasting predictions for the abundance and clustering of collapsed objects
with the observed abundance and clustering of galaxies, groups, and clusters
has been among the most powerful validation tests of structure
formation models \citep[e.g.,][]{press_schechter74,blumenthal_etal84,kaiser84,kaiser86}.

Although real clusters are usually characterized by some quantity
derived from observations (an {\it observable}), such as the X-ray
luminosity, such quantities are generally harder to predict {\it ab
  initio} in theoretical models because they are sensitive to
uncertain physical processes affecting the properties of cluster
galaxies and intracluster gas. Therefore, the predictions for the
abundance of collapsed objects are usually quantified as a function of
their mass, i.e., in terms of the mass function $dn(M,z)$ defined
as the comoving volume number density of halos in the mass interval
$[M,M+dM]$ at a given redshift $z$. The predicted mass function is
then connected to the abundance of clusters as a function of an
observable using a calibrated mass-observable relation (discussed in
\S~\ref{sec:regul} below). Below we review theoretical models for halo abundance and underlying reasons for its approximate universality.

\subsubsection{The mass function and its universality.}  
 
The first statistical model for the abundance of collapsed objects as
a function of their mass was developed by
\citet{press_schechter74}. The main powerful principle underlying this
model is that the mass function of objects resulting from nonlinear
collapse can be tied directly and uniquely to the statistical
properties of the initial {\it linear\/} density contrast field
$\delta({\bx})$.

Statistically, one can define the probability $F(M)$
that a given region within the initial overdensity field smoothed on a
mass scale $M$, $\delta_{\rm M}({\bf x})$, will collapse into a halo
of mass $M$ or larger: 
\be 
F(M)=\int^{\infty}_{-1}p(\delta)C_{\rm coll}(\delta) d\delta,
\label{eq:FM}
\ee
where $p(\delta)d\delta$ is the PDF of $\delta_{\rm M}({\bf x})$, which is given by Equation~\ref{eq:pGaussian} for the Gaussian initial density field, and $C_{\rm coll}$ is the probability that any given point ${\bf x}$ with local overdensity $\delta_{\rm M}({\bf x})$ will actually collapse. The mass function can then be derived 
as a fraction of the total volume collapsing into halos of mass $(M,M+dM)$, i.e., $dF/dM$, divided by the comoving volume within the initial density field occupied by each such halo, i.e., $M/\bar{\rho}_{\rm}$:
\be
\frac{dn(M)}{dM}=\frac{\bar{\rho}_{\rm m}}{M}\left\vert\frac{dF}{dM}\right\vert.
\label{eq:dndMdef}
\ee

In their pioneering model, \citet{press_schechter74} have adopted the
ans{\"a}tz motivated by the spherical collapse model (see \S~\ref{sec:collbasics})  that any point in space with $\delta_{\rm M}({\bf x})D_{+0}(z)\geq \delta_c$ will collapse into a halo of mass $\geq M$ by redshift $z$: i.e., $C_{\rm coll}(\delta)=\Theta(\delta-\delta_c)$, where $\Theta$ is the Heaviside step function. Note that $\delta_{\rm M}({\bf x})$ used above is not the actual initial overdensity, but the initial overdensity evolved to $z=0$ with the linear growth rate. One can easily check that for a Gaussian initial density field this assumption gives $F(M)=\frac{1}{2}{\rm erfc}[\delta_c/(\sqrt{2}\sigma(M,z))]=F(\nu)$. This line of arguments and assumptions thus leads to an important conclusion that {\it the abundance of halos of mass $M$ at redshift $z$ is a universal function of only their peak height} $\nu(M,z)\equiv\delta_c/\sigma(M,z)$. 
In particular, the fraction of mass in halos per logarithmic interval of mass in such a model is: 
\begin{equation}
\frac{dn(M)}{d\ln M}=\frac{\bar{\rho}_{\rm m}}{M}\left\vert\frac{dF}{d\ln M}\right\vert=\frac{\bar{\rho}_{\rm m}}{M}\left\vert\frac{d\ln\nu}{d\ln M}\frac{\partial F}{\partial\ln\nu}\right\vert\equiv\frac{\bar{\rho}_{\rm m}}{M}\left\vert\frac{d\ln\nu}{d\ln M}\right\vert\,g(\nu)\equiv \frac{\bar{\rho}_{\rm m}}{M}\,\psi(\nu).
\label{eq:PSmf}
\end{equation}

\begin{figure}
 \centerline{ 
\psfig{file=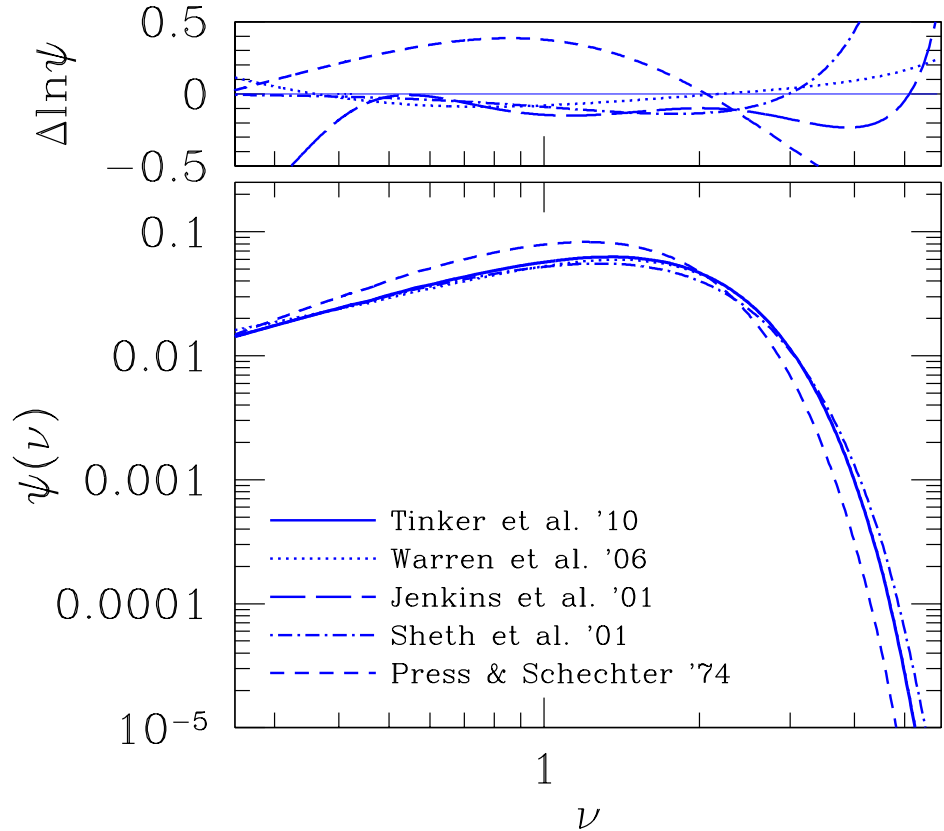,width=14.truecm}
}
\vspace{-1.5cm}
\caption{The function $\psi(\nu)$ defining the comoving abundance of collapsed halos via $dn/d\ln M=(\bar{\rho}_{\rm m}/M)\psi(\nu)$ as a function of $\nu$ from different models and simulation-based calibrations. The upper panel shows deviations of specific models and calibrations for $z=0$ from  \citet{tinker_etal10} based on a suite of $\Lambda$CDM cosmological simulations. }
\label{fig:fnu}
\end{figure}

\begin{figure}
 \centerline{ 
\psfig{file=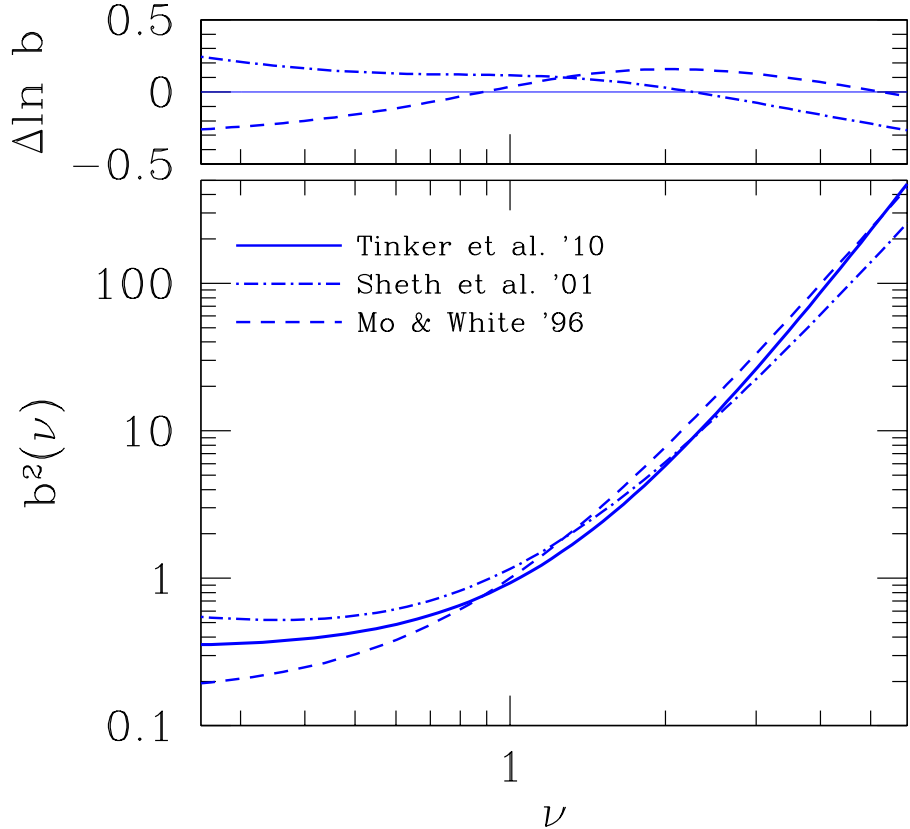,width=14.truecm}
}
\vspace{-1.5cm}
\caption{The square of the bias, $b^2(\nu)$ as a function of peak height $\nu$ corresponding to halos of mass $M_{200m}$ in the bias model based excursion set and  spherical collapse barrier \protect\citep[dashed line][]{mo_white96}, in the excursion set model based on model of ellipsoidal collapse \protect\citep[dot-dashed line][]{sheth_etal01}, and the bias function calibrated using $\Lambda$CDM cosmological simulations \protect\citep[solid line][]{tinker_etal10} for SO halos defined using overdensity of $\Delta=200$ with respect to the mean density. The upper panel shows deviations of excursion set models from the calibration of \citet{tinker_etal10}.}
\label{fig:bnu}
\end{figure}

Clearly, the shape $\psi(\nu)$ in such models is set by the
assumptions of the collapse model. Numerical studies based on
cosmological simulations have eventually revealed that the shape
$\psi_{\rm PS}(\nu)$ predicted by the \citet{press_schechter74} model
deviates by $\gtrsim 50\%$ from the actual shape measured in
cosmological simulations
\citep[e.g.,][]{klypin_etal95,gross_etal98,tormen98,lee_shandarin99,sheth_tormen99,jenkins_etal01}.

A number of modifications to the original ans{\"a}tz have been proposed, which result in $\psi(\nu)$ that more accurately describes simulation results. Such
modifications are based on the collapse conditions that take into
account asphericity of the peaks in the initial density field
\citep{monaco95,audit_etal97,lee_shandarin98,sheth_tormen02,desjacques08}
and stochasticity due to the dependence of the collapse condition on
peak properties other than $\nu$ or shape
\citep[e.g.,][]{desjacques08,maggiore_riotto10,desimone_etal11,ma_etal11,corasaniti_achitouv10}. The
more sophisticated excursion set models match the simulations more
closely, albeit at the expense of more assumptions and
parameters. There may be also inherent limitations in the accuracy of
such models given that they rely on the strong assumption that one can
parameterize all the factors influencing collapse of any given point
in the initial overdensity field in a relatively compact form. In the
face of complications to a simple picture of peak collapse, as discussed
in Section~\ref{sec:realcoll}, one can indeed expect that the
excursion set ans\"atze are limited in how accurately they can
ultimately describe the halo mass function.

\subsubsection{Calibrations of halo mass function in cosmological simulations.}
An alternative route to derive predictions for halo abundance
accurately is to calibrate it using large cosmological simulations of
structure formation. Simulations have generally confirmed the
remarkable fact that the abundance of halos can be parameterized via
a universal function of peak height $\nu$
\citep{sheth_tormen99,jenkins_etal01,evrard_etal02,white02,warren_etal06,reed_etal07,lukic_etal07,tinker_etal08,crocce_etal10,courtin_etal11,bhattacharya_etal11b}. 
Note that in many studies the linear overdensity for collapse is assumed to be constant across redshifts
and cosmologies and the mass function is therefore quantified as a
function of $\sigma^{-1}$ -- the quantity proportional to
$\nu$. However, as pointed out by \citet{courtin_etal11} it is
necessary to include the redshift and cosmology dependence of
$\delta_c(z)$ for an accurate description of the mass function across
cosmologies. Even though $\delta_c$ varies only by $\sim 1-2\%$, it
enters into halo abundance via an exponent and such small variations
can  result in variations in the mass function of several
per cent or more.
  
The main efforts with simulations have thus been aimed at improving
the accuracy of the $\psi(\nu)$ functional form, assessing
systematic uncertainties related to the mass definition, and
quantifying deviations from the universality of $\psi(\nu)$ for
different redshifts and cosmologies.  The mass function, and
especially its exponential tail, is sensitive to the specifics of
halo mass definition, a point emphasized strongly in a number of
studies \citep{jenkins_etal01,white02,tinker_etal08,cohn_white08,klypin_etal11}. Thus,
in precision cosmological analyses using an observed cluster abundance,
care must be taken to ensure that the cluster mass definition matches
that used in the calibration of the halo mass function.

Predictions for the halo abundance as a function of the SO mass for a
variety of overdensities used to define the SO boundaries, accurate to
better than $\approx 5-10\%$ over the redshift interval $z=[0,2]$,
were presented by \citet{tinker_etal08}. In Figure~\ref{fig:fnu} we
compare the form of the function $\psi(\nu)$ calibrated through
simulations by different researchers and compared to $\psi(\nu)$ predicted
by the Press-Schechter model, and to the calibration of the functional
form based on the ellipsoidal collapse ans{\"a}tz by \citet{sheth_etal01}.

These calibrations of the mass function through N--body simulations
provide the basis for the use of galaxy clusters as tools to constrain
cosmological models through the growth rate of perturbations (see the
recent reviews by \citealt{allen_etal11} and
\citealt{weinberg_etal12}). As we discuss in Section~\ref{sec:cosmo}
below, similar calibrations can be extended to models with
non-Gaussian initial density field and models of modified gravity.

Future cluster surveys promise to provide tight constraints on
cosmological parameters, thanks to the large statistics of clusters
with accurately inferred masses. The potential of such surveys clearly
requires a precise calibration of the mass function, which currently
represents a challenge.  Deviations from universality at the level of
up to $\sim 10\%$ have been reported
\citep{reed_etal07,lukic_etal07,tinker_etal08,cohn_white08,crocce_etal10,courtin_etal11}. In
principle, a precise calibration of the mass function is a challenging
but tractable technical problem, as long as it only requires a large
suite of {\it dissipationless} simulations for a given set of
cosmological parameters, and an optimal interpolation procedure
\citep[e.g.,][]{lawrence_etal10}.

A more serious challenge is the modelling of uncertain effects of baryon
physics: baryon collapse,
dissipation, and dynamical evolution, as well as feedback effects
related to energy release by the SNe and AGN, may lead to subtle
redistribution of mass in halos. Such redistribution can affect halo mass
and thereby halo mass function at the level of a few per cent
\citep{rudd_etal08,stanek_etal09,cui_etal11}, although the exact magnitude of the effect is not yet certain due to
uncertainties in our understanding of the physics of galaxy formation in
general, and the process of condensation and dynamical evolution in
clusters in particular. 

\subsection{Clustering of halos}
\label{sec:bias}

Galaxy clusters are clustered much more strongly than galaxies
themselves. It is this strong clustering discovered in the early 1980s
\citep{klypin_kopylov83,bahcall_soneira83} that led to the development
of the concept of bias in the context of Gaussian initial density
perturbation field \citep{kaiser84}. Linear bias of halos is the coefficient between the overdensity of halos within a given sufficiently large region and the overdensity of matter
in that region: $\delta_{\rm h}=b\delta_{\rm}$, with $b$ defined as
the bias parameter. For the 
Gaussian perturbation fields, local linear bias is independent of
scale \citep{scherrer_weinberg98}, such that
the large--scale power spectrum and correlation function on
large scales can be expressed in terms of the corresponding quantities
for the underlying matter distribution as $P_{\rm
  hh}(k)=b^2P_{\rm mm}(k)$ and $\xi_{\rm hh}(r)=b^2\xi_{\rm mm}(r)$, respectively. As we discuss in Section \ref{sec:cosmo}, this
is not true for non-Gaussian initial perturbation fields
\citep{dalal_etal08} or for models with scale--dependent linear growth
rate \citep{parfrey_etal11}, in which cases the
linear bias is generally scale-dependent.

In the context of the hierarchical structure formation, halo bias is
closely related to the overall abundance of halos discussed above, as
illustrated by the ``peak-background split'' framework
\citep{kaiser84,cole_kaiser89,mo_white96,sheth_tormen99}, in which the
linear halo bias is obtained by considering a Lagrangian patch of
volume $V_0$, mass $M_0$, and overdensity $\delta_0$ at some early
redshift $z_0$. The bias is calculated by requiring that the abundance
of collapsed density peaks within such a patch is described by the same
function $\psi(\nu_p)$ as the mean abundance of halos in the Universe,
but with peak height $\nu_p$ appropriately rescaled with respect to
the overdensity of the patch and relative to the rms fluctuations on
the scale of the patch. Thus, the functional form of the bias
dependence on halo mass, $b_h(M)$, depends on the functional form of
the mass function explicitly in this framework. Simulations show that
the peak-background split model provides a fairly accurate (to $\sim
20\%$) prediction for the linear halo bias \citep{tinker_etal10}.

Another line of argument illustrating the close connection between the halo abundance and bias is the fact that if one assumes that all of the mass is in the collapsed halos, as is done for example in the halo models \citep{cooray_sheth02}, the requirement that matter in the Universe is not biased against itself implies that $\int b(\nu)g(\nu)d\ln\nu=1$ \citep[e.g.,][]{seljak00}, where $g(\nu)\equiv \vert d\ln\nu/d\ln M\vert^{-1}\psi(\nu)$ (see eq.~\ref{eq:PSmf}). This integral constraint requires that the form of the bias function $b(\nu)$ is changed whenever $\psi(\nu)$ changes. 
Incidentally, the close connection between $b(\nu)$ and $\psi(\nu)$ implies that if $\psi(\nu)$ is a universal function, then the bias $b(\nu)$ should be a universal function as well. 

The function $b(\nu)$ recently calibrated for the SO-defined halos of different overdensities using a suite of large
cosmological simulations with accuracy $\lesssim 5\%$ and satisfying
the integral constraint \citep{tinker_etal10} is shown in
Figure~\ref{fig:bnu} for halos defined using an overdensity of $\Delta=200$ with respect to the mean. This calibration of the bias is compared to the corresponding prediction of the \citet{press_schechter74} and the \citet{sheth_etal01}
ans\"atze. The figure shows that $b(\nu)$ is a rather weak function of
$\nu$ at $\nu<1$, but steepens substantially for rare peaks of
$\nu>1$. It also shows that the rarest clusters ($\nu \sim 5$) in the Universe can have the amplitude of the correlation function or power spectrum that is two orders of magnitude larger than the clustering amplitude of the galaxy-sized halos ($\nu\lesssim 1$).

%
\subsection{Self--similar evolution of galaxy clusters}
\label{sec:selfsimilar}

In the previous sections we have considered processes that govern the
collapse of matter during cluster formation, the transition to
equilibrium and the equilibrium structure of matter distribution in
collapsed halos. In the following sections, we consider baryonic
processes that shape the observable properties of clusters, such as
their X-ray luminosity or the temperature of the ICM. However, before
we delve into the complexities of such physical processes, it is
instructive to introduce the simplest models based on assumptions of
self-similarity, in which the number of control parameters is minimal.
We discuss the assumptions and predictions of the self-similar
model in some detail because parametric scalings that it predicts 
are in wide use to interpret results from both cosmological
simulations of cluster formation and observations.

\subsubsection{Self-similar model: assumptions and basic expectations.}
\label{sec:ssm}

The self-similar model developed by \citet{kaiser86} makes three key
assumptions. The first assumption is that clusters form via
gravitational collapse from peaks in the initial density field in 
an Einstein--de-Sitter Universe, $\Omega_m=1$.
Gravitational collapse in such a 
Universe is scale-free, or {\it self-similar}. The second assumption
is that the amplitude of density fluctuations is a power-law function of
their size, $\Delta(k)\propto k^{3+n}$. This means that initial
perturbations also do not have a preferred scale (i.e., they are
scale-free or self-similar). The third assumption is that the physical
processes that shape the properties of forming clusters do not introduce
new scales in the problem. With these assumptions the problem has only
two control parameters: the normalization of the power spectrum of the
initial density perturbations at an initial time, $t_{\rm i}$, and its
slope, $n$. Properties of the density field and halo population at $t>t_{\rm i}$ (or corresponding redshift
$z<z_{\rm i}$), such as typical halo masses that collapse or halo abundance as a function of mass, depend only on these two parameters. One can choose any suitable variable that depends on these two parameters as a characteristic variable for a given problem. For evolution of halos and their abundance, the commonly used choice is to define the characteristic nonlinear mass, $\Mnl$
(see Section \ref{sec:collbasics}), which encapsulates such dependence. The halo properties and halo abundance then become universal functions of $\mu\equiv M/\Mnl$ in such model.  Thus, for
example, clusters with masses $M_1(z_1)$ and $M_2(z_2)$ that
correspond to the same ratio $M_1(z_1)/\Mnl(z_1)=M_2(z_2)/\Mnl(z_2)$
 have the same dimensionless properties, such as gas fraction or 
concentration of their mass
distribution. 

As we have discussed above, in more general cosmologies the halo
properties and mass function should be universal functions of the peak
height $\nu$, which encapsulates the dependence on the shape and
normalization of the power spectra for general, non power-law shapes
of the fluctuation spectrum.

\subsubsection{The Kaiser model for cluster scaling relations.}
\label{sec:ssrel}

In the this Section, we define cluster mass to be the mass within the sphere of radius $R$, encompassing characteristic density contrast, $\Delta$, with respect to some reference density $\rho_r$ (usually either $\rhoc$ or $\rhom$): $M=(4\pi/3)\Delta\rho_r R^3$.  In this definition, radius and mass are directly related and interchangeable. The model assumes spherical symmetry and that the ICM is in equilibrium within the cluster gravitational potential, so that the HE equation (eq.~\ref{eq:m_he}) holds. The mass $M(<R)$ derived from the HE equation is proportional to $T(R)R$ and the sum of the logarithmic slopes of the gas density and temperature profiles at $R$. 
{\it In addition} to the assumptions about self-similarity discussed
above, a key assumption made in the model by \citet{kaiser86} is that
these slopes are independent of $M$, so that 
\be
T\propto\frac{M}{R}\propto (\Delta\rho_r)^{\frac{1}{3}}M^{\frac{2}{3}}. 
\label{eq:TMo}
\ee 
Note that formally the quantity $T$ appearing in the above
equation is the temperature measured at $R$, whereas some average
temperature at smaller radii is usually measured in
observations. However, if we parameterize the temperature profile as
$T(r)=\Ts\tT(x)$, where $\Ts$ is the characteristic temperature and
$\tT$ is the dimensionless profile as a function of dimensionless radius
$x\equiv r/R$, and we assume that $\tT(x)$ is independent of $M$, any
temperature averaged over the same fraction of radial range
$[x_1,x_2]$ will scale as $\propto \Ts\propto T(R)$. 
The latter is not strictly true for the ``spectroscopic'' temperature,
$\Tx$, derived by fitting an observed X--ray spectrum to a
single--temperature bremsstrahlung model
\citep{mazzotta_etal04,vikhlinin06}, although in practice deviations
of $\Tx$ from the expected behavior for $\Ts$ are small.

The gas mass within $R$ can be computed by integrating over the gas density profile, which,  by analogy with temperature, we parameterize as
  $\rhog(r)=\rhogs\trhog(x)$, where $\rhogs$ is the characteristic
  density and $\trhog$ is the dimensionless profile. The gas mass within $R$ can then be expressed as
\be
\Mg(<R)=4\pi\rhogs R^3\int^{1}_0x^2\trhog(x)dx=3M\frac{\rhogs}{\Delta\rho_r} I_{\rhog}\propto M(<R).
\label{eq:MgMo}
\ee 
The latter proportionality is assumed in the \citet{kaiser86} model, which means that $\rhogs$ and $I_{\rhog}$ are assumed to be independent of $M$. Note that $\rhogs\propto \Delta\rho_r$, so $\Delta\rho_r$ does not enter the $\Mg$-$M$ relation. 

Using the scalings of $\Mg$ and $T$ with mass, we can construct other
cluster properties of interest, such as the luminosity of ICM emitted
due to its radiative cooling. Assuming that the ICM emission is due to the free-free radiation and neglecting the weak logarithmic dependence of the Gaunt factor on temperature, the bolometric luminosity can be written as \citep[e.g.,][]{sarazin86}:
\be
\Lbol\propto \rhog^2T^{\frac{1}{2}}V\propto \frac{\Mg^2}{V} T^{\frac{1}{2}}\propto \Delta^{\frac{7}{6}}M^{\frac{4}{3}}.
\label{eq:LbolM}
\ee
We omit $\rho_r$ in these equations for clarity; it suffices to remember that $\rho_r$ enters into the scaling relations exactly as $\Delta$. Note that the bolometric luminosity of a cluster is not observable directly, and the X-ray luminosity in soft band (e.g., $0.5-2$~keV), $\Lsx$, is frequently used. Such soft band X-ray luminosity is almost insensitive to temperature at $T>2$~keV \citep[][as can be easily verified with a plasma emission code]{fabricant_gorenstein83}, so that its temperature dependence can be neglected. $\Lsx$ then scales as:
\be
\Lsx\propto \rhog^2 V\propto\frac{\Mg^2}{V}\propto \Delta M.
\ee
At temperatures $T<2$~keV temperature dependence is more complicated
both for the bolometric and soft X-ray emissivity due to significant
flux in emission lines. 
Therefore, strictly speaking, for lower mass systems the above $L-M$ scaling relations are not applicable, and scaling of the emissivity with
temperature needs to be calibrated separately taking also into account
the ICM metallicity. The same is true for luminosity defined in some other energy band or for the bolometric luminosity.
 
Another quantity of interest is the ICM ``entropy'' defined in X-ray analyses as
\be
K_{\rm X}\equiv \frac{k_{\rm B}\Tx}{n_e^{2/3}}\propto \rhog^{-2/3}T\propto\Delta^{-1/3}M^{2/3}\,.
\label{eq:KMo}
\ee
where $n_e$ is the electron number density. 
Finally, the quantity, $Y=M_{\rm g}T$ where gas mass and temperature
are measured within a certain range of radii scaled to $R_\Delta$, 
is used  to characterize the ICM in the analyses of
SZ and X-ray observations. This
quantity is expected to be a particularly robust proxy of the cluster
mass \citep[e.g.,][see also discussion in
Section \ref{sec:regul}]{dasilva_etal04,motl_etal05,nagai06,kravtsov_etal06,fabjan_etal11}
because it is proportional to the global thermal energy of ICM. Using
Equations \ref{eq:MgMo} and \ref{eq:TMo} the scaling of $Y$ with mass
in the self-similar model is
\be 
Y\equiv \Mg T\propto \Delta^{1/3}M^{5/3}.
\label{eq:YMo}
\ee

Note that the redshift dependence in the normalization of the scaling
relations introduced above {\it is due solely to the particular SO
  definition of mass} and associated redshift dependence of
$\Delta\rho_r$. In $\Omm\ne 1$ cosmologies, there is a choice of
either defining the mass relative to the mean matter density or
critical density (Section~\ref{sec:so}).  This specific, {\it
  arbitrary} choice determines the specific redshift dependence of
the observable-mass relations. It is clear that this evolution due to
$\Delta(z)$ factors has no deep physical meaning. However,
the {\it absence} of any additional redshift dependence in the
normalization of the scaling relations is just the consequence of the
assumptions of the \citet{kaiser86} model and is a physical reflection
of these assumptions.

Extra evolution can, therefore, be expected if one or more of the
assumptions of the self--similar model is violated. This can be due to
either actual physical processes that break self-similarity {\it or}
 the fact that some of the model assumptions are not accurate. We
 discuss physical processes that lead to the breaking of
self-similarity in subsequent sections. Here below we first
consider possible deviations that may arise because  assumptions of
the Kaiser model do not hold exactly, i.e. deviations not
ascribed to physical processes that explicitly violate self-similarity. 

\subsubsection{Extensions of the Kaiser model.}

Going back to equations~\ref{eq:TMo} and \ref{eq:MgMo}, we note that
the specific scaling of  $T\propto M^{2/3}$ and $\Mg\propto M$ will only
hold, if the assumption that the dimensionless temperature and gas
density profiles, $\tT(x)$ and $\trhog(x)$, are independent of $M$
holds. In practice, however, some mass dependence of these profiles is
expected. For example, if the concentration of the gas distribution
depends on mass similarly to the concentration of the DM
profile \citep{ascasibar_etal06}, the weak mass dependence of DM concentration implies weak mass dependence of $\trhog(x)$
and $\tT(x)$. Indeed, concentration depends on mass even in purely
self-similar models \citep{cole_lacey96,navarro_etal97}. These
dependencies imply that predictions of the Kaiser model may not
describe accurately even the purely self-similar evolution. This is
evidenced by deviations of scaling relation evolution from these
predictions in hydrodynamical simulations of cluster formation even in
the absence of any physical processes that can break self-similarity
\citep[e.g.,][]{nagai06,stanek_etal10}.

In addition, the characteristic gas density, $\rhogs$, may be mildly
modified by a mass-dependent, non self-similar process during some
early stage of evolution. If such a process does not introduce a
pronounced mass scale and is confined to some early epochs (e.g., owing
to shutting off of star formation in cluster galaxies due to AGN
feedback and gas accretion suppression), then subsequent evolution may
still be described well by the self-similar model. The Kaiser model is
just the simplest {\it specific} example of a more general class of
self-similar models, and can therefore be extended to take into
account deviations described above.

For simplicity, let us assume that the scalings of gas mass and gas
mass fraction against total mass can be expressed as a power law of mass: 
\be
\Mgas=\Cg M^{1+\ag},\ \ \ f_{\rm g}\equiv \frac{\Mgas}{M}=\Cg M^{\ag}=\Cg \Mnl^{\ag}\mu^{\ag}\,.
\label{eq:fga}
\ee 
This does not violate the self-similarity of the problem per se, as long as dimensionless properties of an object,
such as $\fgas$, remain  a function of only the dimensionless mass
$\mu(z)\equiv M/\Mnl(z)$. This means that the normalization of the
$\Mg-M$ relation must scale as $\Cg\propto \Mnl^{-\ag}$ during the
self-similar stages of evolution, such that 
\be 
\Mg=\Cgo \Mnl^{-\ag}(z)M^{1+\ag}=\Cgo M\mu^{\ag}.
\label{eq:MgMa}
\ee
Note that self-similarity requires that the slope $\ag$ does not evolve with
  redshift.  

By analogy with the $\Mg-M$ relation, we can assume that the $T-M$ relation can be well described by a power law of the form
\be
T=\CT\Delta^{\frac{1}{3}}M^{\frac{2}{3}+\aT},
\label{eq:TME}
\ee
where $\aT$ describes mild deviation from the scaling due, e.g., to mild dependence of gas and temperature profile slopes in the HE equation. The dimensionless quantities involving temperature $T$ can be constructed using ratios of $T$ with $T_{\rm NL}=(\mu m_p/k_B)G\Mnl/R_{\rm NL}\propto \Delta^{1/3}\Mnl^{2/3}$. As before for the gas fraction, requirement that such dimensionless ratio depends only on $\mu$ requires $\CT\propto\Mnl^{-\aT}$ so that 
\be
T=C_{\rm To}\Mnl^{-\aT}\Delta^{\frac{1}{3}}M^{\frac{2}{3}+\aT}=C_{\rm To}\Delta^{\frac{1}{3}}M^{\frac{2}{3}}\mu^{\aT},
\label{eq:TME2}
\ee

Other observable-mass scaling relations can be constructed in the
manner similar to the derivation of the original relations
above. These are summarized below for the specific choice of
$\Delta\rho_r\equiv \Dc\rhoc(z)\propto h^2E^2(z)$:
\begin{eqnarray}
\Mg&\propto& \Mnl^{-\ag}(z)\Mdc^{1+\ag},\label{eq:MgMEa}\\
T&\propto& E(z)^{2/3}\Mnl^{-\aT}\Mdc^{2/3+\aT},\label{eq:TMEa}\\
\Lbol&\propto& E(z)^{7/3}\Mnl^{-2\ag-\aT/2}(z)\Mdc^{4/3+2\ag+\frac{\aT}{2}},\label{eq:LbolMEa}\\
K&\propto&E^{-2/3}(z)\Mnl^{\frac{2}{3}\ag-\aT}(z)\Mdc^{\frac{2}{3}(1-\ag)+\aT},\label{eq:SMEa}\\
Y&\propto& E(z)^{2/3}(z)\Mnl^{-\ag-\aT}(z)\Mdc^{5/3+\ag+\aT}.
\label{eq:YMEa}
\end{eqnarray}
In all of the relations one can, of course, recover the original relations by setting $\ag=\aT=0$. 
The observable-mass relations can be used to predict
observable-observable relations by eliminating mass from the corresponding relations above.

Note that the evolution of scaling relations in this extended model
arises both from the redshift dependence of $\Delta(z)\rho_r(z)$ and
from the extra redshift dependence due to factors involving $\Mnl$.
The practical implication is that if measurements show that $\ag\ne 0$
and/or $\aT\ne 0$ at some redshift, the original Kaiser scaling
relations are not expected to describe the evolution, even if the
evolution is self-similar.  Instead, relations given by
Equations~\ref{eq:MgMEa}-\ref{eq:YMEa} should be used. Note that at
$z\approx 0$, observations indicate that within the radius
$r_{500}$ enclosing overdensity $\Delta_c=500$, $\ag\approx 0.1-0.2$, while
$\aT\approx 0.-0.1$. Therefore, the extra evolution compared to the
Kaiser model predictions due to factors involving $\ag$ and, to a
lesser degree, factors involving $\aT$ is expected. Such evolution,
consistent with predictions of the above equations, is indeed observed
both in simulations \citep[see, e.g., Fig. 10 in][]{vikhlinin_etal09a}
and in observations \citep[][although see
\citeauthor{boehringer_etal11}
\citeyear{boehringer_etal11}]{lin_etal12}.

In practice, evolution of the scaling relations can be quite a bit more complicated than the evolution predicted by the above equations. The complication is not due to any deviation from self-similarity but rather due to specific mass definition and the fact that cluster formation is an extended process that is not characterized by a single collapse epoch. Some clusters evolve only mildly after their last major merger. However, the mass of such clusters will change with $z$ even if their potential does not change, simply because mass definition is tied to a reference density that changes with expansion of the Universe and because density profiles of clusters extends smoothly well beyond the virial radius. Any observable property of clusters that has radial profile differing from the mass profile, but which is measured within the same $R_{\Delta}$, will change differently than mass with redshift. As a simplistic toy model, consider a population of clusters that does not evolve from $z=1$ to $z=0$. Their X-ray luminosity is mostly due to the ICM in the central regions of clusters and it is not sensitive to the outer boundary of integration as long as it is sufficiently large. Thus, X-ray luminosity of such a non-evolving population will not change with $z$, but masses $M_{\Delta}$ of clusters will increase with decreasing $z$ as the reference density used to define the cluster boundary decreases. Normalization of the $L_{\rm X}-M_{\Delta}$ relation will thus decrease with decreasing redshift simply due to the definition of mass. The strength of the evolution will be determined by the slope of the mass profile around $R_{\Delta}$, which is weakly dependent on mass. Such  an effect may, thus, result in the  evolution of both the slope and normalization of the relation. In this respect, quantities that have radial profiles most similar to the total mass profile (e.g., $\Mg$, $Y$) will suffer the least from such spurious evolution.  

Finally, we note again that in principle for general non power-law initial perturbation spectra of the CDM models the scaling with $M/\Mnl$ needs to be replaced with scaling with the peak height $\nu$. For clusters within a limited mass range, however, the power spectrum can be approximated by a power law and thus a characteristic mass similar to $\Mnl$ can be constructed, although such mass should be within the typical mass range of the clusters. The latter is not true for $\Mnl$, which is considerably smaller than typical cluster mass at all $z$.  
 
\subsubsection{Practical implications for observational calibrations of scaling relations.}
In observational calibrations of the cluster scaling relations, it is often necessary to rescale between different redshifts either to bring results from the different $z$ to a common redshift, or because the scaling relation is evaluated using clusters from a wide range of redshifts due to small sample size. It is customary to use predictions of the Kaiser model to carry out such rescaling to take into account the redshift dependence of $\Delta(z)$. In this context, one should keep in mind that these predictions are approximate due to the approximate nature of some of the assumptions of the model, as discussed above. Inaccuracies introduced by such scalings may, for example, then be incorrectly interpreted as intrinsic scatter about the scaling relation. 

In addition, because the $\Delta(z)$ factors are a result of an
arbitrary mass definition, they should not be interpreted as
physically meaningful factors describing evolution of mass. For
example, in the $T$-$M$ relation, the $\Delta^{1/3}$ factor in
Equation~\ref{eq:TMo} arises due to the {\it dimensional} $M/R$ factor
of the HE equation. As such, this factor does not change even if the power-law index of the $T$-$M$ relation deviates from $2/3$, in which case the relation has the form $T\propto \Delta^{1/3}M^{2/3+\alpha_T}$. In other words, if one fits for the parameters of this relation, such as normalization $A$ and slope $B$, using measurements of temperatures and masses for a sample of clusters spanning a range of redshifts, the proper parameterization of the fit should be
\be
\frac{T}{T_{\rm p}}=A\Delta^{1/3}\left(\frac{M}{M_{\rm p}}\right)^B,\ \ \mathrm{or}\ \ 
\frac{T}{T_{\rm p}}=A\Delta^{1/3}\left(\frac{M}{M_{\rm p}}\right)^{2/3+B},
\ee
where $T_{\rm p}$ and $M_{\rm p}$ are appropriately chosen pivots. The parameterization
$T/T_{\rm p}=A(\Delta^{1/3}M/M_{\rm p})^B$ that is sometimes adopted in observational analyses {\it is not correct} in the context of the self-similar model. In other words, only the observable quantities should be rescaled by the $\Delta\rho_r$ factors, and not the mass. 
Likewise, only the $\Delta\rho_r$ factors actually predicted by the Kaiser model should be present in the scalings. For example, no such factor is predicted for the $\Mg$-$M$ relation and therefore the gas and total masses of clusters at different redshifts should not be scaled by $\Delta\rho_r$ factors in the fits of this relation. 

Finally, we note that observational calibrations of the
observable-mass scaling relations generally depend on the distances to clusters and are therefore cosmology dependent. Such dependence arises 
because distances are used to convert observed angular scale to physical
scale within which an ``observable'' is defined, 
$R=\theta d_A(z)\propto
\theta h^{-1}$, or to convert observed flux $f$ to luminosity, $L=4\pi fd_L(z)$,
where $d_A(z)$ and $d_L(z)=d_A(z)(1+z)^2$ are the angular diameter
distance and luminosity distance, respectively.
Thus, if the total mass $M$ of a cluster is measured using the HSE equation, we have $M_{\rm HE}\propto TR\propto d_A\propto h^{-1}$. The same scaling is expected for the mass derived from the weak lensing shear profile measurements. 

If the gas mass is measured from the X-ray flux from a volume
$V\propto R^3\propto \theta^3d_A^3$, which scales as $f=\Lx/(4\pi
d_L^2)\propto \rhog^2 V/d_L^2\propto \Mg^2/(Vd_L^2)\propto
\Mg^2/(\theta d_L^2 d_A^3)$ and where $f$ and $\theta$ are
observables, gas mass then scales with distance as $\Mg\propto d_L
d_A^{3/2}\propto h^{5/2}$. This dependence can be exploited to
constrain cosmological parameters, as in the case of X--ray
measurements of gas fractions in clusters
\citep{ettori_etal03,allen_etal04,laroque_etal06,allen_etal08,ettori_etal09}
or abundance evolution of clusters as a function of their observable
\citep[e.g.,][]{vikhlinin_etal09b}.  In this respect, the $\Mg-M$
relation has the strongest scaling with distance and cosmology, whereas
the scaling of the $T-M$ relation is the weakest \citep[e.g., see
discussion by][]{vikhlinin_etal09a}.

\subsection{Cluster formation and Thermodynamics of the Intra-cluster
gas}
\label{sec:thermo}

Gravity that drives the collapse of the initial large-scale density
peaks affects not only the properties of the cluster DM halos, but
also the thermodynamic properties of the intra-cluster plasma. The
latter are also affected by processes related to galaxy formation,
such as cooling and feedback. Below, we discuss the thermodynamic
properties of the ICM resulting from gravitational heating, radiative
cooling, and stellar and AGN feedback during cluster formation.

\subsubsection{Gravitational collapse of the intra-cluster gas.}  
\begin{figure}
\centerline{ 
\psfig{file=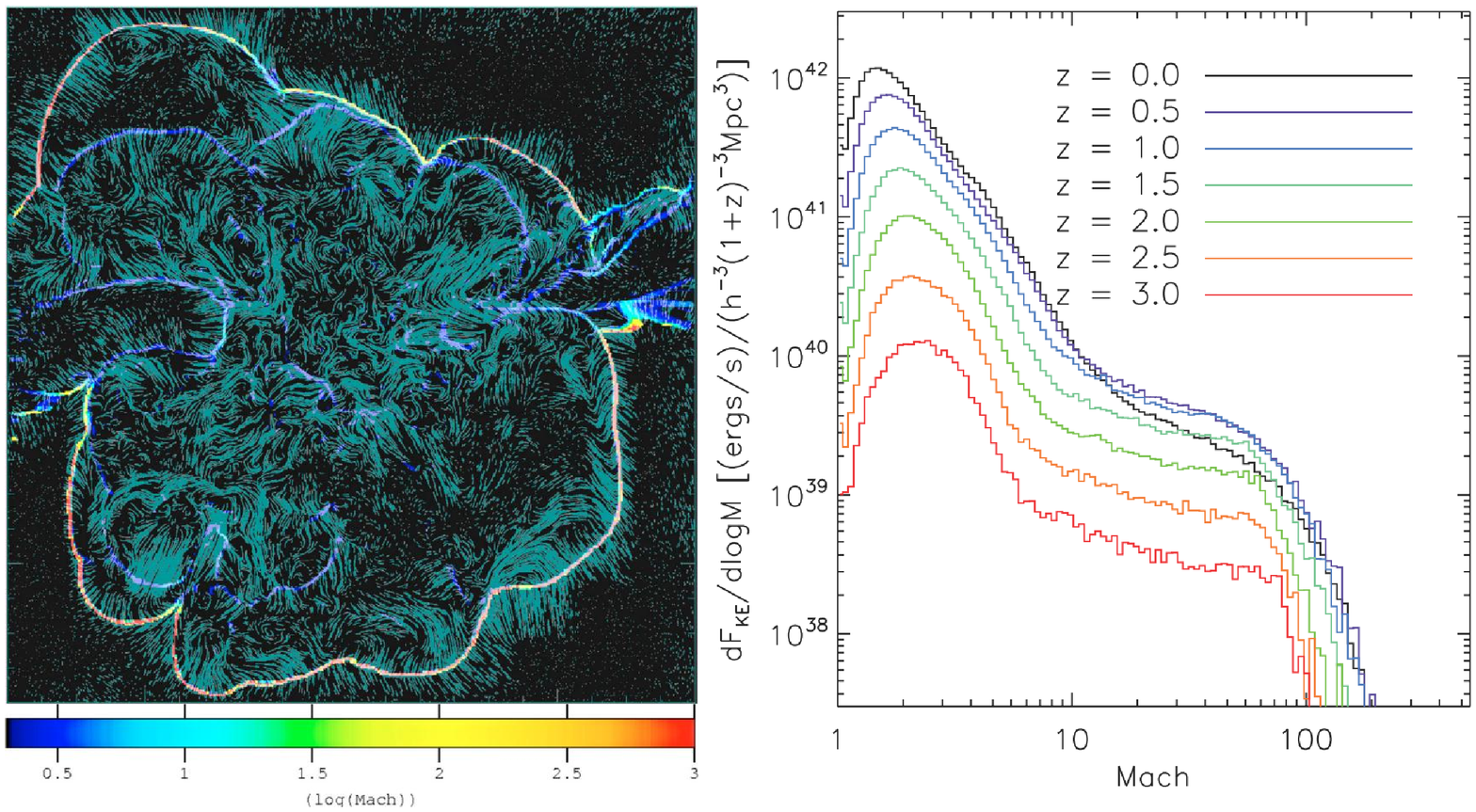,width=17.truecm,angle=-90} }
\vspace{-2cm}
\caption{Left panel: map of the shocked cells identified by the
  divergence of velocity colored by the local shock Mach number and
  turbulent gas velocity ﬁeld (streamlines) in a slice of the
  simulation box 7.5 Mpc on a
  side and depth of 18 kpc at $z=0.6$, for a simulated cluster that
  reaches a mass of $\sim 2\times 10^{14}M_\odot$ by $z=0$
  \protect\citep[from][]{vazza_etal09}. Right panel: redshift
  evolution of the distribution of the kinetic energy processed by
  shocks, as a function of the Mack number ${\mathcal M}$ in a
  cosmological simulation
  \protect\citep[from][]{skillman_etal08}. Results shown in both
  panels are based on the adaptive mesh refinement ENZO code
  \protect\citep[][]{oshea_etal04}.}
\label{fig:shocks}
\end{figure}

The diffuse gas infalling onto the DM-dominated potential wells of
clusters converts the kinetic energy acquired during the collapse into
thermal energy via adiabatic compression and shocks. As gas
settles into HE, its temperature approaches
values close to the virial temperature corresponding to the cluster
mass.  In the spherically symmetric collapse model of
\cite{1985ApJS...58...39B}, supersonic accretion gives rise to the
expanding shock at the interface of the inner hydrostatic gas with a
cooler, adiabatically compressed, external medium. Real
three-dimensional collapse of clusters is more complicated and
exhibits large deviations from spherical symmetry, as accretion
proceeds both in a quasi-spherical fashion from low-density regions
and along relatively narrow filaments. The gas accreting along the
latter penetrates much deeper into the cluster virial region and does
not undergo a shock at the virial radius (see
Fig.~\ref{fig:shocks}). The strong shocks are driven not just by the
accretion of gas from the outside but also ``inside-out'' during major
mergers \citep[e.g.,][]{poole_etal07}.

The shocks arising during cluster formation can be classified into two
broad categories: strong external shocks surrounding filaments and the
virialized regions of DM halos and weaker internal shocks, located
within the cluster virial radius
\citep[e.g.,][]{pfrommer_etal06,skillman_etal08,vazza_etal09}. The
strong shocks arise in the high-Mach number flows of the intergalactic
gas, whereas weak shocks arise in the relatively low-Mach number flows
of gas in filaments and accreting groups, which was pre-heated at
earlier epochs by the strong shocks surrounding filaments or external
groups. The left panel of Figure \ref{fig:shocks} shows these two
types of shocks in a map of the shocked cells identified in a
cosmological adaptive mesh refinement simulation of a region
surrounding a galaxy cluster \citep[from][]{vazza_etal09}, along with
the gas velocity field.  This map highlights the strong external
shocks, characterized by high Mach numbers ${\mathcal M}> 30$, surrounding
the cluster at several virial radii from the cluster center, and
weaker internal shocks, with ${\mathcal M}\mincir 2-3$. The cluster is
shown at the epoch immediately following a major merger, which
generated substantial velocities of gas within virial radius. As we
discuss in \S~\ref{sec:regul} below, incomplete thermalization of
these gas motions is one of the main sources of non--thermal pressure
support in the ICM.

The right panel of Figure \ref{fig:shocks} shows the distribution of
the kinetic energy processed by shocks, as a function of the local
shock Mach number, for different redshifts
\citep{skillman_etal08}. The figure shows that a large fraction of the
kinetic energy is processed by weak internal shocks and this fraction
increases with decreasing redshift as more and more of the accreting
gas is pre-heated in filaments.  Yet, the left panel of
Fig. \ref{fig:shocks} highlights that large--${\mathcal M}$ shocks
surround virialized halos in such a way that most gas particles
accreted in a galaxy cluster must have experienced at least one strong
shock in their past.

Becasuse gravity does not have a characteristic length scale, we
expect the predictions of the self--similar model, presented in
Section \ref{sec:selfsimilar}, to apply when gravitational gas
accretion determines the thermal properties of the ICM.  The scaling
relations and their evolution predicted by the self-similar model are
indeed broadly confirmed by the non-radiative hydrodynamical
simulations that include only gravitational heating
\citep[e.g.,][]{navarro_etal95,eke_etal98,nagai_etal07b}, although
some small deviations arising due to small differences in the dynamics
of baryons and DM were also found
\citep{ascasibar_etal06,nagai06,stanek_etal10}.

As discussed in Section \ref{sec:obsprop}, observations carried out
with the {\sl Chandra} and {\sl XMM}--Newton telescopes during the
past decade showed that the outer regions of clusters ($r\gtrsim
r_{2500}$) exhibit self-similar scaling, whereas the core regions
exhibit strong deviations from self-similarity. In particular, gas
density in the core regions of small-mass clusters is lower than
expected from self-similar scaling of large-mass systems.  These
results indicate that some additional non-gravitational processes are
shaping properties of the ICM. We review some of these processes
studied in cluster formation models below.

\subsubsection{Phenomenological pre-heating models.}
The first proposed mechanism to break self--similarity was
high-redshift ($z_h\magcir 3$) pre--heating by non--gravitational
sources of energy, presumably by a combined action of the AGN and
stellar feedback \citep{kaiser91,evrard_henry91}.  The specific extra
heating energy per unit mass, $E_h$, defines the temperature scale
$T^*\propto E_h/k_B$, such that clusters with virial temperature
$T_{\rm vir}>T^*$ should be left almost unaffected by the extra
heating, whereas in smaller clusters with $T_{\rm vir}<T^*$ gas
accretion is suppressed. As a result, gas density is relatively
lower in lower massive systems, especially at smaller radii, while
their entropy will be higher.

Both analytical models
\citep[e.g.][]{tozzi_norman01,babul_etal02,voit_etal03} and
hydrodynamical simulations
\citep[e.g.][]{bialek_etal01,borgani_etal02,muanwong_etal02} have
demonstrated that with a suitable pre-heating prescription and typical
heating injection of $E_h\sim 0.5$--1 keV per gas particle
self--similarity can be broken to the degree required to reproduce
observed scaling relations. Studies of the possible feedback
mechanisms show that such amounts of energy cannot be provided by SNe
\citep[e.g.,][]{kravtsov_yepes00,renzini00,borgani_etal04,kay_etal07,henning_etal09},
and must be injected by the AGN population
\citep[e.g.,][]{wu_etal00,lapi_etal05,bower_etal08} or by some other
unknown source.

However, regardless of the actual sources of heating, strong
widespread heating at high redshifts would conflict with the observed
statistical properties of the Lyman--$\alpha$ forest
\citep{shang_etal07,borgani_viel09}. Moreover, hydrodynamical
simulations have demonstrated that simple pre--heating models predict
large isentropic cores \citep[e.g.,][]{borgani_etal05,younger_bryan07}
and shallow pressure profiles \citep{kay_etal12}. This is at odds
with the entropy and pressure profiles of real clusters which exhibit
smoothly declining entropy down to $r\sim 10-20$~kpc
\citep[e.g.,][]{cavagnolo_etal09,arnaud_etal10}.

\subsubsection{The role of radiative cooling.}
\begin{figure}
 \centerline{ 
\hbox{
\psfig{file=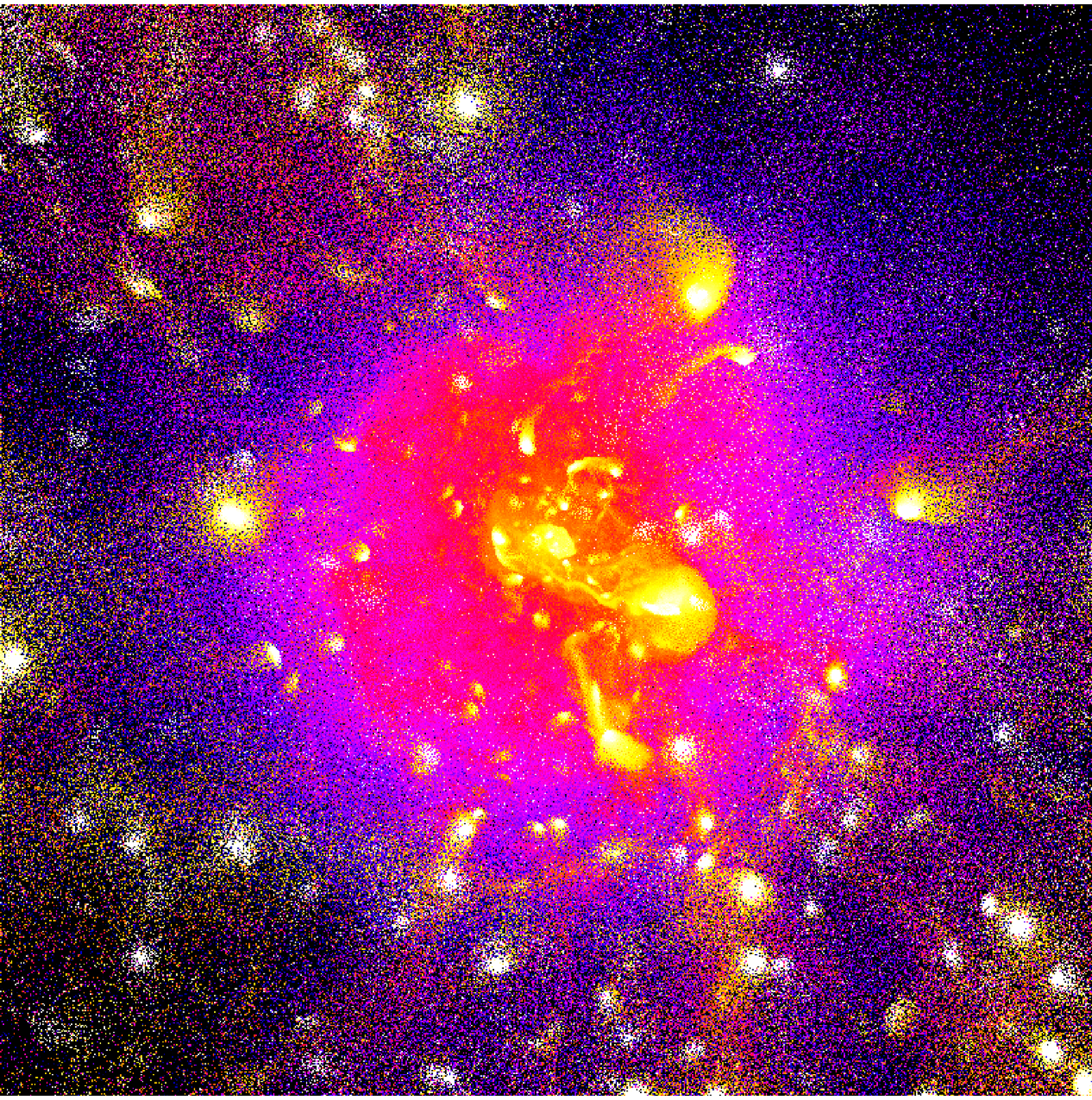,width=7.truecm}
\psfig{file=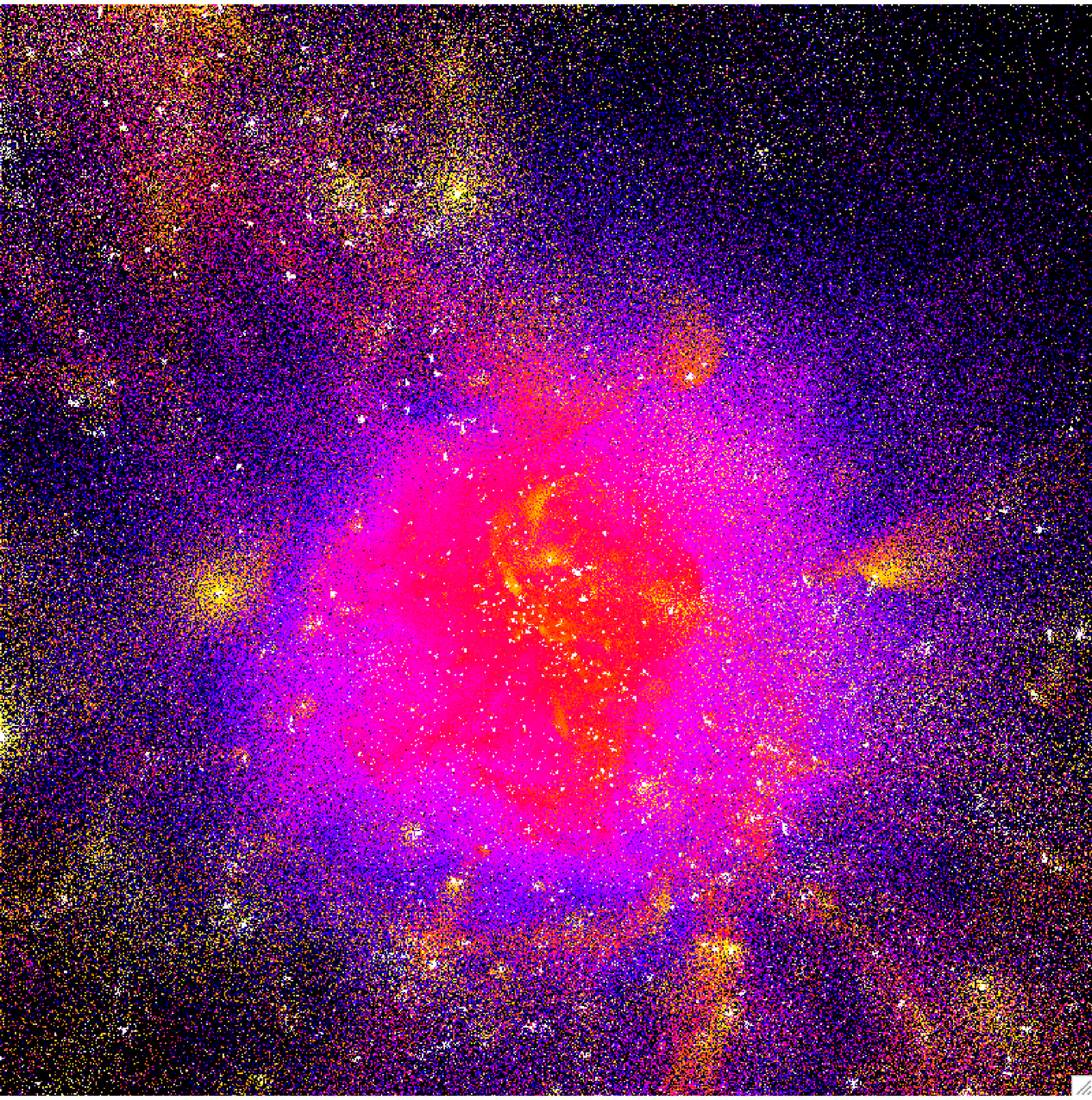,width=7.truecm}
}
}
\vspace{-1cm}
\caption{Maps of entropy in cosmological hydrodynamical simulations of
  a galaxy cluster of mass $M_{500}\simeq 10^{15}h^{-1}M_\odot$ at
  $z=0$, carried out without (left panel) and with (right panel)
  radiative cooling.  Brighter colors correspond to lower gas entropy.
  Each panel encompasses a physical scale of $6.5\hm$, which
  corresponds to $\approx 2.5$ virial radii for this cluster. The
  simulations have been carried out using the GADGET-3 smoothed
  particle hydrodynamics code
  \protect\cite{springel05}.}
\label{fig:entrmaps}
\end{figure}

The presence of galaxies in clusters and low levels of the ICM entropy
in cluster cores are a testament that radiative cooling has operated
during cluster formation in the past and is an important process
shaping thermodynamics of the core gas at present. Therefore, in
general radiative cooling cannot be neglected in realistic models of
cluster formation. Given that cooling generally introduces new scales,
it can break self--similarity of the ICM even in the absence of
heating \citep{voit_bryan01}. In particular, cooling removes
low--entropy gas from the hot ICM phase in the cluster cores, which is
replaced by higher entropy gas from larger radii. Somewhat
paradoxically, the cooling thus leads to an entropy increase of the
hot, X-ray emitting ICM phase. This effect is illustrated in Figure
\ref{fig:entrmaps}, which shows the entropy maps in the simulations of
the same cluster with and without cooling.  In the absence of cooling
(left panel), the innermost region of the cluster is filled by
low--entropy gas. Merging substructures also carry low--entropy gas,
which generates comet--like features by ram--pressure stripping, and
is hardly mixed in the hotter ambient of the main halo. In the
simulation with radiative cooling (right panel), most of the
low--entropy gas associated with substructures and the central cluster
region is absent, and most of the ICM has a relatively high entropy.

\begin{figure}
 \centerline{ 
\psfig{file=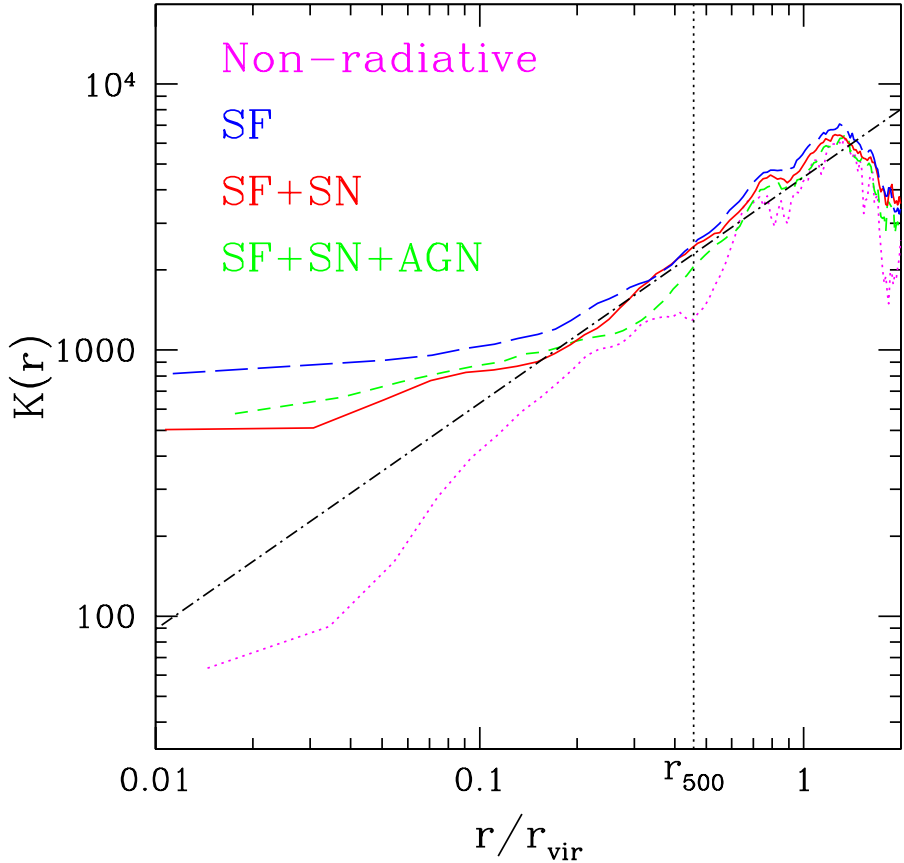,width=13.truecm}
}
\caption{Radial profiles of entropy (in units of
  kiloelectronvolt--centimeters squared) for the same simulations
  whose entropy maps are shown in Figure
  \protect\ref{fig:entrmaps}. Magenta dotted, long-dashed blue,
  continuous red, and short-dashed green curves refer to the
  non-radiative simulation and to the three radiative simulations
  including only cooling and star formation, including also the effect
  of galactic ejecta from supernova, and including also the effect of
  AGN feedback, respectively.  The dot-dashed line shows the power-law
  entropy profile with slope $K(r)\propto r^{1.1}$, whereas the
  vertical dotted line marks the position of $r_{500}$.  }
\label{fig:simul_profs}
\end{figure}

A more quantitative analysis of the entropy distribution for these
simulated clusters is shown in Figure \ref{fig:simul_profs}, in which
the entropy profiles of clusters simulated with inclusion of different
physical processes are compared with the baseline analytic spherical
accretion model; this model predicts the power-law entropy profile
$K(r)\propto r^{1.1}$ \citep[e.g.][]{tozzi_norman01,voit05}. The
figure shows that the entropy profile in the simulation with radiative
cooling is significantly higher than that of the non-radiative
simulation. The difference in entropy is as large as an order of
magnitude in the inner regions of the cluster and is greater by a
factor of two even at $r_{500}$.

Interestingly, the predicted level of entropy at $r\sim
r_{2500}-r_{500}$ in the simulations with cooling (but no significant
heating) is consistent with the ICM entropy inferred from X-ray
observations.  However, this agreement is likely to be spurious
because it is achieved with the amount of cooling that results in
conversion of $\approx 40\%$ of the baryon mass in clusters into stars
and cold gas, which is inconsistent with observational measurements of
cold fraction varying from $\simeq 20-30\%$ for small-mass,
X-ray-emitting clusters to $\mincir 10\%$ for massive clusters (see
\S~\ref{sec:obsprop}).

Finally, note that inclusion of cooling in simulations with
pre-heating discussed above usually results in problematic
star-formation histories. In fact, if pre-heating takes place at a
sufficiently high redshift, clusters exhibit excessive cooling at
lower redshifts, as pre-heated gas collapses and cools at later epochs
compared to the simulations without pre-heating
\citep[e.g.][]{tornatore_etal03}.  These results highlight the
necessity to treat cooling and heating processes simultaneously using
heating prescriptions that can realistically reproduce the heating
rate of the ICM gas as a function of cosmic time. We discuss
efforts in this direction next.

\subsubsection{Thermodynamics of the intracluster medium with stellar
  and active galactic nuclei feedback.}
\label{sec:feedb}

The results discussed above strongly indicate that, in order to
reproduce the overall properties of clusters, cooling should be
modelled together with a realistic prescription for non-gravitational
heating. This is particularly apparent in the cluster cores, where a
steady heating is required to offset the ongoing radiative cooling
observed in the form of strong X-ray emission \citep[see,
e.g.,][]{peterson_fabian06}. Studies of the feedback processes in
clusters is one of the frontiers in cluster formation
modelling. Although we do not yet have a complete picture of the ICM
heating, a number of interesting and promising results have been
obtained.

In Figure \ref{fig:simul_profs}, the solid line shows the effect of
the SN feedback on the entropy profile. In these simulations, the
kinetic feedback of SNe is included in the form of galactic winds
carrying the kinetic energy comparable to all of the energy released
by Type-II SNe expected to occur according to star formation in the
simulation. This energy partially compensates for the radiative losses in
the central regions, which leads to a lower level of entropy in the
core.  However, the core ICM entropy in these simulations is still
considerably higher than observed \citep[e.g.,][]{sun_etal09}. The
inefficiency of the SN feedback in offsetting the cooling sufficiently is also
evidenced by temperature profiles.

Figure \ref{fig:Tprofs} \cite[from ][]{leccardi_molendi08} compares
the observed temperature profiles of a sample of local clusters with
results from simulations that include the SN feedback. The figure
shows that simulations reproduce the observed temperature profile at
$r\magcir 0.2r_{180}$. The overall shape of the profile at these large
radii is reproduced by simulations including a wide range of physical
processes, including non-radiative simulations
\citep[e.g.,][]{loken_etal02,borgani_etal04,nagai_etal07b}. At large
radii, however, the observed and predicted profiles do not match: The
profiles in simulated clusters continue to increase to the smallest
resolved radii, whereas the observed profiles reach a maximum
temperature $T_{\rm max}\approx 2T_{180}$ and then decrease with
decreasing radius to temperatures of $\sim 0.1-0.3T_{\rm max}$. The
high temperatures of the central gas reflects its high entropy and is
due to the processes affecting the entropy, as discussed above.

Another indication that the SN feedback alone is insufficient is the
fact that the stellar mass of the BCGs in simulations that include
only the feedback from SNe is a factor of two to three larger than the
observed stellar masses. For example, in the simulated clusters shown
in figure \ref{fig:entrmaps}, the baryon fraction in stars within
$r_{500}$ decreases from $\simeq 40\%$ in simulations without SN
feedback to $\simeq 30\%$, which is still a factor of two larger than
observational measurements. The overestimate of the stellar mass is
reflected in the overestimate of the ICM metallicity in cluster cores
\citep[e.g.,][and references therein]{borgani_etal08a}.

\begin{figure}
 \centerline{ 
\psfig{file=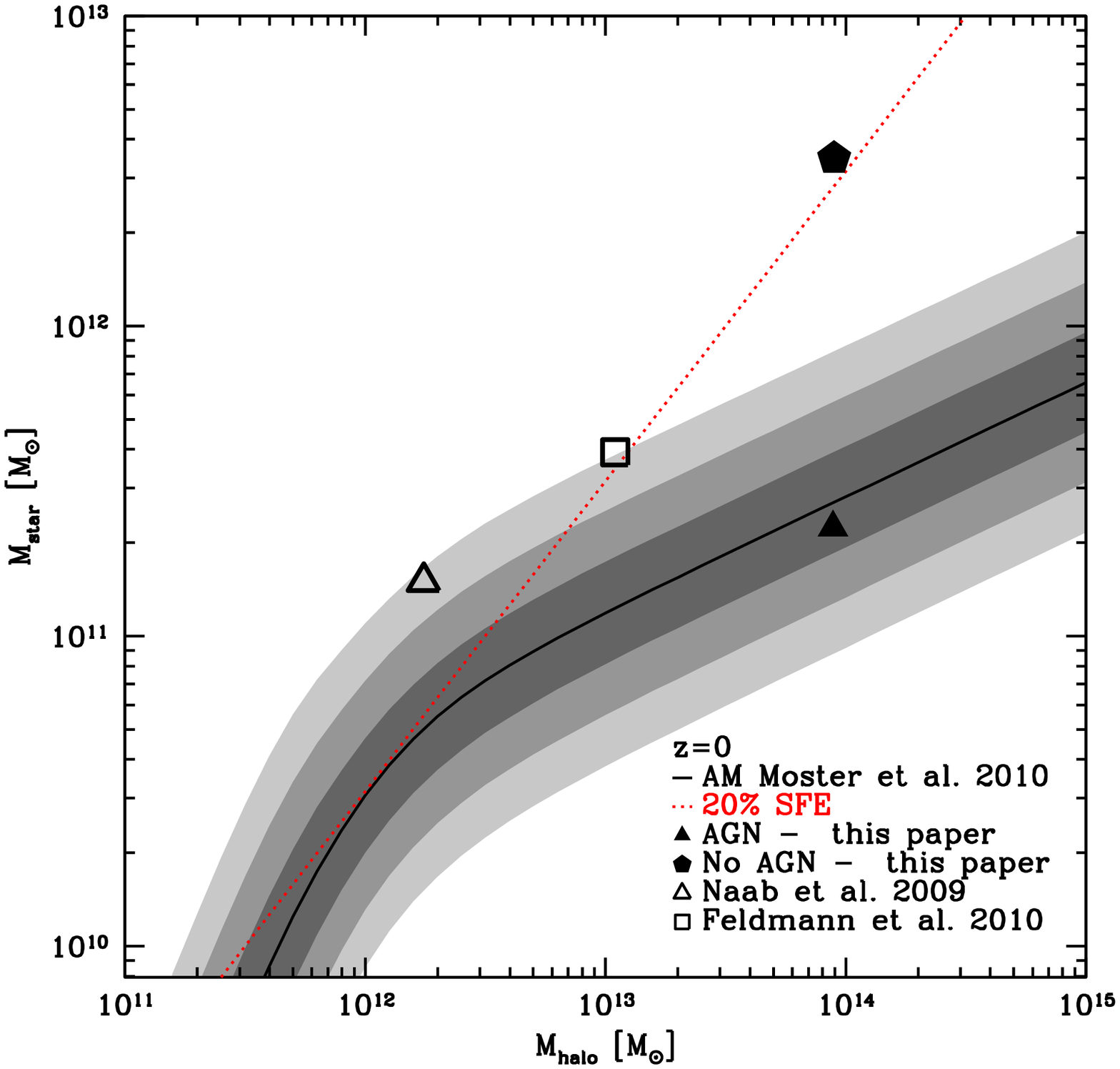,width=13.truecm}
}
\caption{Comparison of the relation between stellar mass and total
  halo mass as predicted by cosmological hydrodynamical simulations of
  four early-type galaxies (symbols) (from
  \protect\citealt{martizzi_etal11}). The open triangle and square
  refer to the simulations presented by \protect\cite{naab_etal09} and
  by \protect\cite{feldman_etal10}, both based on the smoothed
  particle hydrodynamics codes and not including AGN feedback. The
  filled symbols refer to the simulations by
  \protect\cite{martizzi_etal11} with the brightest cluster galaxies
  forming at the center of a relatively poor cluster carried out with
  an AMR code, both including (triangle) and excluding (pentagon) AGN
  feedback. The red dotted line represents the relation expected for
  $20\%$ efficiency in the conversion of baryons into stars. The solid
  black line is the prediction from \protect\cite{moster_etal10} of a
  model in which dark matter halos are populated with stars in such a
  way as to reproduce the observed stellar mass function. The grey shaded
  areas represent the 1-, 2- and 3-$\sigma$ scatter around the average
  relation.}
\label{fig:abund_match}
\end{figure}

Different lines of evidence indicate that energy input from the AGN in
the central cluster galaxies can provide most of the energy required
to offset cooling \citep[see][for a comprehensive
review]{mcnamara_nulsen07}. Because the spatial and temporal scales
resolved in cosmological simulations are larger than those relevant
for gas accretion and energy input, the AGN energy feedback can only
be included via a phenomenological prescription. Such prescriptions
generally model the feedback energy input rate by assuming the Bondi
gas accretion rate onto the SMBHs, included as the sink particles, and
incorporate a number of phenomenological parameters, such as the
radiative efficiency and the feedback efficiency, which quantify the
fraction of the radiated energy that thermally couples to the
surrounding gas \citep[e.g.,][]{springel_etal05b}. The values of these
parameters are adjusted so that simulations reproduce the observed
relation between black hole mass and the velocity dispersion of the
host stellar bulge \citep[e.g.,][]{marconi_hunt03}. An alternative way
of implementing the AGN energy injection is the AGN--driven winds,
which shock and heat the surrounding gas
\citep[e.g.,][]{omma_etal04,dubois_etal11,gaspari_etal11}.

In general, simulations of galaxy clusters based on different variants
of these models have shown that the AGN feedback can reduce star
formation in massive cluster galaxies and reduce the hot gas content
in the poor clusters and groups, thereby improving agreement with the
observed relation between X--ray luminosity and temperature
\citep[e.g.,][]{sijacki_etal07,puchwein_etal08}. Figure
\ref{fig:abund_match} (from \citealt{martizzi_etal11}) shows that
simulations with the AGN feedback results in stellar masses of the
BCGs that agree with the masses required to match observed stellar
masses of galaxies and masses of their DM halos predicted by the
models. The figure also shows that stellar masses are over-predicted
in the simulations without the AGN feedback. Incidentally, the
large-scale winds at high redshifts and stirring of the ICM in cluster
cores by the AGN feedback also help to bring the metallicity profiles
into cluster simulations in better agreement with observations
\citep[][]{fabjan_etal10,mccarthy_etal10}.

Although results of simulations with the AGN feedback are promising,
simulations so far have not been able to convincingly reproduce the
observed thermal structure of cool cores. As an example, Figure
\ref{fig:simul_profs} shows that the entropy profiles in such
simulations still develop large constant entropy core inconsistent
with observed profiles.  Interestingly, the adaptive mesh refinement
simulations with jet-driven AGN feedback by \citet{dubois_etal11}
reproduce the monotonically decreasing entropy profiles inferred from
observations. However, such agreement only exists if radiative cooling
does not account for the metallicity of the ICM; in simulations that
take into account the metallicity dependence of the cooling rates the
entropy profile still develosp a large constant entropy core.

The presence of a population of relativistic particles in AGN--driven
high--entropy bubbles has been suggested as a possible solution to
this problem \citep{guo_oh08,sijacki_etal08}.  A relativistic plasma
increases the pressure support available at a fixed temperature and
can, therefore, help to reproduce the observed temperature and entropy
profiles in core regions.
However, it remains to be seen whether the required population of the
cosmic rays is consistent with available constraints inferred from
$\gamma$ and radio observations of clusters \citep[e.g.,][and
references therein]{brunetti11}. A number of additional processes,
such as thermal conduction \citep[e.g.,][]{narayan_medvedev01} or
dynamical friction heating by galaxies \citep{elzant_etal04} have been
proposed. Generally, these processes cannot effectively regulate
cooling in clusters by themselves
\citep[e.g.,][]{dolag_etal04,conroy_ostriker08}, but they may play an
important role when operating in concert with the AGN feedback
\citep{voit11} or instabilities in plasma
\citep[e.g.,][]{sharma_etal12}.

In summary, results of the theoretical studies discussed above indicate
that the AGN energy feedback is the most likely energy source regulating
the stellar masses of cluster galaxies throughout their evolution and
suppressing cooling in cluster cores at low redshifts. The latter
likely requires an interplay between the AGN feedback and a number of
other physical processes: e.g., injection of the cosmic rays in the 
high--entropy bubbles, buoyancy of these bubbles stabilized by
magnetic fields, dissipation of their mechanical energy through
turbulence, thermal conduction, and thermal instabilities.  Although
details of the interplay are not yet understood, it is clear that it
must result in a robust self-regulating feedback cycle in which
cooling immediately leads to the AGN activity that suppresses further
cooling for a certain period of time.

\section{Regularity of the cluster populations}
\label{sec:regul}
Processes operating during cluster formation and evolution discussed
in the previous section are complex and nonlinear. However, it is now
also clear that most of the complexity is confined to cluster cores
and affects a small fraction of volume and mass of the clusters. In
this regime, clusters' observational properties exhibit strong
deviations from the self--similar scalings described in Section
\ref{sec:selfsimilar} (see also \citealt{voit05}). At larger radii,
ICM is remarkably regular. In this section, we discuss the
origins of such highly regular behaviour and the range of radii where
it can be expected. We argue that the existence of this radial range
allows us to define integral observational quantities, which have low
scatter for clusters of a given mass that are not sensitive to the
astrophysical processes operating during cluster formation and
evolution. This fact is especially important for the current and
future uses of clusters as cosmological probes
\citep{allen_etal11,weinberg_etal12}.

\subsection{Characterizing regularity}
\label{sec:chareg}

A number of observational evidences, based on X--ray measurements of
gas density \citep[e.g.][]{croston_etal08} and temperature profiles
\citep[][]{vikhlinin_etal06,pratt_etal07,leccardi_molendi08}, and the
combination of the two in the form of entropy profile
\citep{cavagnolo_etal09}, demonstrate that clusters have a variety of
behaviors in central regions, depending on the presence and prominence
of cool cores. As discussed in Section \ref{sec:obsprop}, outside of
core regions, clusters behave as a more homogeneous population and
obey assumptions and expectations of the self--similar model
(discussed above in \ref{sec:ssm}). For instance, Figure
\ref{fig:obsICM} shows that the ICM density is nearly independent of
temperature once measured outside of core regions $r\magcir r_{2500}$,
at least for relatively hot systems with $T\magcir 3$ keV. Quite
remarkably, observed and simulated temperature profiles agree with
each other within this same radial range (see Figure
\ref{fig:Tprofs}).

\begin{figure}
 \centerline{ 
\psfig{file=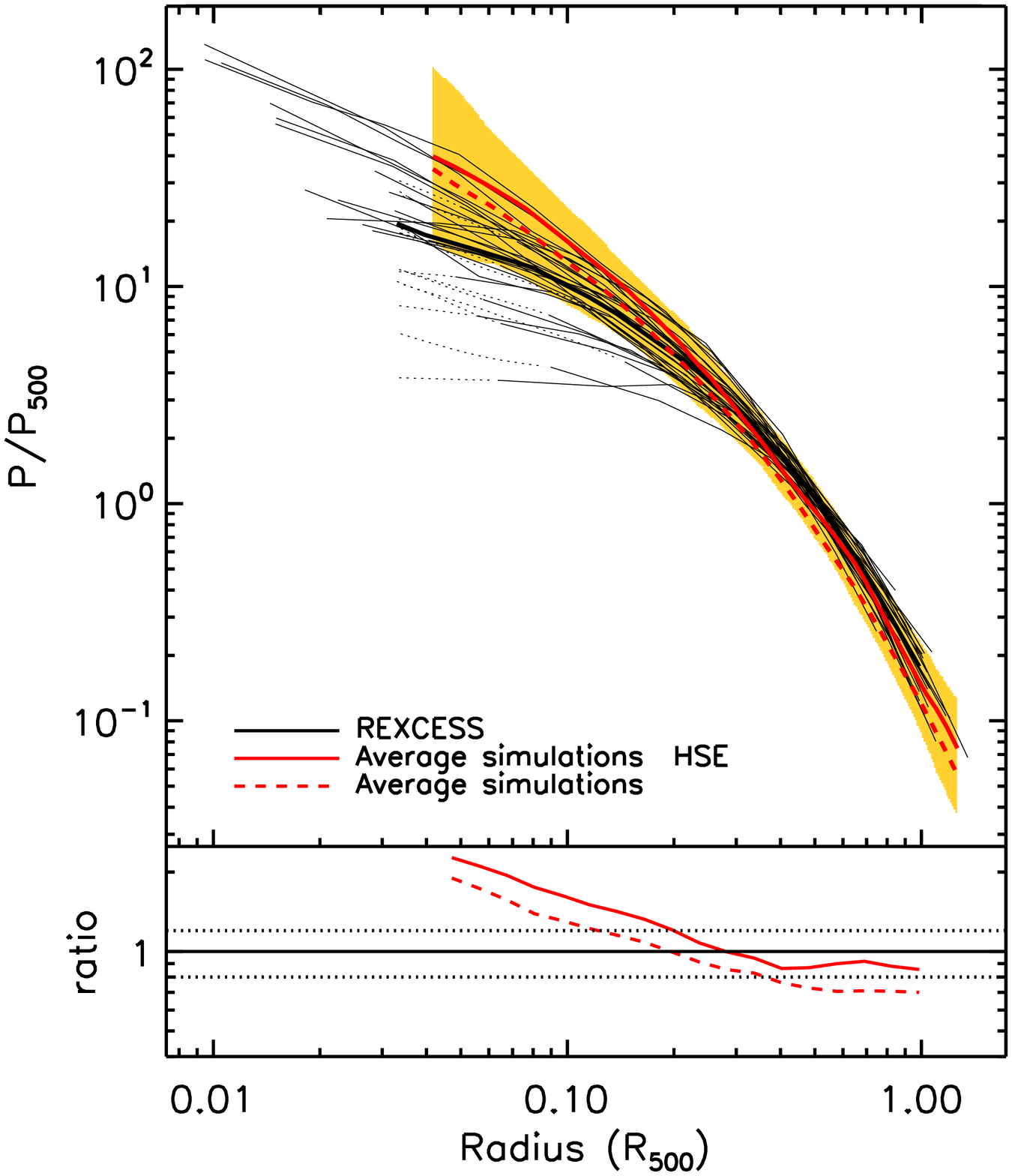,width=13.truecm}
}
\vspace{-2cm}
\caption{Comparison between observed (black lines) and simulated (red
  lines with orange shaded area) pressure profiles (from
  \protect\citealt{arnaud_etal10}). Observational data refer to the
  Rrepresentative XMM-Newton Cluster Structure Survey (REXCESS) sample
  of nearby clusters \protect\citep{boehringer_etal07} observed with
  XMM--Newton. Simulation results are obtained by combining different
  sets of clusters simulated with both smoothed particle hydrodynamics
  and adaptive mesh refinement codes (see
  \protect\citealt{arnaud_etal10} for details). The continuous red
  line corresponds to the average profile from simulations, after
  rescaling profiles according to the values of $R_{500}$ and
  $M_{500}$ predicted by hydrostatic equilibrium, with the orange area
  showing the corresponding rms scatter. The red dotted line shows
  the simulation results when using instead the true $M_{500}$
  value. The bottom panel shows the ratio between average simulation
  profiles and average observed profiles.}
\label{fig:press_arnaud}
\end{figure}

A good illustration of the regularity of the ICM properties is
represented by the pressure profiles shown in Figure
\ref{fig:press_arnaud} (from \citealt{arnaud_etal10}, but see also
\citealt{sun_etal11}) rescaled to the values of radius and pressure at
$r_{500}$. The perfectly regular, self-similar behavior would
correspond to a single line in this plot for clusters of all
masses. The pressure profiles shown in this figure are derived from
X--ray observations and are defined as the product of electron number
density and temperature profiles. Similar profiles are now derived
from SZ observations, which probe pressure more directly
\citep[e.g.,][]{bonamente_etal12}. Quite remarkably, the observed
pressure profiles at $r\magcir 0.2r_{500}$ follow a nearly universal
profile \citep[see also][]{nagai_etal07b}, exhibiting fractional
scatter of $\lesssim 30\%$ at $r\sim 0.2r_{500}$ and even smaller
scatter of $\sim 10-15\%$ at $r\sim 0.5r_{500}$. At smaller radii the
scatter of pressure profiles is much larger, with steep profiles
corresponding to the cool core clusters and flatter profiles for
disturbed clusters.  Figure \ref{fig:press_arnaud} shows that
simulated and observed pressure profiles agree well with each other
for $r\magcir 0.2r_{500}$, i.e., in the regime where the cluster
population has a more regular behaviour. At smaller radii the profiles
from simulations are on average steeper than observed, and exhibit a
lower degree of diversity between cool core and non-cool core
clusters.

The scatter in the cluster radial profiles can be used to define the
following three radial regimes.
\begin{itemize}
\item[1.] Cluster cores, $r\lesssim r_{2500}$, which exhibit the
  largest scatter and where scaling with mass differs significantly
  from the self-similar scaling expectation. We do not yet have a
  complete and adequate theoretical understanding of the observed
  properties of the ICM and their diversity in the cluster cores. This
  is one of the areas of active ongoing theoretical and observational
  research.
\item[2.] Intermediate radii, $r_{2500}\lesssim r\lesssim r_{500}$,
  which exhibit the smallest scatter and scaling with mass close to
  the self-similar scaling. Although the processes affecting
  thermodynamics of these regions are not yet fully understood, the
  simple scaling and regular behavior make observable properties of
  clusters at these radii useful for connecting them to the total
  cluster mass.
\item[3.] Cluster outskirts $r>r_{500}$, where scatter is increasing
  with radius and scaling with mass can be expected to be close to
  self-similar on theoretical grounds, but have not yet been
  constrained observationally. In this regime, clusters are
  dynamically younger, characterized by recent mergers, departures
  from equilibrium, and a significant degree of gas
  clumping. Significant progress is expected in the near future due to
  a combination of high--sensitivity SZ and X--ray observations using
  the next generation of instruments.
\end{itemize}
The physical origin of the regular scaling with mass is the fact that
cluster mass is the key control variable of cluster formation, which
sets the amount of gas mass, the average temperature of the ICM, etc.
It is important to note that the close to self-similar scaling with
mass outside the cluster core does not imply that the non-gravitational
physical processes are negligible in this regime. For instance,
\cite{sun_etal09} showed that entropy measured at $r_{500}$ has a
scaling with temperature quite close to the self--similar prediction,
yet its level is higher than expected from a simple model in which
only gravity determines the evolution of the intra--cluster
baryons. This implies that whatever mechanism one invokes to account
for such an entropy excess, it must operate in such a way as to not
violate the self--similar scaling. The scatter around the mean profile
exhibited by clusters at different radii can be due to a number of
factors. In particular, the small scatter at intermediate radii is a
non-trivial fact, given that different clusters of the same mass are
in different stages of their dynamical evolution and physical
processes affecting their profiles may have operated differently due
to different formation histories.

One of the interesting implications of the small scatter in the
pressure profiles is that it provides an upper limit on the
contribution of non--thermal pressure support or, at least, on its
cluster-by-cluster variation. A well-known source of non--thermal
pressure is represented by residual gas motions induced by mergers,
galaxy motions, and gas inflow along large-scale
filaments. Cosmological hydrodynamical simulations of cluster
formation have been extensively used to quantify the pressure support
contributed by gas motions and the corresponding level of violation of
HE
\citep[e.g.,][]{rasia_etal04,nagai_etal07b,ameglio_etal09,piffaretti_valda08,lau_etal09,biffi_etal11a}.
All these analyses consistently found that ICM velocity fields
contribute a pressure support of about $5\%$ to $15\%$ per cent of the
thermal one, the exact amount depending on the radial range considered
(being larger at larger radii) and on the dynamical state of the
clusters. Currently, there are only indirect indications for turbulent
motions in the ICM of the real clusters from fluctuations of gas
density measured in X--ray observations
\citep[e.g.,][]{schuecker_etal04,churazov_etal12}. Direct measurements
or upper limits on gas velocities and characterization of their
statistical properties should be feasible with future high--resolution
spectroscopic and polarimetric instruments on the next-generation
X--ray telescopes
\citep[e.g.,][]{inogamov_sunyaev03,zhuravleva_etal10}.

The galaxies and groups orbiting or infalling onto clusters not only
stir the gas, but also make the ICM clumpier. The dense inner regions
of clusters ram-pressure strip the gas on a fairly short time scale,
so that the clumping is fairly small near cluster cores. However, it is
substantial in the outskirts in cluster simulations where 
orbital times are longer and accretion of new galaxies
and groups is ongoing. Given that the X-ray emissivity scales as the square of
the local gas density, the clumpiness can bias the measurement of gas
density from X--ray surface brightness profiles toward higher values if it is not
accounted for. Because clumping is expected to increase with increasing
cluster-centric radius, the inferred slope of gas density profiles can
be underestimated, thus affecting the resulting pressure profile and
hydrostatic mass estimates.  Furthermore, gas clumping also
affects X--ray temperature which is measured by fitting the X--ray
spectrum to a single--temperature plasma model
\citep{mazzotta_etal04,vikhlinin06}. Clumping can therefore contribute
to the scatter of pressure profiles at large radii, especially at
$r>r_{500}$ \citep[e.g.,][]{nagai_lau11}.

Indirect detections of gas clumping through X--ray observations out to
$r_{200}$ have been recently claimed, based on {\em Suzaku}
observations of a flattening in the X--ray surface brightness profiles
at such large radii (\citealt{simionescu_etal11}). However, these
results are prone to significant systematic uncertainties
\citep{ettori_molendi11}. Independent analyses based on the ROSAT data
\citep[e.g.][]{eckert_etal11}) show the
surface brightness profiles steepens beyond $r_{500}$ \citep[see
also][]{vikhlinin_etal99,neumann_etal05}, which is inconsistent with the degree of 
gas clumping inferred from the {\em
  Suzaku} data, but consistent with predictions of
hydrodynamical simulations. 

Clearly, the clumpiness of the ICM depends on a number of uncertain
physical processes, such as efficient feedback, which removes gas from
merging structures, or thermal conduction, which homogenizes the ICM
temperatures \citep[e.g.,][]{dolag_etal04}.  The degree of gas
clumping in density and temperature is therefore currently uncertain
in both theoretical models and observations. Future high-sensitivity
SZ observations of galaxy clusters with improved angular resolution
will allow a direct measurement of projected pressure profiles. Their
comparison with X--ray derived profiles will help in understanding the
impact of gas clumping on the thermal complexity of the ICM.

Additional non-thermal pressure support can be provided by the
magnetic fields and relativistic cosmic rays, the presence of which in
the ICM is demonstrated by radio observations of the radio halos:
diffuse and faint radio sources filling the central Mpc$^3$ region of
many galaxy clusters \citep[e.g.,][]{giovannini_etal09,venturi_etal08}
arising due to the synchrotron emission of highly relativistic
electrons moving in the ICM magnetic fields. The origin of these
relativistic particles still needs to be understood, although several
models have been proposed. Shocks and turbulence associated with
merger events are expected to compress and amplify magnetic fields and
accelerate relativistic electrons (see, e.g., \citealt{ferrari_etal08}
and \citealt{dolag_etal08} for reviews).  Numerical simulations
including injection of cosmic rays from accretion shocks and SN
explosions \citep[e.g.,][]{pfrommer_etal07,vazza_etal12} indicate that
cosmic rays contribute a pressure support, which can be as high as $\sim
10\%$ for relaxed clusters and $\sim 20\%$ for unrelaxed clusters at
the outskirts. At smaller radii, the pressure contribution of cosmic rays in
these models becomes small ($\lesssim 3\%$ at $r\lesssim 0.1r_{\rm
  vir}$), which is consistent with the upper limits from $\gamma$--ray
observations by the {\em Fermi Gamma-ray Telescope} \citep[e.g.,][]{ackermann_etal10}.

The role of intracluster magnetic fields have been investigated in a
number of studies using cosmological simulations \citep[see][for a
review]{dolag_etal08}.  The general result is that pressure support
from magnetic fields should be limited to $\mincir 5\%$, which is
consistent with observational constraints on the magnetic field
strength ($\sim \mu G$) \citep[e.g.,][]{vogt_ensslin05,govoni_etal10}
and upper limits on the contribution of magnetic fields to non-thermal
pressure support \citep[e.g.][]{lagana_etal10}.

As a summary, the scatter in cluster profiles in the cluster cores is
mainly driven by differences in the physical processes such as cooling
and heating by AGN feedback and different merger activity that
different clusters experienced during their evolution. At intermediate
radii, the scatter is small because the ICM is generally in good
HE within cluster gravitational potential and
because processes that shaped its thermodynamic processes have not
introduced new mass scale so that self-similar scaling is not
broken. In the cluster outskirts, the scatter is expected to be driven
by deviations from HE and other sources of
non-thermal pressure support such as the cosmic rays, as well as 
by a rapid increase of ICM clumpiness with increasing radius.

\subsection{Scaling relations}
\label{sec:scal}

Existence of the radial range, where ICM properties scale with mass
similarly to the self-similar expectation with a small scatter,
implies that we can define integral observable quantities within this
range that will obey tight scaling relations among themselves and with
the total cluster mass. Furthermore, these scaling relations are also
expected to be weakly sensitive to the cluster dynamical state, given
that relaxed and unrelaxed clusters have similar profiles at these
intermediate radii. Indeed, as we showed in \S~\ref{sec:obsprop} (see
Fig. \ref{fig:ly_rexcess}), X--ray luminosity measured within the
radial range $[0.15-1]r_{500}$ exhibits a tight scaling against the
total ICM thermal content measured by the $Y_X$ parameter, with
relaxed and unrelaxed clusters following the same relation. Here $Y_X$
is defined as the product of gas mass and X--ray temperature, both
measured within $r_{500}$ but like $L_{\rm X}$ temperature is measured
after excising the contribution from $r<0.15r_{500}$.

\begin{figure}
 \centerline{ 
\psfig{file=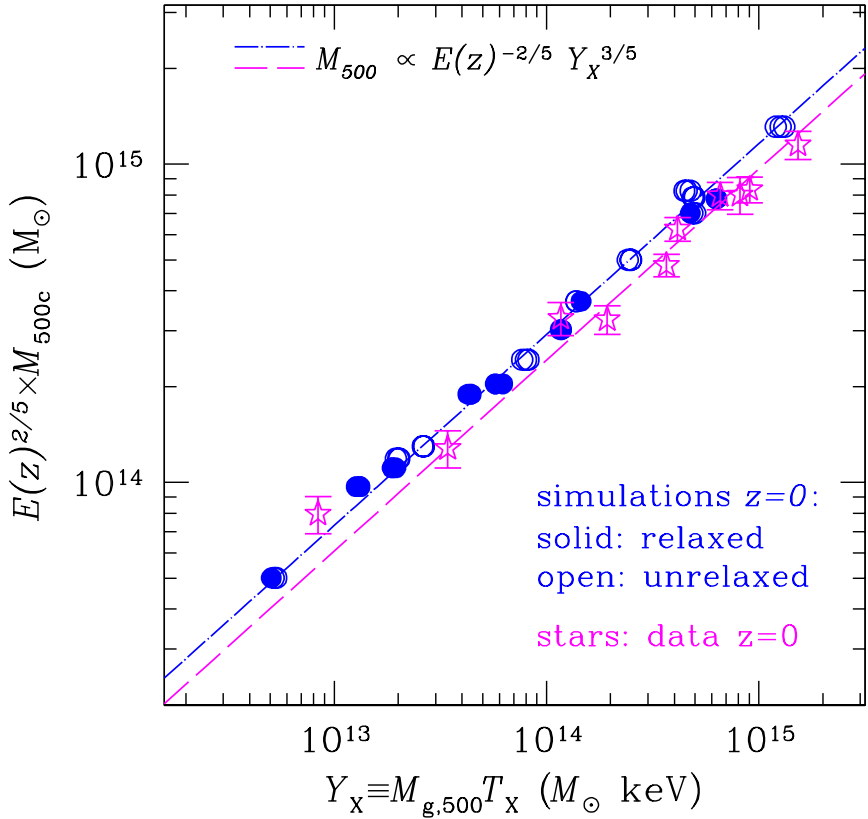,width=13.truecm}
}
\caption{The $Y_X$--$M_{500}$ relation for a set of simulated clusters
  at z = 0 (circles) and for a sample of relaxed Chandra clusters from
  \protect\cite{vikhlinin_etal06} (stars with errorbars). Filled and
  open circles refer to simulated clusters which are classified as
  relaxed and unrelaxed, respectively. Core regions inside
  $0.15r_{500}$ are excised in the measurement of the X-–ray
  temperature entering into the computation of $Y_X$, for both
  simulated and real clusters. True and hydrostatic masses are shown
  for simulated and observed clusters, respectively.  The dot-dashed
  line shows the best-fit power-law relation for the simulated
  clusters with the slope fixed to the self--similar value of $alpha =
  3/5$. The dashed line shows the same best-fit power-law relation to
  simulations, but with the normalization scaled down by 15\%, which
  takes into account the putative effect of hydrostatic mass bias due
  to residual gas motions. From \protect\cite{kravtsov_etal06}.}
\label{fig:yx_k06}
\end{figure}

As discussed in \S~\ref{sec:selfsimilar}, the gas temperature $T$, gas
mass $M_{\rm gas}$ and total thermal content of the ICM $Y=M_{\rm
  gas}T$, are commonly used examples of integral observational
quantities whose scaling relations with cluster mass are predicted by
the self--similar model and for which calibrations based on X--ray and
SZ observations (or their combinations) and simulations are available.
For example, Figure \ref{fig:yx_k06} shows the scaling relation
between $Y_X$, and $M_{500}$ for simulated clusters and for a set of
clusters with detailed Chandra observations from a study by
\cite{kravtsov_etal06}, where $Y_{\rm X}$ was introduced and defined
specifically to use the temperature estimated only at
$0.15r_{500}<r<r_{500}$ in order to minimize the scatter. The relation
of $Y_{\rm X}$ with $M_{500}$ in simulations has scatter of only
$\approx 8\%$ when both relaxed and unrelaxed clusters are included
and evolution of its normalization with redshift is consistent with
expectations of the self--similar model. The insensitivity of the
relation the dynamical state of clusters is not trivial and is due to
the fact that during mergers clusters move almost exactly along the
relation \citep[e.g.,][]{poole_etal07,rasia_etal11}.  In addition, the
slope and normalization of the $Y_{\rm X}-M_{500}$ relation is also
not sensitive to specific assumptions in modelling cooling and
feedback heating processes in simulations
\citep{stanek_etal10,fabjan_etal11}, which makes them more robust
theoretically.

The $Y_{SZ}$--$M$ relation also exhibits a comparably low--scatter and
the slope and evolution of normalization are close to the predictions
of the self--similar model \citep{dasilva_etal04,motl_etal05}, which
is not surprising given the similarity between the $Y_{\rm X}$ and the
integrated $Y_{\rm SZ}$ measured from SZ observations. Its
normalization changes by up to 30--40\% depending on the
interplay between radiative cooling and feedback processes included in
the simulations \citep[e.g.][and references
therein]{nagai06,bonaldi_etal07,battaglia_etal11}. At the same time,
simulation analysis is also shedding light on the effect of projection
\citep[e.g.][]{kay_etal12} and mergers \citep[e.g.][]{krause_etal12}
on the scatter in the $Y_{SZ}$--$M$ scaling.

The tight relation of integral quantities such as $Y_{\rm X}$, $Y_{\rm
  SZ}$, $\Mg$, or core-excised X-ray luminosity with the total mass
makes them good proxies for observational estimates of cluster mass,
which can be used at high redshifts even with a relatively small
number of X-ray photons. For instance, integral measurements of gas
mass or temperature requires $\sim 10^3$ photons, which is feasible
for statistically complete cluster samples out to $z\sim 1$
\citep[e.g.,][]{maughan07,vikhlinin_etal09,mantz_etal10a} or even
beyond.  This makes these integral quantities very useful as ``mass
proxies'' in cosmological analyses of cluster populations
\cite[e.g.,][]{allen_etal11}. Clearly, the relation of such mass
proxies with the actual mass needs to be calibrated both via detailed
observations of small controlled cluster samples and in cosmological
simulations of cluster formation.

\begin{figure}
\centerline{ 
\psfig{file=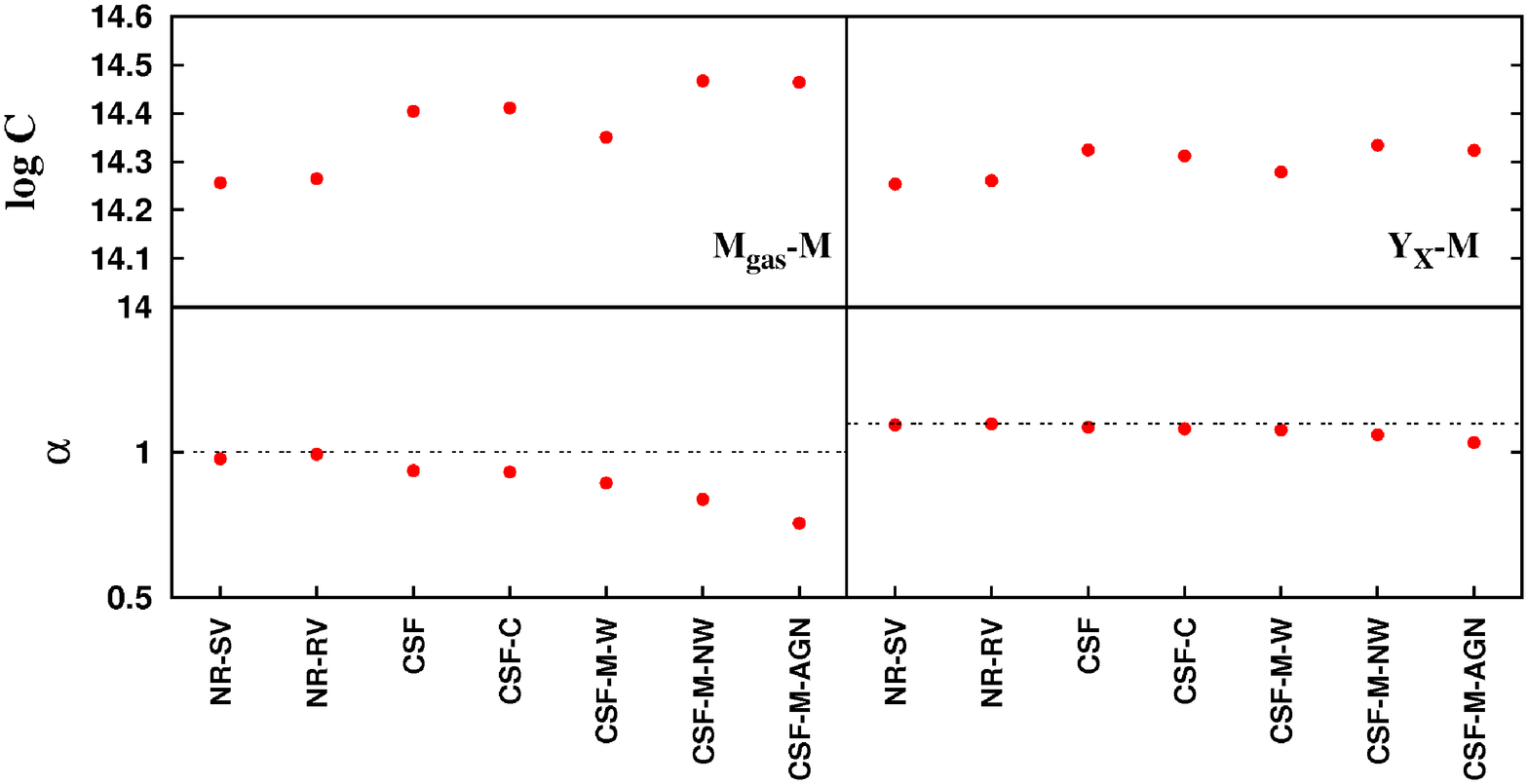,width=14.truecm,angle=-90}
}
\vspace{-0.5cm}
\caption{Sensitivity of different mass proxies on the physical
  description of the intracluster medium included in cosmological
  hydrodynamical simulations for a set of galaxy clusters
  \protect\citep[from][]{fabjan_etal11}. Results for the scaling
  relation of $M_{500}$ with gas mass $M_{gas}$ and $Y_X$ are shown in
  the left and right panels, respectively. Best fitting normalization
  $C$ and slope $\alpha$ of the scaling relations of these three mass
  proxies are shown in the upper and lower panels, respectively. Here,
  $T_{\rm mw}$ is the mass-weighted temperature computed excluding the
  central cluster regions within $0.15r_{500}$. In the lower panels,
  the horizontal dashed lines mark the values of the slope of the
  scaling relations predicted by the self--similar model. Results are
  shown for simulations only including non–-radiative hydrodynamics
  with standard (NR-SV) and reduced artiﬁcal viscosity (NR-RV);
  cooling and star formation without (CSF) and with thermal conduction
  (CSF-C); cooling and star formation with metal enrichment, with
  (CSF-M-W) and without (CSF-M-NW) galactic winds from SN explosions;
  and cooling and star formations with the eﬀect of AGN feedback
  (CSF-M-AGN) (see \protect\citealt{fabjan_etal11} for further
  details).}
\label{fig:scal_f11}
\end{figure}

The potential danger of relying on simulations for this calibration is
that results could be sensitive to the details of the physical
processes included.  This implies that a mass proxy is required to
have not only a low scatter in its scaling with mass, but also to be
robust against changing the uncertain description of the ICM
physics. As we noted above, $Y_{\rm X}$ is quite robust to changes
within a wide range of assumptions about cooling and heating processes
affecting the ICM. This is illustrated in Figure \ref{fig:scal_f11}
(taken from \citealt{fabjan_etal11}) which shows how the normalization
and slope of the scaling relation of gas mass and $Y_X$ versus
$M_{500}$ change with the physical processes included. The evolution
of the $Y_{\rm X}-M_{500}$ relation with redshift is also consistent
with self-similar expectations for different models of cooling and
feedback.  Other quantities, such as $\Mg$, often exhibit a similar or
even smaller degree of scatter compared to $Y_{\rm X}$ but are more
sensitive to the choice of physical processes included in
simulations. An additional practical consideration is that theoretical
models should consider observables derived from mock observations of
simulated clusters that take into account instrumental effects of
detectors and projection effects
\citep[e.g.,][]{rasia_etal06,nagai_etal07a,biffi_etal11b}.

Ultimately, calibration of mass proxies for precision use should be
obtained via independent observational mass measurements, using the
weak lensing analysis, HE, or velocity dispersions of member
galaxies. The combination of future large, wide-area X--ray, SZ, and
optical/near-IR surveys should provide a significant progress in this
direction.

\section{Cluster formation in alternative cosmological models} 
\label{sec:cosmo}

In previous sections, we have discussed the main elements of
cluster formation in the standard $\Lambda$CDM cosmology. Although
this model is very successful in explaining a wide variety of
observations, some of its key assumptions and ingredients are not yet
fully tested. This provides motivation to explore different assumptions
and alternative models.  

As discussed in Section \ref{sec:mf}, the halo mass function for a
Gaussian random field is uniquely specified by the peak height
$\nu=\delta_c/\sigma(R,z)$, where $R$ is the filtering scale
corresponding to the cluster mass scale $M$. For sufficiently large
mass, that is rare peaks with $\nu\gg 1$, the mass function becomes
exponentially sensitive to the value of $\nu$. At the same time, the
mass function also determines the halo bias (see Section
\ref{sec:bias}). Again, for $\nu \gg 1$ and Gaussian perturbations,
the bias function scales as $b(\nu)\sim
\nu^2/\delta_c=\nu/\sigma(R,z)$. Therefore, the cluster 2--point
correlation function can be written as $\xi_{\rm
  cl}(r)=\nu^2(\xi_R(r)/\sigma_R^2)$, where $\xi_R(r)$ is the
correlation function of the smoothed fluctuation field (see Section
\ref{sec:collbasics}). Once the peak height $\nu$ is constrained by
requiring a model to predict the observed cluster abundance, the
value of the cluster correlation function at a single scale $r$ provides a
measurement of the shape of the power spectrum through the ratio of
the clustering strength at the scale $r$ and at the cluster
characteristic scale $R$. These predictions are only valid under two
assumptions, namely Gaussianity of primordial density perturbations
and scale independence of the linear growth function $D(z)$, as
predicted by the standard theory of gravity. Therefore, the combination of
number counts and large--scale clustering studies offers a powerful means
to constrain the possible violation of either one of these two
assumptions that hold for the $\Lambda$CDM model.

In this section, we briefly review the specifics of cluster formation in
models with non-Gaussian initial density field and with
non-standard gravity, the most frequently discussed modifications to
the standard structure formation paradigm.

\subsection{Mass function and bias of clusters in 
  non-Gaussian models}
\label{sec:nong}
One of the key assumptions of the standard model of structure
formation is that initial density perturbations are described by a
Gaussian random field (see Section~\ref{sec:collbasics}).  The
simplest single-field, slow-roll inflation models predict nearly
Gaussian initial density fields.  However, deviations from Gaussianity
are expected in a broad range of inflation models that violate
slow-roll approximation, and have multiple fields, or modified kinetic
terms \citep[see][for a review]{bartolo_etal04}. Given that there is
no single preferred inflation model, we do not know which specific
form of non-Gaussianity is possibly realized in nature.  Deviations
from Gaussianity are parameterized using a heuristic functional
form. One of the simplest and most common choices for such a form is the
local non-Gaussian potential given by $\Psi_{\rm NG}(\bx) =
-(\phi_{\rm G}(\bx) +f_{\rm NL}[\phi_{\rm G}(\bx)^2 - \langle\phi_{\rm
  G}^2\rangle])$, where $\Psi_{\rm NG}$ is the usual Newtonian
potential, $\phi_{\rm G}$ is the Gaussian random field with zero mean,
and the parameter $f_{\rm NL}=\rm const$ controls the degree and
nature of non--Gaussianity
\citep[e.g.,][]{salopek_bond90,matarrese_etal00,komatsu_spergel01}. The
simplest inflation models predict $f_{\rm NL}\approx 10^{-2}$
\citep[e.g.,][]{maldacena03}, but a number of models that predict much
larger degree of non-Gaussianity exist as well
\citep{bartolo_etal04}. The current CMB constraint on
scale-independent non-Gaussianity is $f_{\rm NL}=30\pm 20$ at the 68\%
confidence level
\citep[e.g.,][]{komatsu10} and there is thus still room for existence
of sizable deviations from Gaussianity.

The non-Gaussian fields with $f_{\rm NL}<0$ have a PDF of the
potential field that is skewed toward positive values and the
abundance of peaks that seed the collapse of halos is reduced compared
to Gaussian initial conditions. Conversely, the PDF of the potential
field in models with $f_{\rm NL}>0$ has negative skewness, and hence
an increased number of potential minima (density peaks). This would
result in an enhanced abundance of rare objects, such as massive
distant clusters, relative to the Gaussian case \citep[see, e.g.,
figure 1 in][for an illustration of the effect of $f_{\rm NL}$ on the
large-scale structure that forms]{dalal_etal08}. The suppression or
enhancement of abundance of halos increases with increasing peak
height.

The mass functions resulting from non-Gaussian initial
conditions have been studied both analytically
\citep[e.g.,][]{chiu_etal98,matarrese_etal00,loverde_etal08,afshordi_tolley08}
and using cosmological simulations
\citep{grossi_etal07,dalal_etal08,loverde_etal08,loverde_smith11,wagner_verde11}. These
studies showed that accurate formulae for the halo abundance from the
initial linear density field exist for the non-Gaussian models as
well. The general result is that the fractional change in the abundance of the rarest peaks is of order unity for the initial fields with $\vert f_{\rm NL}\vert\sim 100$. The abundance of clusters is thus only mildly sensitive to deviations of Gaussianity within the currently constrained limits
\citep{scoccimarro_etal04,sefusatti_etal07,sartoris_etal10,cunha_etal10}.
In contrast,  primordial non-Gaussianity may also leave an imprint in the spatial
distribution of clusters in the form of a scale-dependence of
large-scale linear bias.
  
As was discovered by \citet{dalal_etal08} and confirmed in subsequent
analytical
\citep{matarrese_verde08,mcdonald08,afshordi_tolley08,taruya_etal08,slosar_etal08}
and numerical studies
\citep{desjacques_etal09,pillepich_etal10,grossi_etal09,shandera_etal11},
the linear bias of collapsed objects in the models with {\it local}
non-Gaussianity can be described as a function of wavenumber $k$ by
$b_{\rm NG}=b_{\rm G}+ f_{\rm NL}\times{\rm\, const}/k^2$, where
$b_{\rm G}$ is the linear bias in the corresponding cosmological model
with the Gaussian initial conditions discussed in
\S~\ref{sec:bias}. This scale dependence arises because in the
non-Gaussian models the large-scale modes that boost the abundance of
peaks are correlated with the peaks themselves, which enhances (or
suppresses) the peak amplitudes by a factor proportional to $f_{\rm
  NL}\phi\propto f_{\rm NL}\delta/k^2$. Because this effect of
modulation increases with increasing peak height,
$\nu=\delta_c/\sigma(M,z)$, the scale--dependence of bias increases
with increasing halo mass. This unique signature can be used as a
powerful constraint on deviations from Gaussianity (at least for
models with {\it local} non-Gaussianity) in large samples of clusters
in which the power spectrum or correlation function can be measured on
large scales.

\subsection{Formation of clusters in modified gravity models}
\label{sec:nongr}

Recently, there has been a renewed interest in modifications to the
standard GR theory of gravity \citep[e.g., see][for recent
reviews]{capozziello_delaurentis11,durrer_maartens08,silvestri_trodden09}.
These models have implications not only for cosmic expansion, but also
for the evolution of density perturbations and, therefore, for the
formation of galaxy clusters.

For instance, in the class of the $f(R)$ models, cosmic acceleration
arises from a modification of gravity law given by the addition of a
general function $f(R)$ of the Ricci curvature scalar $R$ in the
Einstein-Hilbert action \citep[see, e.g.,][for recent
reviews]{sotiriou_faraoni10,jain_khoury10}. Such modifications result
in enhancements of gravitational forces on scales relevant for
structure formation in such a way that the resulting linear
perturbation growth rate $D$ becomes scale dependent; whereas on very
large scales gravity behaves similarly to GR gravity, on smaller
scales it is enhanced compared to GR and the rate of structure
formation is thereby also enhanced. The nonlinear halo collapse and
growth are also faster in $f(R)$ models, which leads to enhanced
abundance of massive clusters
\citep{schmidt_etal09a,ferraro_etal11,zhao_etal11} compared to the
predictions of the models with GR gravity and identical cosmological
parameters. Likewise, the peaks collapsing by a given $z$ have lower
peak height $\nu$ in the modified gravity models compared to the peak
height in the standard gravity model. This results in the reduced bias
of clusters of a given mass compared to the standard
model. 
Furthermore, the scale dependence of the linear growth also induces a
scale dependence of bias, thus offering another route to detect
modifications of gravity \citep{parfrey_etal11}.
Qualitatively similar effects on cluster abundance and bias are
expected in the braneworld-modified gravity models based on higher
dimensions, such as the Dvali-Gabadadze-Porrati \citep[DGP,][]{DGP}
gravity model
\citep{schaefer_koyama08,khoury_wyman09,schmidt09,schmidt_etal10} and
its successors with similar LSS phenomenology consistent with current
observational constraints, such as models of ghost-free massive
gravity \citep{derham_etal11,damico_etal11}.

A general consequence of modifying gravity is that the Birkhoff
theorem no longer holds, which does not allow a straightforward
extension of the spherical collapse model described in Section
\ref{sec:sphcoll} to a generic model of modified
gravity. Nevertheless, numerical calculations of spherical collapse
have been presented for a number of specific models
\citep[e.g.,][]{schaefer_koyama08,schmidt_etal09a,schmidt_etal10,martino_etal09}.
For both the $f(R)$ and the DGP classes of models, the results of
simulations obtained so far suggest that halo mass function and bias
can still be described by the universal functions of peak height, in
which the threshold for collapse and the linear growth rate are modified
appropriately from their standard model values
\citep{schmidt_etal09a,schmidt_etal10}. This implies that it should be
possible to calibrate mass function and bias of halos in the modified
gravity models with the accuracy comparable to that in the standard
structure formation models.

\begin{figure}[tb]
\centerline{
\psfig{file=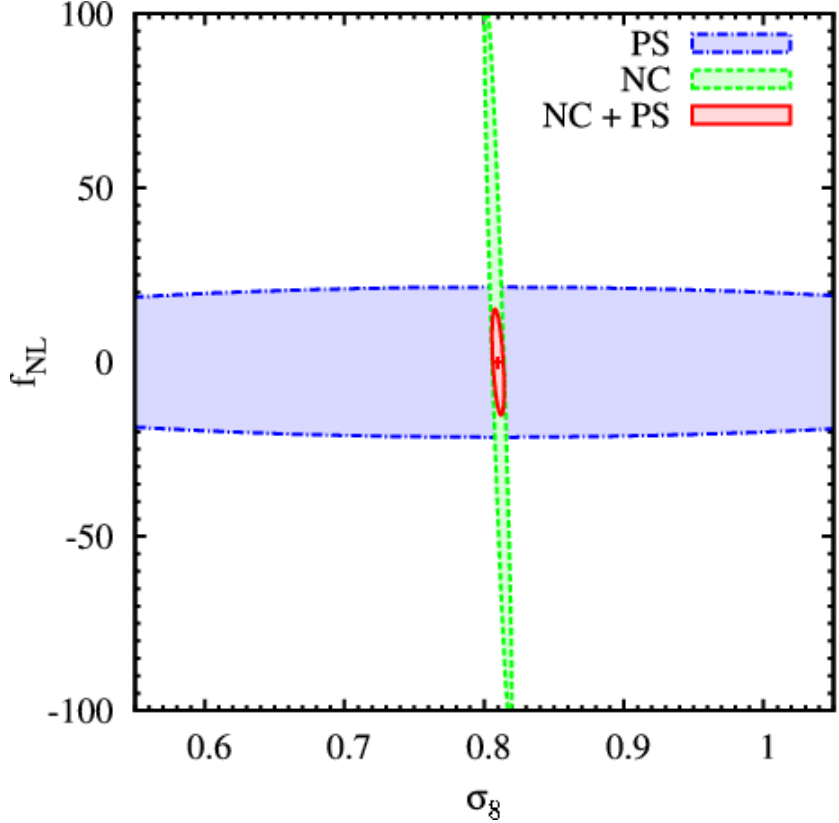,width=11.truecm}}
\caption{The potential of future cluster X--ray surveys to constrain
  deviations from Gaussian density perturbations (adapted from
  \protect\citealt{sartoris_etal10}). The figure shows constraints on
  the power--spectrum normalization, $\sigma_8$, and non--Gaussianity
  parameter, $f_{\rm NL}$, expected from surveys of galaxy clusters to
  be carried out with the next--generation Wide Field X--ray
  Telescope. Dot-dashed blue curve and dashed green curve show the
  68\% confidence regions provided by the evolution of power spectrum
  (PS) of the cluster distribution and cluster number counts (NC),
  respectively. The solid red ellipse shows the constraints obtained
  by combining number counts and power spectrum information. Cocmic
  Microwave Background {\em Planck} priors for Gaussian perturbations
  have been included in the analysis.}
\label{fig:nong}
\end{figure}

\begin{figure}
\centerline{
\psfig{file=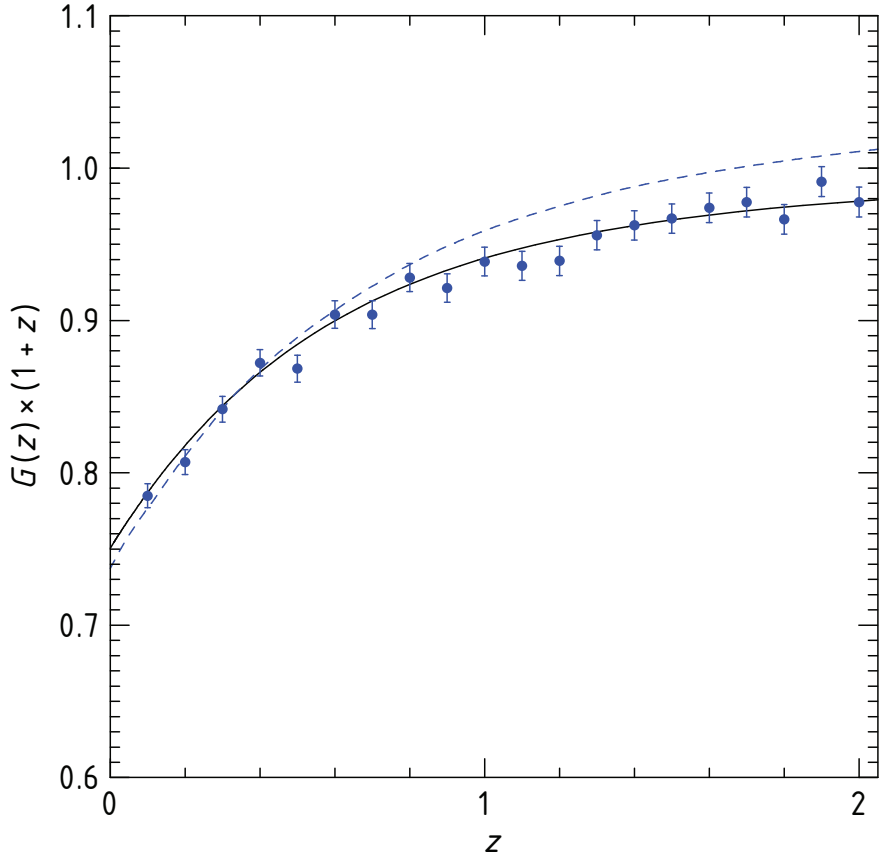,width=13.truecm}}
\caption{The potential of future cluster surveys to constrain
  deviations from General Relativity (from
  \protect\citealt{vikhlinin_wfxt09}). The linear growth factor of
  density perturbations, $G(z)=D(z)$ (not normalized to unity at
  $z=0$), recovered from 2000 clusters, distributed in 20 redshift
  bins, each containing 100 massive clusters, identified in a
  high-sensitivity X--ray cluster survey. The solid black line
  indicates the evolution of the linear growth factor for a
  $\Lambda$CDM model, whereas the dashed blue curve is the prediction
  of a modified gravity model (the brane world model by
  \protect\citealt{DGP}), having the same expansion history of the
  $\Lambda$CDM model.}
\label{fig:modgrav}
\end{figure}

\section{Summary and outlook}

All of the main elements of the overall narrative of how clusters form
and evolve discussed in this review have been established over the
past four decades. The remarkable progress in our understanding of
cluster formation has been accompanied by great progress in
multi-wavelength observations of clusters and our knowledge of
the properties of the main mass constituents of clusters: stars, hot
intracluster gas, and gravitationally dominant DM.

Formation of galaxy clusters is a complicated, non-linear process
accompanied by a host of physical phenomena on a wide range of
scales. Yet, some aspects of clusters exhibit remarkable regularity,
and their internal structure, abundance, and spatial distribution
carry an indelible memory of the initial linear density perturbation
field and the cosmic expansion history.  This is manifested both by
tight scaling relations between cluster properties and the total mass,
as well as by the approximate universality of the cluster mass
function and bias, when expressed as a function of the peak height
$\nu$.
   
Likewise, there is abundant observational evidence that complex
processes -- in the form of a non-linear, self-regulating cycle of gas
cooling and accretion onto the SMBHs and associated
feedback -- have been operating in the central regions of clusters.  In
addition, the ICM is stirred by continuing accretion of the
intergalactic gas, motion of cluster galaxies, and AGN bubbles.
Studies of cluster cores provide a unique window into the interplay
between the evolution of the most massive galaxies, taking place under
extreme environmental conditions, and the physics of the diffuse hot
baryons. At the same time, processes accompanying galaxy formation also
leave a mark on the ICM properties at larger radii. In these
regions, the gas entropy measured from observations is considerably
higher than predicted by simple models that do not include such
processes, and the ICM is also significantly enriched by heavy
elements. This highlights that the ICM properties are the end
product of the past interaction between the galaxy evolution processes
and the intergalactic medium.
Nevertheless,  at intermediate radii, 
$r_{2500}\lesssim r\lesssim r_{500}$, the scaling of the radial profiles of gas
density, temperature, and pressure with the total mass is close to simple, self-similar
expectations for clusters of sufficiently large mass (corresponding to
$kT\gtrsim 2-3$~keV). This implies that the baryon processes affecting the
ICM during cluster formation do not introduce a new mass scale. Such
regular behaviour of the ICM profiles provides a basis for the
definition of integrated quantities, such as the core-excised X-ray
luminosity and temperature, gas mass, or integrated pressure, which
are tightly correlated with each other and with the total cluster mass.

The low-scatter scaling relations are used to interpret abundance and
spatial distribution of clusters and derive cosmological constraints
(see \citealt{allen_etal11} and \citealt{weinberg_etal12} for recent
reviews). Currently, cluster counts measured at high redshifts provide
interesting constraints on cosmological parameters complementary to
other methods
\citep[e.g.,][]{vikhlinin_etal09b,mantz_etal10b,rozo_etal10} and a
crucial test of the entire class of $\Lambda$CDM and quintessence
models
\citep[e.g.,][]{jee_etal11,benson_etal11,mortonson_etal11}. Although
the statistical power of large future cluster surveys will put
increasingly more stringent requirements on the theoretical
uncertainties associated with cluster scaling relations and mass
function \citep{cunha_evrard10,wu_etal10}, future cluster samples can
provide competitive constraints on the non-Gaussianity in the initial
density field and deviations from GR gravity.

A combination of cluster abundance and large--scale clustering
measurements can be used to derive stringent constraints on
cosmological parameters and possible deviations from the standard
$\Lambda$CDM paradigm. As an example, Figure \ref{fig:nong} shows the
constraints on the normalization of the power spectrum and the $f_{\rm
  NL}$ parameter, \citep[from][]{sartoris_etal10} expected for a
future high--sensitivity X-ray cluster survey.  It shows that future
cluster surveys can achieve a precision of $\sigma_{f_{\rm NL}}\approx
5-10$ \citep[see also][]{cunha_etal10,pillepich_etal12}, thus
complementing at smaller scales constraints on non--Gaussianity, which
are to be provided on larger scales by observations of CMB
anisotropies from the Planck satellite.

Although a variety of methods will provide constraints on the equation
of state of DE and other cosmological parameters
\citep[e.g.,][]{weinberg_etal12}, clusters will remain one of the most
powerful ways to probe deviations from the GR gravity
\cite[e.g.,][]{lombriser_etal09}. Even now, the strongest constraints
on deviations from the GR on the Hubble horizon scales are derived
from the combination of the measured redshift evolution of cluster
number counts and geometrical probes of cosmic expansion
\citep{schmidt_etal09}. Figure \ref{fig:modgrav} illustrates the
potential constraints on the linear rate of perturbation growth that
can be derived from a future high--sensitivity X--ray cluster survey
using similar analysis. The figure shows that a sample of about 2000
clusters at $z< 2$ with well-calibrated mass measurements would allow
one to distinguish the standard $\Lambda$CDM model from a
braneworld--modified gravity model with the identical expansion
history at a high confidence level.

The construction of such large, homogeneous samples of clusters will
be aided in the next decade by a number of cluster surveys both in the
optical/near-IR (e.g., DES, PanSTARRS, EUCLID) and X-ray (e.g.,
eROSITA, WFXT) bands. At the same time, the combination of higher
resolution numerical simulations including more sophisticated
treatment of galaxy formation processes and high--sensitivity
multi-wavelength observations of clusters should help to unveil the
nature of the physical processes driving the evolution of clusters and
provide accurate calibrations of their masses. The cluster studies
thus will remain a vibrant and fascinating area of modern cosmology
for years to come.

\bigskip
{\bf{\Large{Acknowledgments}}}

\noindent
We are grateful to Brad Benson, Klaus Dolag, Surhud More, Piero
Rosati, Elena Rasia, Ming Sun, Paolo Tozzi, Alexey Vikhlinin, Mark Voit, and Mark Wyman for
useful discussions and comments, and to John Carlstrom for a careful
reading of the manuscript. We thank Dunja Fabjan and Barbara Sartoris
for their help in producing Fig. \ref{fig:scal_f11} and
Fig. \ref{fig:nong}, respectively. The authors wish to thank the Kavli
Institute for Theoretical Physics (KITP) in Santa Barbara for
hospitality during the early phase of preparation of this review and
participants of the KITP workshop ``Galaxy clusters: crossroads of
astrophysics and cosmology'' for many stimulating discussions. AK was
supported by NSF grants AST-0507596 and AST-0807444, NASA grant
NAG5-13274, and by the Kavli Institute for Cosmological Physics at the
University of Chicago through grants NSF PHY-0551142 and
PHY-1125897. SB acknowledges partial support by the European
Commission’s FP7 Marie Curie Initial Training Network CosmoComp
(PITN-GA-2009-238356), by the PRIN-INAF09 project ``Towards an Italian
Network for Computational Cosmology'', by the PRIN-MIUR09 ``Tracing
the growth of structures in the Universe'' and by the PD51 INFN grant.

\clearpage
\bibliography{araa_final}

\end{document}